\documentclass[11pt]{article}
\pdfoutput=1
\usepackage{jheppub}
\usepackage{graphicx}
\usepackage{color}
\usepackage{bm}

\graphicspath{{./figures/}}

\newcommand{\pd}{{\partial}}
\newcommand{\ol}{\overline}

\newcommand{\mb}{\mathbf}

\newcommand{\half}{I\!\!I}

\newcommand{\be}{\begin{equation}}
\newcommand{\ee}{\end{equation}}
\newcommand{\bp}{\begin{pmatrix}}
\newcommand{\ep}{\end{pmatrix}}
\newcommand{\bsp}{\left(\begin{smallmatrix}}
\newcommand{\esp}{\end{smallmatrix}\right)}

\newcommand{\C}{{\mathbb C}}
\newcommand{\Z}{{\mathbb Z}}

\newcommand{\CA}{{\mathfrak I}}

\newcommand{\CB}{{\mathcal B}}

\newcommand{\CD}{{\mathcal D}}

\newcommand{\CI}{{\mathcal I}}

\newcommand{\CN}{{\mathcal N}}

\newcommand{\CT}{{\mathcal T}}
\newcommand{\CV}{{\mathcal V}}

\newcommand{\CZ}{{\mathcal Z}}
\newcommand{\CL}{{\mathcal L}}

\title{(0,2) Dualities and the 4-Simplex}
\author[1]{Tudor Dimofte}
\author[2]{Natalie M. Paquette}

\affiliation[1]{Department of Mathematics and Center for Quantum Mathematics and Physics (QMAP), University of California, Davis, CA 95616, USA}
\affiliation[2]{Walter Burke Institute for Theoretical Physics, California Institute of Technology,
Pasadena, CA 91125, USA}

\dedicated{CALT-TH-2019-016}

\abstract{We propose that a simple, Lagrangian 2d $\CN=(0, 2)$ duality interface between the 3d $\CN=2$ XYZ model and 3d $\CN=2$ SQED can be associated to the simplest triangulated 4-manifold: the 4-simplex. We then begin to flesh out a dictionary between more general triangulated 4-manifolds with boundary and 2d $\CN=(0, 2)$ interfaces. In particular, we identify IR dualities of interfaces associated to local changes of 4d triangulation, governed by the (3,3), (2,4), and (4,2) Pachner moves. We check these dualities using supersymmetric half-indices. We also describe how to produce stand-alone 2d theories (as opposed to interfaces) capturing the geometry of 4-simplices and Pachner moves by making additional field-theoretic choices, and find in this context that the Pachner moves recover abelian $\CN=(0,2)$ trialities of Gadde-Gukov-Putrov.
Our work provides new, explicit tools to explore the interplay between 2d dualities and 4-manifold geometry that has been developed in recent years.}

\begin{document}
\today
\maketitle


\section{Introduction}

The past decade has seen incredible progress in physics and mathematics inspired by compactifications of the 6d $(2,0)$ theory (\emph{a.k.a.} the M5-brane worldvolume theory) on manifolds of various dimensions. Some of the first results, concerning compactification on Riemann surfaces, produced new webs of dualities in 4d $\CN=2$ theories, interconnected with wall crossing and cluster algebras \cite{Gaiotto-dualities, GMN, GMN-Hitchin}. Compactifications on 3-manifolds, initially studied in \cite{DGH, Yamazaki, DGG, CCV, DGG-index}, connected 3d $\CN=2$ theories and their dualities to three-dimensional geometry and topology (in particular, hyperbolic geometry), and produced new quantum three-manifold invariants. 
In recent years, this led to a definition of Khovanov-like homological invariants of compact three-manifolds \cite{GPV, GPPV}
(inspired by the original work on knot homology and BPS states in M-theory \cite{GSV}, and its field-theoretic counterpart \cite{Witten-5branes}).
Compactifications on 4-manifolds \cite{GGP} (generalized in \cite{SN4}) incited the discovery of new, fundamental 2d $\CN=(0,2)$ dualities \cite{GGP-trialities}, and have given new structure to Vafa-Witten and Seiberg-Witten invariants \cite{DGP}.

In this paper, we add a short chapter to the 4-manifold story: we describe 2d $\CN=(0,2)$ theories associated to the simple \emph{pieces} of 4-manifolds, namely to ideal 4-simplices --- also known as (ideal) pentachora. We will argue that the pentachoron theory is
\be \label{TD4} \begin{array}{rcl}T[\Delta^4] &\simeq &  \text{two free 2d $\CN=(0,2)$ fermi multiplets} \\[.1cm]
 && =\;\text{two 2d chiral fermions}\,. \end{array} \ee
This is rather similar in form to the 3d $\CN=2$ theory associated to an ideal tetrahedron \cite{DGG}
\be \label{TD3} T[\Delta^3] \;\simeq\; \text{one free 3d $\CN=2$ chiral multiplet}\,. \ee

The identifications \eqref{TD4}--\eqref{TD3} are only interesting insofar as they are supplemented with \emph{gluing rules}, explaining how free fields should be coupled together when the corresponding simplices are glued together. We describe a partial set of gluing rules, and check that two different gluings related by four-dimensional Pachner moves of type (3,3), (2,4), and (4,2) lead to two $\text{IR-dual}$ 2d $\CN=(0,2)$ theories. In particular, we verify the dualities by computing identities of elliptic genera.

Again, this parallels the three-dimensional story quite closely. 
A full set of rules for gluing tetrahedron theories was spelled out in \cite{DGG}. It was found there that the basic Pachner move of type (3,2), which relates ideal triangulations of 3-manifolds, corresponds to the fundamental duality \cite{AHISS} between the 3d  XYZ model (a theory of three chirals with a cubic superpotential) and 3d $\CN=2$ SQED. 

\begin{figure}[htb]
\centering
\includegraphics[width=5.4in]{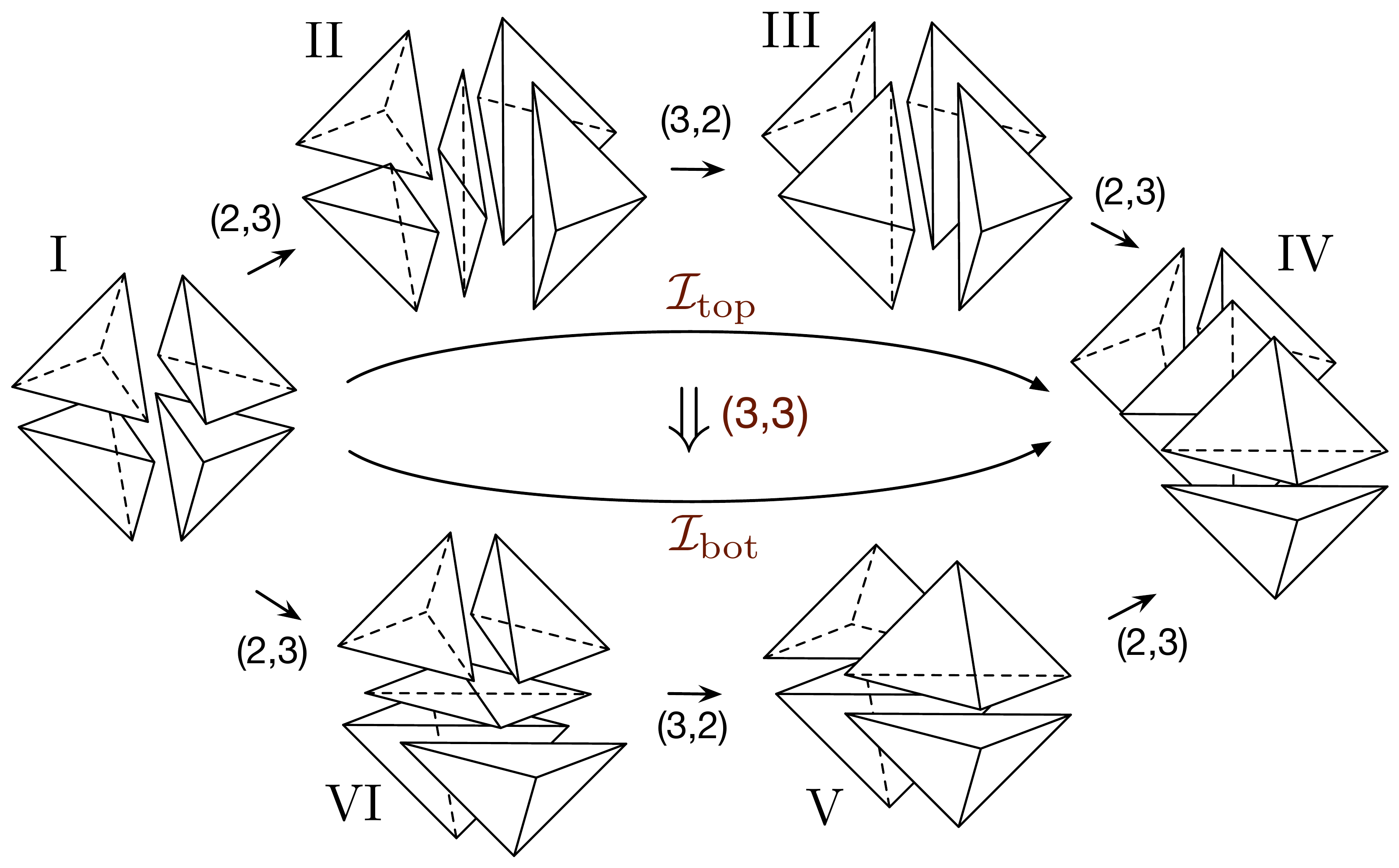}
\caption{The (3,3) Pachner move relates two sequences of (2,3) and (3,2) Pachner moves applied to triangulated octahedra. In field theory, it relates two composite interfaces $\CI_{\rm top}$ and $\CI_{\rm bot}$  between the 3d $\CN=2$ theories I and IV associated to octahedra on the far left and right of this figure.}
\label{fig:oct-intro}
\end{figure}

Amusingly, the 2d $\CN=(0,2)$ dualities that we find associated to 4d Pachner moves recover abelian versions of the more general $\CN=(0,2)$ trialities of \cite{GGP-trialities}. The latter were motivated by applying fundamental moves to handle decompositions of 4-manifolds, as opposed to triangulations.
In particular, the (3,3) Pachner move is based on a simple duality between 1) $\CT_{U(1)}$, a 2d $\CN=(0,2)$ $U(1)$ gauge theory  with two chiral multiplets, three fermi multiplets, and a cubic superpotential; and 2) $\CT_{FF}$, two free fermi multiplets. The Pachner move itself relates two copies of $\CT_{U(1)}$ to two other copies,
\be \CT_{U(1)}\otimes   \CT_{U(1)}' \;\simeq\;  \CT_{U(1)}'' \otimes  \CT_{U(1)}'''\,. \ee
Each side flows to four fermi multiplets $\CT_{FF}\otimes \CT_{FF}$, but the two sides are related by a nontrivial permutation of flavor symmetries.

One may well wonder what it means to ``compactify'' the 6d (2,0) theory on a simplex, which has boundaries, and corners, and corners of corners$\ldots$\,.  In the three-dimensional case, an answer was carefully described in \cite{DGG, CCV, DGV}: the theory $T[\Delta^3]$ is not really an isolated 3d $\CN=2$ theory, but a boundary or interface in a 4d $\CN=2$ theory $T[\pd \Delta^3]$. Additional choices are required to make $T[\Delta^3]$ truly three-dimensional; such choices were implicitly involved in the description \eqref{TD3}. Similarly, the pentachoron theory most naturally appears as a 2d $\CN=(0,2)$ interface between 3d $\CN=2$ theories; namely, it is the duality interface for the (2,3) move
\be \label{TD4-int} T[\Delta^4]\;\simeq\; \text{duality interface between 3d $\CN=2$ XYZ and 3d $\CN=2$ SQED}\,. \ee

We explain this perspective further in Section \ref{sec:inter}.
Geometrically, \eqref{TD4-int} corresponds to the fact that the boundary of a pentachoron contains five tetrahedra, which may be split into two clusters containing 3 and 2 tetrahedra each, exactly as in the XYZ-SQED duality.
In order to obtain a purely 2d description \eqref{TD4} of $T[\Delta^4]$, additional choices must be made.

The physical properties of the duality interface \eqref{TD4-int} were studied in detail in our recent work \cite{DGP-bdy}, on dualities of boundary conditions in 3d $\CN=2$ theories. This was actually our starting point for the current paper. The duality interface is also closely related to the transformation walls of \cite{GGP-param}, studied via holomorphic blocks \cite{BDP}. Once the duality interface is identified as $T[\Delta^4]$, the ensuing analysis of gluing and 4d Pachner moves becomes systematic. For example, the (3,3) Pachner move translates to a relation between two sequences of duality interfaces among 3d $\CN=2$ theories associated to triangulated octahedra, depicted schematically in Figure \ref{fig:oct-intro}. We will explain this picture in detail in Sections \ref{sec:triang} and \ref{sec:33}.

The results of this paper are just the first steps in a (potential) program of associating 2d $(0,2)$ theories to more general 4-manifolds, via their ideal triangulations. Specifically, one may hope to construct oriented 4-manifolds $M$ with ``ideal boundary'' that are complements of embedded surfaces in closed 4-manifolds,
\be M = \ol M \, \backslash \, \Sigma\,. \ee
This is the 4d analogue of a knot or link complement.
The gluing rules we describe  are only sufficient to build a \emph{restricted} class of four-manifolds, which in particular are bundles over $S^1$ (so-called {mapping tori}).
This is because the interface interpretation \eqref{TD4-int} necessarily separates the boundary of a pentachoron into 3+2 tetrahedra, and our current rules are based on this particular splitting.
The rules must be generalized a bit, treating pentachora more symmetrically, to construct more general four-manifolds. We hope to pursue this in future work.%
\footnote{Even so, 2d (0,2) theories associated to ideal triangulations of 4-manifolds $M$ may not capture the full SCFT (as in \cite{GGP, DGP}) obtained by compactifying the 6d (2,0) theory on $M$. Such a discrepancy arose already in constructions of 3d theories associated to triangulated 3-manifolds, \emph{cf.} \cite{CDGS, GPV}, and has not yet been fully resolved. Thus, both pitfalls and possibilities await; but we try not to get ahead of ourselves by making predictions for theories that don't yet exist!}

Several other future directions include:
\begin{enumerate}
\item Exploring the (homological) chiral algebras associated to pentachoron theories $T[\Delta^4]$ and their gluings. Such chiral algebras have been studied in \cite{GGP, DGP, CDG}; they ``categorify'' the elliptic genera computed in this paper.

\item Compactifying setups such as \eqref{TD4-int} on a Riemann surface $\Sigma$. In the case of \eqref{TD4-int}, the 4d theory $T[\Delta^4]$ should correspond to a chain equivalence that acts on the BPS Hilbert space of the XYZ model to give the BPS Hilbert space of SQED. More generally, one expects theories associated to triangulated 4d cobordisms to act as chain maps on the BPS Hilbert spaces of theories associated to triangulated 3-manifolds.
The latter Hilbert spaces have already been studied extensively in \cite{GPV, Bullimore}.

\item The interpretation of $T[\Delta^4]$ as an interface in a 3d theory can naturally be lifted all the way to four dimensions. Namely, since the 3d XYZ and SQED theories are themselves interfaces in a 4d $\CN=2$ theory \cite{CCV} (the Argyres-Douglas $A_2$ theory), the pentachoron theory $T[\Delta^4]$ lifts all the way to a \emph{2d defect} lying at a \emph{junction of 3d interfaces} in 4d $\CN=2$ theory. It should be very similar to the duality defects of \cite{GGP-defects}, and the more recent work of \cite{SN1, SN2, SN3, GaiottoOkazaki}.

\end{enumerate}


\section{Ideal triangulations and Pachner moves in $d$ dimensions}\label{sec:triang}

We begin by reviewing some standard notions from topology.

In any dimension $d$, one can consider the category of PL manifolds. These are topological manifolds equipped with a piecewise linear structure, meaning a set of open charts whose transition functions are piecewise linear maps. Two PL manifolds are deemed equivalent if there exists a piecewise linear homeomorphism between them. 

In dimensions $d\leq 3$, the category of PL manifolds is equivalent to the category TOP of topological manifolds \cite{moise, rado} (up to homeomorphism) and the category DIFF of smooth differentiable manifolds (up to diffeomorphism). In dimension $d=4$, PL is equivalent to DIFF, but distinct from TOP \cite{freedman1982topology}. For $d\geq 5$, all three categories are distinct \cite{kirby1977foundational}.\footnote{More precisely, every PL manifold of dimension $d\leq 6$ possesses a \textit{unique} compatible differentiable structure---see \cite{Milnor} and references therein. PL is a stronger condition than triangulable, and this distinction leads to many foundational results (see, e.g. \cite{manolescu2016pin}), but we will focus on PL-manifolds in this note.}

From now on, we will work exclusively within the realm of PL manifolds. By ``homeomorphism'' we mean PL-homeomorphism.

A \emph{triangulation} of a PL manifold $M$ is a decomposition into $d$-simplices $\Delta^d$
\be M = \bigcup_{i\in I} \Delta_i^d\,, \ee
such that the $(d-1)$-dimensional faces of the simplices are identified with each other pairwise, and the interiors of simplices remain disjoint. This is also called a \emph{tiling} by $d$-simplices. Pachner \cite{pachner} proved that any two triangulations of a PL manifold could be related by a finite sequence of moves --- now commonly referred to as \emph{Pachner moves}.

The Pachner moves in dimension $d$ admit a uniform description as cobordisms through a $(d+1)$-simplex $\Delta^{d+1}$. We explain this in some detail, as it will be useful for constructing the field-theoretic analogue of a 4-dimensional Pachner move later in the paper. 

Let us number the vertices of a simplex $\Delta^{d+1}$, as $0,1,...,d+1$, and denote the simplex itself as $\Delta^{d+1}=[012...(d+1)]$. The simplex $\Delta^{d+1}$ has $d+2$ distinct faces $(\pd \Delta^{d+1})_{(i)}$, each obtained by deleting the $i$-th vertex, and constructing the convex hull of the remaining vertices:
\be (\pd \Delta^{d+1})_{(i)} = [01...\hat i...(d+1)] \simeq \Delta^d\,.\ee
As indicated, each face is (PL-)homeomorphic to $d$-simplex $\Delta^d$. Altogether, the boundary of $\Delta^{d+1}$, which is homeomorphic to a sphere $S^{d}$, is triangulated by the faces:
\be \pd \Delta^{d+1} =   \bigcup_{i=0}^{d+1} (\pd \Delta^{d+1})_{(i)}\,. \ee

Now, to define a Pachner move, one chooses an integer $1\leq n \leq d+1$, and splits the boundary of $\Delta^{d+1}$ into two parts,
\be (\pd \Delta^{d+1})_- =   \bigcup_{i=0}^{n-1} (\pd \Delta^{d+1})_{(i)}\,,\qquad  (\pd \Delta^{d+1})_+ =   \bigcup_{i=n}^{d+1} (\pd \Delta^{d+1})_{(i)}\,. \ee
Note that both $(\pd \Delta^{d+1})_-$ and $(\pd \Delta^{d+1})_+$ are homeomorphic to discs $D^d$. Moreover, any other splitting of the boundary into two collections of $n$ and $d+2-n$ faces is related to the one above by a reordering of the vertices (which is a symmetry of $\Delta^{d+1}$).

Given a triangulated $d$-manifold $M$, a $(n,d+2-n)$ \emph{Pachner move} acts by choosing a collection of $n$ simplices in the triangulation of $M$ that are glued together the same way as $(\pd \Delta^{d+1})_-$, and replacing them with the $d+2-n$ simplices that appear in $(\pd \Delta^{d+1})_+$. Alternatively, working in $d+1$ dimensions, we may think of the Pachner move as gluing $\Delta^{d+1}$ to $M$ along $(\pd \Delta^{d+1})_-$, to obtain a new triangulation of $M$ that includes $(\pd \Delta^{d+1})_+$. This can also be thought of as a cobordism through $\Delta^{d+1}$. Since $(\pd \Delta^{d+1})_-$ and $(\pd \Delta^{d+1})_+$ are PL-homeomorphic to each other (and both homeomorphic to standard discs), the underlying PL structure of $M$ is left unchanged.

Altogether, in dimension $d$, there are $d+1$ distinct Pachner moves, labelled by the integer $1\leq n \leq d+1$. 
They are of type $(1,d+1),\,(2,d),\,(3,d-1),\ldots,(d+1,1)$. The moves of types $(n,d+2-n)$ and $(d+2-n,n)$ are inverses of each other.

\bigskip
\noindent\textbf{Examples:}

In dimension $d=2$, triangulations are literally tilings by triangles. There are three Pachner moves, of types (1,3), (2,2), and (3,1). They each correspond to a cobordism through $\Delta^3 \simeq$ a tetrahedron. Namely, the (1,3) move corresponds to splitting the boundary of the tetrahedron into 1+3 triangles; it replaces a single triangle in the triangulation of a surface with three triangles that all share a common vertex

The (3,1) move does the reverse: it replaces three triangles that share a common vertex with a single triangle. The (2,2) move, often called the \emph{flip}, corresponds to splitting the boundary of the tetrahedron into 2+2 triangles; in the triangulation of a surface, it replaces two triangles glued along a common edge with two new triangles where the edge has ``flipped'':

\be \includegraphics[width=5.5in]{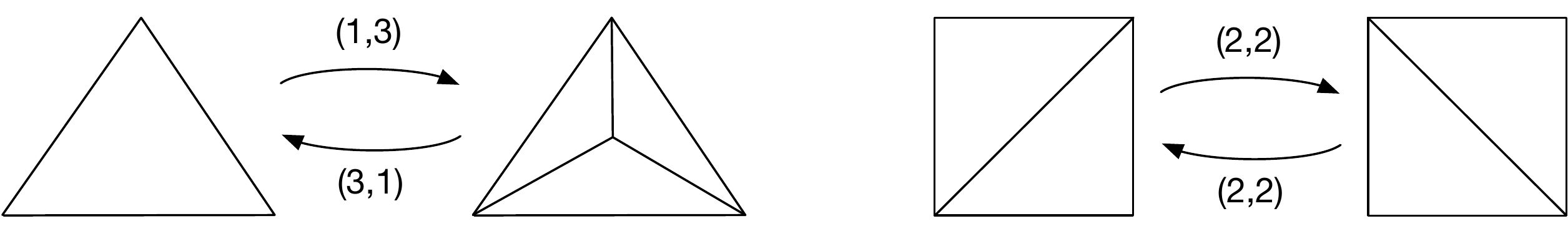} \notag \ee

In dimension $d=3$, triangulations are tilings by tetrahedra, and Pachner moves are cobordisms through a 4-simplex $\Delta^4$, also called a pentachoron. The pentachoron has five tetrahedra on its boundary, and the Pachner moves are of types (1,4), (2,3), (3,2), and (4,1). The (1,4) move takes single tetrahedron and replaces it with four tetrahedra that share a common vertex, while the (4,1) move does the reverse:

\be \includegraphics[width=5.7in]{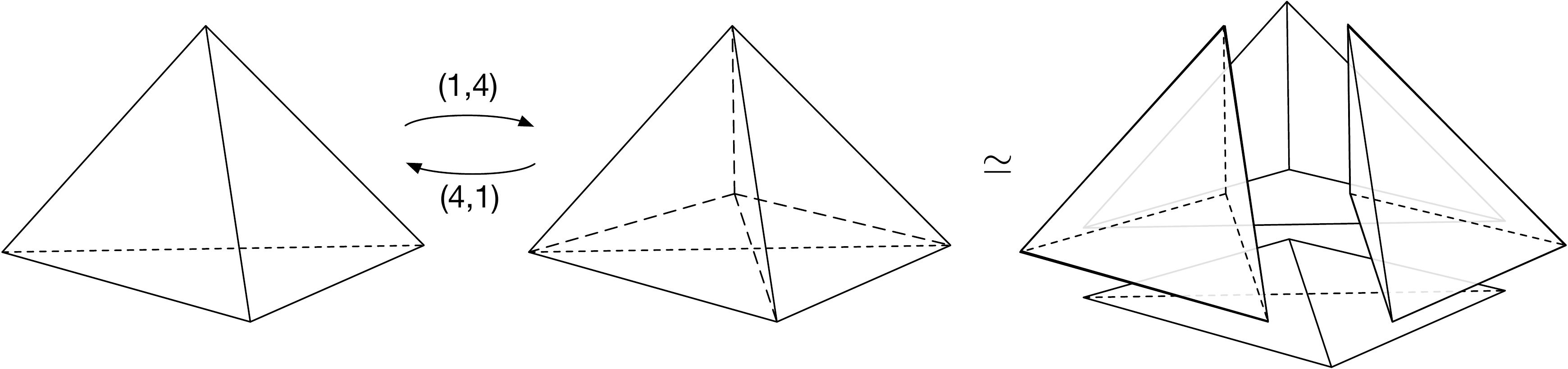} \notag \ee

The (2,3) and (3,2) moves replace two tetrahedra glued along a common face with three tetrahedra that share a common edge:

\be \includegraphics[width=5.5in]{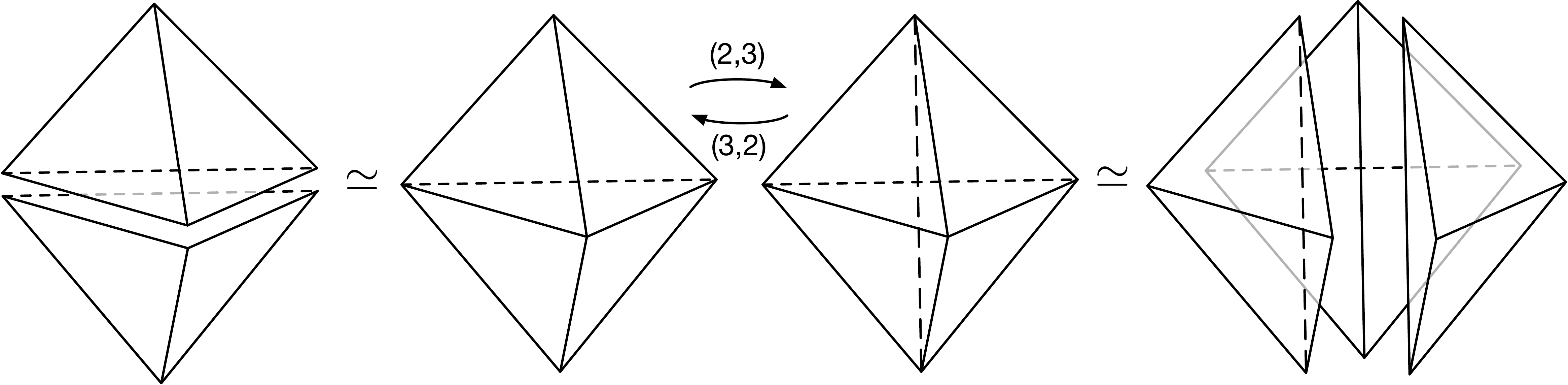} \label{Pach23} \ee

In dimension $d=4$, the Pachner moves act on clusters of pentachora, and correspond to cobordisms through a 5-simplex $\Delta^5$ (whose boundary consists of six pentachora). The moves are thus of type (1,5), (2,4), (3,3), (4,2), and (5,1).

\subsection{A visualization of 4d moves}
\label{sec:vis}

The 4-dimensional Pachner moves are difficult to visualize directly. We thus introduce a standard trick that effectively reduces their dimensionality. The general principle is that

\emph{A Pachner move in $d$ dimensions can be described as a relation between \emph{sequences} of moves in $d-1$ dimensions. In particular, the $(n,d+2-n)$ move relates a sequence of $n$ $(d-1)$-dimensional moves (of various types) to another sequence of $d+2-n$ $(d-1)$-dimensional moves.}

To warm up, we illustrate this idea with 3-dimensional Pachner moves first. Consider the $(2,3)$ move. It relates two different triangulations of a bipyramid, which is homeomorphic to a 3-disc $D^3$ (\emph{i.e.} a ball). The boundary of the bipyramid is topologically a sphere $S^2$, triangulated into five triangles. We choose a splitting of this boundary into two triangulated discs, which each look like a pentagon
\be \begin{array}{c} \pd(\text{bipyramid}) = \text{pentagon}_-\cup \text{pentagon}_+\,;\qquad\text{or topologically}\,,\quad S^2\simeq D^2\cup_{S^1}D^2\,. \\[.2cm]
\includegraphics[width=4in]{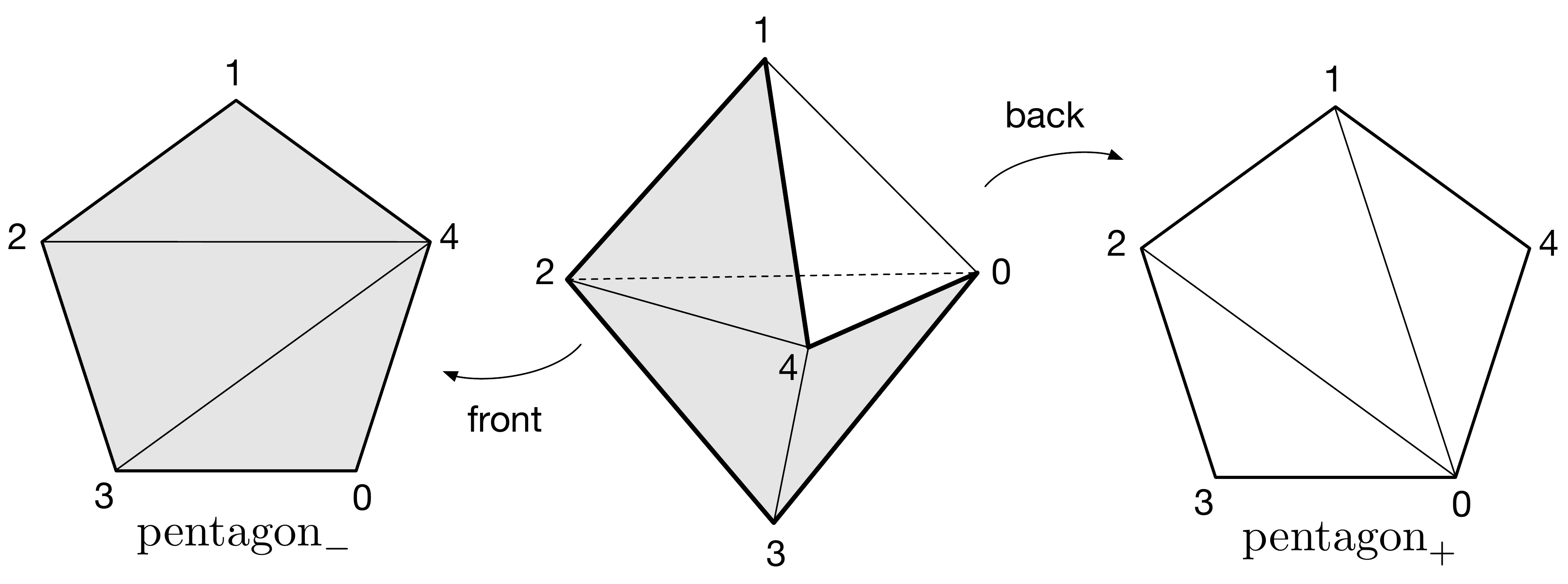} \end{array} \ee

\begin{figure}[htb]
\centering
\includegraphics[width=5.5in]{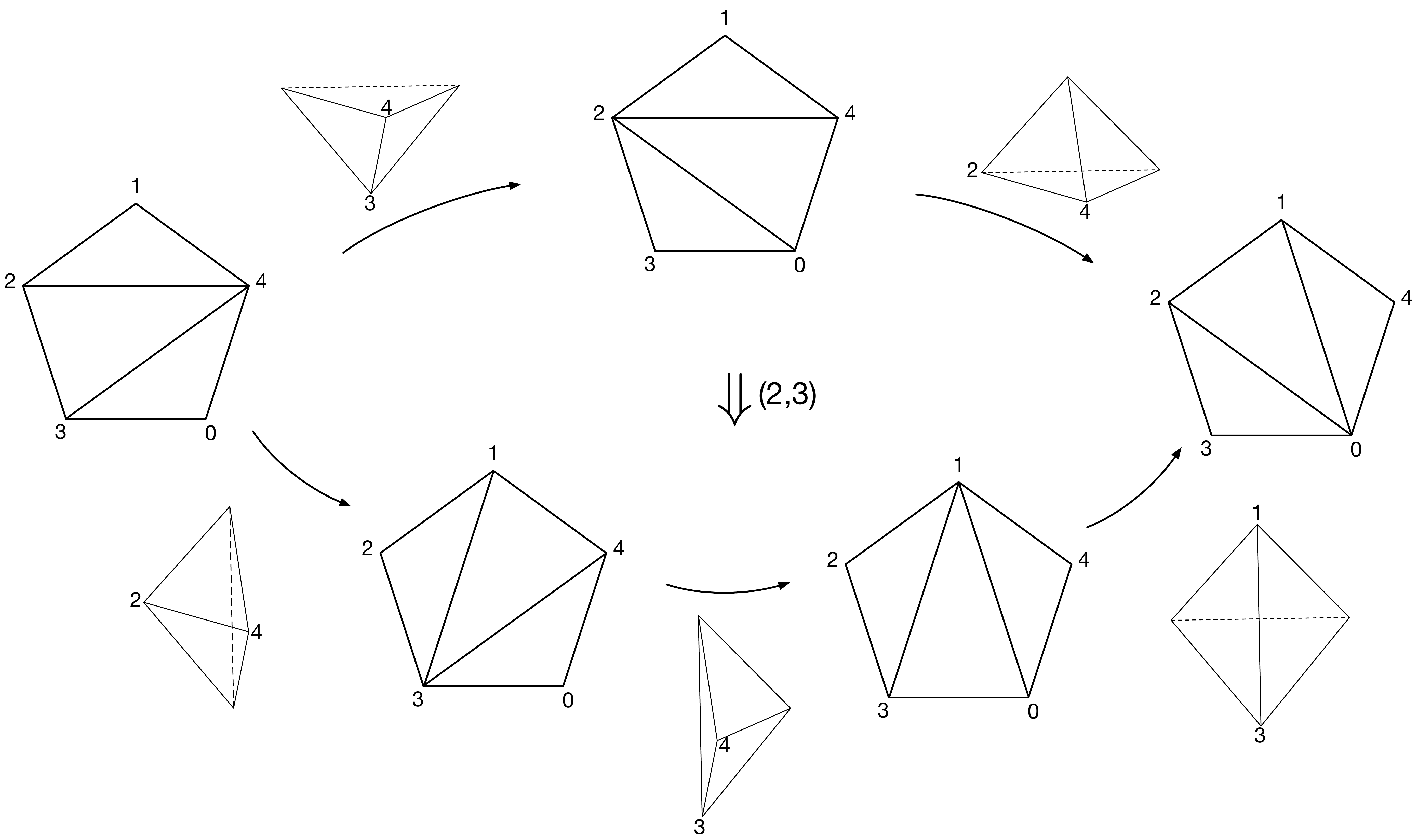}
\caption{The (2,3) Pachner move, interpreted as relating two different sequences of flips.}
\label{fig:Cobord23}
\end{figure}

Then we interpret the bipyramid itself as a cobordism from pentagon$_-$ to pentagon$_+$. When the bipyramid is composed of two tetrahedra (on the ``2'' side of the (2,3) move), the cobordism may be decomposed as two successive flips, \emph{i.e.} two (2,2) moves, one for each tetrahedron. This is illustrated at the top of Figure \ref{fig:Cobord23}. 
 When the bipyramid is composed of three tetrahedra, the cobordism may be decomposed a sequence of three flips, as on the bottom of Figure \ref{fig:Cobord23}. The (2,3) move itself relates/replaces the sequence of flips on the top with the sequence of flips on the bottom.

Now consider a 4-dimensional Pachner move, say of type (3,3). We may interpret it in a similar way, as relating sequences of moves in three dimensions, acting on triangulations of a suitable 3d polyhedron.

To make this precise, let $\Delta^5=[012345]$ be the 5-simplex, with vertices numbered from 0 to 5. To describe the (3,3) move, we first split the boundary of $\Delta^5$ into 3+3 pentachora, which we choose here to be
\be (\pd\Delta^5)_- = [12345]\,\cup\,[01345]\,\cup\,[01235]\,,\qquad (\pd\Delta^5)_+ = [02345] \,\cup\, [01245]\,\cup\,[01234]\,. \label{dD5} \ee
(Thus $(\pd\Delta^5)_-$ omits vertices 0, 2, and 4; while $(\pd\Delta^5)_+$ omits vertices 1, 3, and 5.) Note that $(\pd\Delta^5)_\pm$ are triangulations of the same 4d polyhedron, homeomorphic to a standard 4-disc $D^4$, which plays a role analogous to the bipyramid above.

\begin{figure}[htb]
\centering
\includegraphics[width=5.8in]{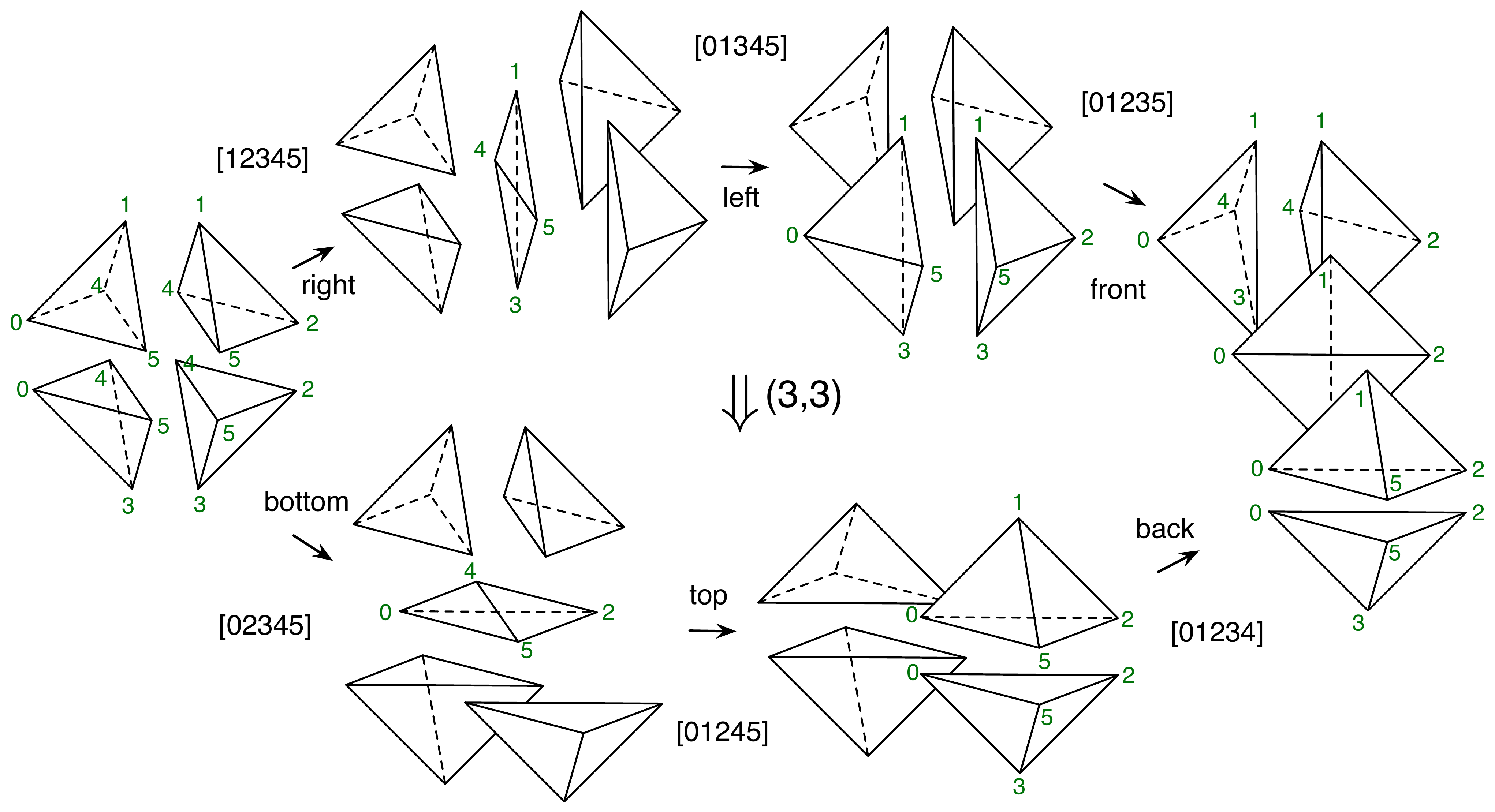}
\caption{The (3,3) Pachner move, interpreted as relation between sequences of (2,3) and (3,2) moves.}
\label{fig:33sec2}
\end{figure}

Next, we claim that the boundary of both $(\pd\Delta^5)_+$ and $(\pd\Delta^5)_-$ can be split symmetrically into two 3-dimensional octahedra:
\be \begin{array}{c} \pd\big( (\pd\Delta^5)_\pm\big) = \text{octahedron}_-\cup \text{octahedron}_+\,;\qquad\text{or topologically,} \quad S^3 \simeq D^3\cup_{S^2} D^3\,.\\[.2cm]
\includegraphics[width=5in]{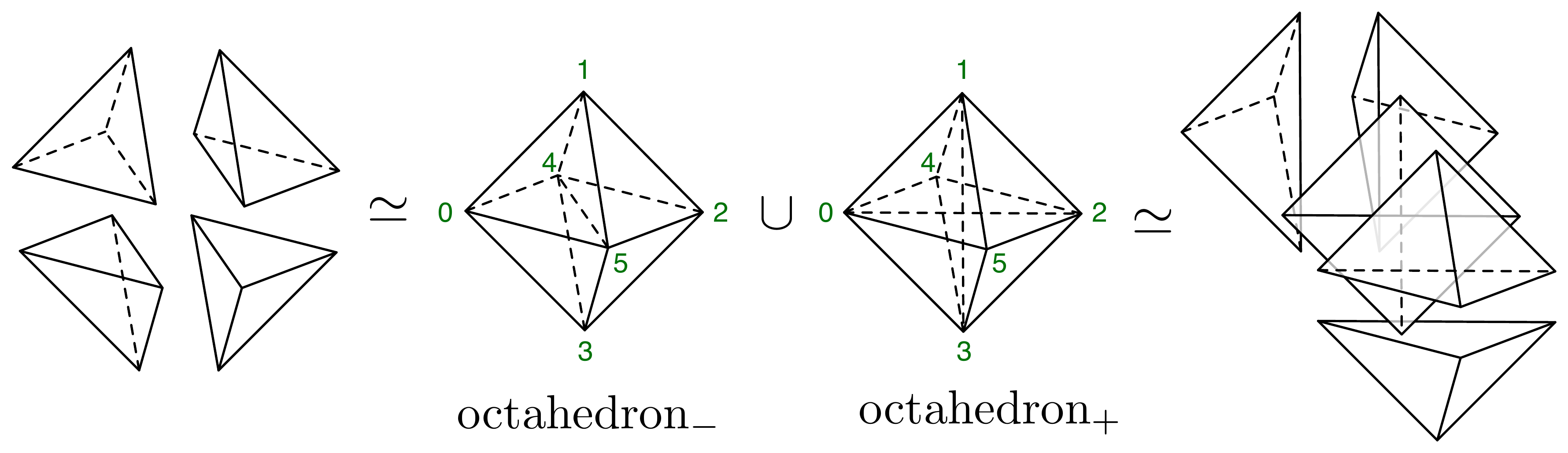} \end{array}
\ee
Combinatorially, the boundary of $(\pd\Delta^5)_-$ consists a collection of tetrahedra obtained by omitting a vertex from any of the three pentachora $[12345],\,[01345],\,[01235]$ that appear in~\eqref{dD5}. We find:
\be \label{bdy3-}
 \begin{array}{ccc}
 \pd [12345] &=& [2345]\,\cup\,\mb{[1345]}\,\cup\,[1245]\,\cup\,\mb{[1235]}\,\cup\,[1234] \\
 \pd [01345] &=& \mb{[1345]}\,\cup\,[0345]\,\cup\,[0145]\,\cup\,\mb{[0135]}\,\cup\,[0134] \\
 \pd [01235] &=& \mb{[1235]}\,\cup\,[0235]\,\cup\,\mb{[0135]}\,\cup\,[0125]\,\cup\,[0123]
 \end{array}
\ee
Of the 15 tetrahedra in this list, three of them (in bold) appear \emph{twice}. The tetrahedra that appear twice are glued together pairwise, and do not contribute to the total boundary $\pd\big((\pd\Delta^5)_-\big)$; put differently, they are internal 3-faces in the triangulation of $(\pd\Delta^5)_-$. The actual boundary $\pd\big((\pd\Delta^5)_-\big)$ consists of the remaining nine tetrahedra,
\be \begin{array}{cclc} \pd\big((\pd\Delta^5)_-\big) &=& \big( [2345]\,\cup\,[1245]\,\cup\,[0345]\,\cup\,[0145]\big)&\leftarrow \;\text{oct}_- \\
 &&\cup\;\big([1234]\,\cup\,[0134]\,\cup\,[0235]\,\cup\,[0125]\,\cup\,[0123]\big) & \leftarrow \;\text{oct}_+ \end{array} \ee
We split the boundary as indicated into two octahedra $\text{oct}_\pm$. The first octahedron, triangulated into four tetrahedra, is shown on the far left of Figure \ref{fig:33sec2}. The second octahedron, triangulated into five tetrahedra, is shown on the far right of Figure \ref{fig:33sec2}.

Successive cobordisms through the three pentachora $[12345],\,[01345],\,[01235]$ may now be interpreted as a sequence of (2,3) and (3,2) Pachner moves that take us from $\text{oct}_-$ to $\text{oct}_+$. This sequence of moves is shown in the top part of Figure \ref{fig:33sec2}. Explicitly, cobordism through $[12345]$ is a (2,3) move; cobordism through $[01345]$ is a (3,2) move; and cobordism through $[01235]$ is another $(2,3)$ move. Note that the three internal tetrahedra $[1345],\,[1235],\,[0135]$ that appeared twice in \eqref{bdy3-} all play a role in Figure \ref{fig:33sec2}. Namely, they are the tetrahedra that are both `created' and subsequently `annihilated' by moves in the top sequence.

Similarly, we may compute the triangulated boundary of $(\pd\Delta^5)_+$ by first considering
\be \label{bdy3+}
 \begin{array}{ccc}
 \pd [02345] &=& [2345]\,\cup\,{[0345]}\,\cup\,\mb{[0245]}\,\cup\,{[0235]}\,\cup\,\mb{[0234]} \\
 \pd [01245] &=& {[1245]}\,\cup\,\mb{[0245]}\,\cup\,[0145]\,\cup\,{[0125]}\,\cup\,\mb{[0124]} \\
 \pd [01234] &=& {[1234]}\,\cup\,\mb{[0234]}\,\cup\,{[0134]}\,\cup\,\mb{[0124]}\,\cup\,[0123]
 \end{array}
\ee
After removing the internal (repeated) tetrahedra, we find
\be \begin{array}{cclc} \pd\big((\pd\Delta^5)_+\big) &=& \big( [2345]\,\cup\,[1245]\,\cup\,[0345]\,\cup\,[0145]\big)&\leftarrow \;\text{oct}_- \\
 &&\cup\;\big([1234]\,\cup\,[0134]\,\cup\,[0235]\,\cup\,[0125]\,\cup\,[0123]\big) & \leftarrow \;\text{oct}_+ \\[.1cm]
  &= &  \pd\big((\pd\Delta^5)_-\big)\,. \end{array} \ee
As required, this boundary is identical to $\pd\big((\pd\Delta^5)_-\big)$. We split it into the \emph{same} two octahedra $\text{oct}_\pm$ as before, and interpret the pentachora $[02345],\,[01245],\,[01234]$ as a sequence of cobordisms --- a sequence of (2,3) and (3,2) moves --- shown along the bottom of Figure \ref{fig:33sec2}. The three internal tetrahedra $[0245],\,[0234],\,[0123]$ now appear at intermediate stages of this bottom sequence, first created and then annihilated. 

Altogether, the (3,3) Pachner move is an operation that replaces a cluster of 3 pentachora represented by the top sequence in Figure \ref{fig:33sec2} by a cluster represented by the bottom sequence.

The remaining 4-dimensional Pachner moves can be given similar interpretations/visualizations. In particular, the (2,4) and (4,2) moves can be visualized in terms of the \emph{same} six triangulated octahedra in Figure \ref{fig:33sec2}! The (2,4) move relates a two-step sequence of moves going around the circle of octahedra in one direction, with a four-step sequence of moves going around the circle of octahedra in the opposite direction. The (4,2) move does the reverse. The (1,5) and (5,1) moves require a different underlying 3d polyhedron, we will not consider them in this paper, because they do not preserve ideal triangulations (see the next section).

This sort of interpretation of 4d Pachner moves in terms of sequences of 3d moves has been used frequently in the mathematics literature. A particularly clear discussion and visualization appeared in \cite{CarterKauffmanSaito} (see also \cite{kashaev2015, kashaev2018}). The choice of splitting of $\pd\big((\pd\Delta^5)_\pm\big)$ described in \cite{CarterKauffmanSaito} and used in \cite{kashaev2015, kashaev2018} differs from the one above; it is less symmetric, but has the advantage of only involving sequences of (2,3) moves, with no (3,2) moves.

\subsection{Comments on ideal triangulations}
\label{sec:ideal}

As noted in the introduction, a bottom-up approach to constructing 2d $\CN=(0,2)$ theories $T[M^4]$ associated to triangulated 4-manifolds is only likely to make sense in the case of \emph{oriented} 4-manifolds with boundary and \emph{ideal} triangulations thereof.

\begin{figure}[htb]
\centering
\includegraphics[width=5.5in]{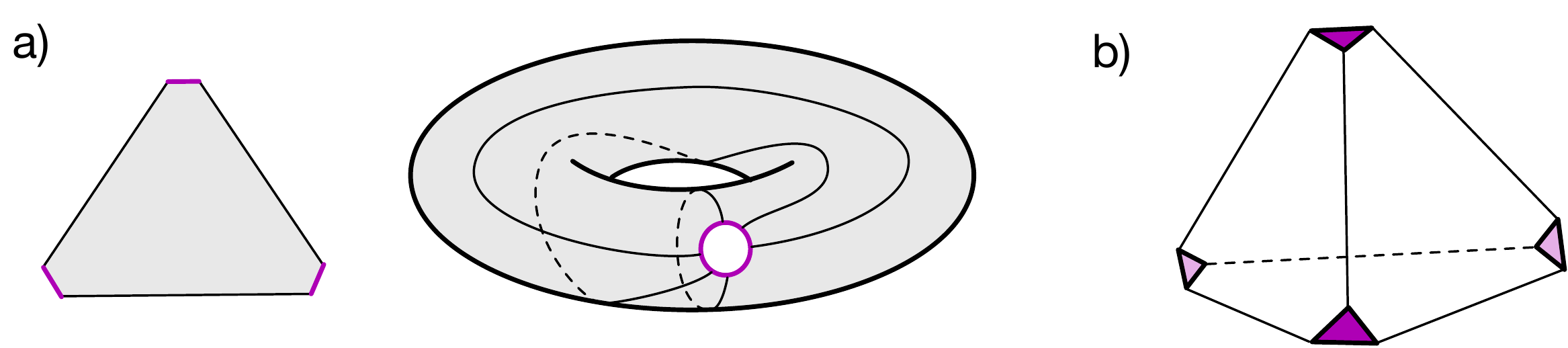}
\caption{a) An ideal triangle, and an ideal triangulation of a punctured 2-torus into two ideal triangles; b) an ideal tetrahedron.}
\label{fig:ideal}
\end{figure}

Topologically, an ideal $d$-simplex is a $d$-simplex whose vertices have been removed.
In the category of PL manifolds, it is more convenient to think of an ideal $d$-simplex $\Delta^{d}_{id}$  as a $d$-simplex with small \emph{neighborhoods} of the vertices removed, \emph{i.e.} a $d$-simplex that has been slightly truncated. Examples of ideal triangles and ideal tetrahedra are shown in Figure \ref{fig:ideal}. Note that the boundary of an ideal simplex is not triangulated; rather, it is tiled by $d+1$ ``big'' faces, each an ideal simplex $\Delta^{d-1}_{id}$, which are truncations of the original faces of $\Delta^d$; and in addition by $d+1$ ``small'' faces, each an ordinary simples $\Delta^{d-1}$, which are the boundaries of the removed vertex neighborhoods. For example, the boundary of an ideal tetrahedron consists of four big hexagons (ideal triangles), and four small triangles. The boundary of an ideal pentachoron consists of five ideal tetrahedra, and five small ordinary tetrahedra.

Given an oriented PL $d$-manifold $M^d$ with boundary $\pd M^d$, an ideal triangulation of $M^d$ is a tiling by ideal (and oriented) $d$-simplices, such that all the big faces are glued pairwise, while all the small faces are left unglued, and compose a standard triangulation of $\pd M^d$.

An example of an ideal triangulation of a 2d surface with small $S^1$ boundaries is shown in Figure \ref{fig:ideal}. In three dimensions, ideal triangulations were popularized by W. Thurston \cite{ThurstonBook}, and are now extremely common (\emph{e.g.}) in the study of knot and link complements. In the case of a knot complement, the underlying 3-manifold $M^3$ is obtained by starting with a knot $K\subset S^3$ and excising a neighborhood of $K$,
\be M^3 = S^3\backslash \text{neighborhood}(K)\,. \ee
Topologically, the boundary of the knot complement is a torus, $\pd M^3 \simeq T^2$. An ideal triangulation of $M^3$ is a tiling by truncated tetrahedra, whose small-triangle boundary components are left unglued, and tile the boundary $T^2$.

For most of the current paper, it will \emph{not} be important to distinguish between ordinary and ideal triangulations. The triangulated manifolds that we study are local: they are formed from clusters of simplices, partially glued along their (big) boundaries to form oriented polyhedra. Whether or not neighborhoods of the vertices are removed is immaterial. 
We will continue writing $\Delta^d$ instead of $\Delta^d_{id}$ to refer to simplices.

One relevant consequence of using ideal triangulations is that they restrict the set of valid Pachner moves. In $d$ dimensions, (truncated versions of) the Pachner moves of types $(n,d+2-n)$ with $2\leq n \leq d$ can all be used to change the local structure of an ideal triangulation. The exceptional moves of type $(1,d+1)$ and $(d+1,1)$ are no longer allowed, since in the context of ideal triangulations they would add or remove a boundary component. For example, in three dimensions, the (1,4) move replaces a single tetrahedron with four tetrahedra that share a new, common vertex; in the context of ideal triangulations, this move would create a new, spherical boundary component. In four dimensions, the Pachner moves that act on ideal triangulations are of types (2,4), (3,3), and~(4,2).

The distinction between ideal and non-ideal triangulations becomes very important globally.
Ultimately, we would hope to be able to associate 2d $\CN=(0,2)$ theories to ideal triangulations of complements of knotted surfaces in 4-manifolds. Namely, given a surface $\Sigma$ (PL-)embedded in a closed (PL) 4-manifold $\ol M$, one can consider the complement
\be M^4  = \ol M \backslash \text{neighborhood}(\Sigma)\,. \ee
The boundary $\pd M^4$ will be an $S^1$-bundle over $\Sigma$ (in particular, it will be a Seifert-fibered 3-manifold). As long as $\pd M^4$ is nonempty, $M^4$ has a chance of admitting an ideal 4d triangulation.

Knotted surfaces and their complements in 4-manifolds have been studied in \emph{e.g.} \cite{CarterSaito, Hillman} (see also references therein) and more recently in the context of cusped hyperbolic 4-manifolds \cite{RT, LR, KolpakovMartelli, martelli2015hyperbolic}.
A few concrete examples of such ideal triangulations, for the complement of knotted $\Sigma\simeq S^2$ in $S^4$, were discussed in \cite{BBH, issa}. At the moment, though, the literature on 4d ideal triangulations is rather less developed than in 3d.

In dimensions two and three, it is known that all ideal triangulations of a manifold with boundary are connected by finite sequences of allowed Pachner moves (type (2,2) in 2d, and types (2,3) and (3,2) in 3d). The result in three dimensions is a nontrivial generalization of Pachner's theorem \cite{Matveev, Piergallini, BenedettiPetronio, Amendola, RST}. One would expect an analogous result to hold in four dimensions as well, though to the best of our knowledge it has not yet been established.


\section{The pentachoron theory}\label{sec:inter}

We saw in the previous section that $(d+1)$-simplices mediate $d$-dimensional Pachner moves. We now combine this geometric construction with the expected physics of compactifications to associate a 2d $\CN=(0,2)$ theory to a 4-simplex, \emph{a.k.a.} a pentachoron.

The abstract reasoning goes as follows.
Let us suppose that the 6d (2,0) SCFT can be compactified on a $d$-manifold $M^d$, with a suitable $d$-dimensional topological twist, to produce an effective $(6-d)$-dimensional theory $T[M^d]$. Moreover, suppose that the data of $T[M^d]$ is explicitly presented in a way that depends on a choice of triangulation $\mb t$ for $M^d$, but that the dependence on this choice disappears in the far infrared. In other words, given any $\mb t$ one can construct a $(6-d)$-dimensional theory $T[M^d,\mb t]$, in such a way that theories $T[M^d,\mb t]$ and $T[M^d,\mb t']$ associated to any pair of triangulations $\mb t,\mb t'$ are infrared dual. 
Then we should be able to identify the $(6-d-1)$-dimensional theory $T[\Delta^{d+1}]$ associated to a $(d+1)$-simplex as a \emph{duality interface} between any pair of theories $T[M^d,\mb t]$ and $T[M^d,\mb t']$, where $\mb t$ and $\mb t'$ differ by a single Pachner move.

We will apply this abstract reasoning to concrete examples of theories $T[M^d,\mb t]$ labelled by triangulations, which descend from the 6d (2,0) theory of type $A_1$. In particular, we are interested in:
\begin{itemize}
\item[$\bullet$\;d=2]  $M^2=\Sigma$ a punctured surface, with $T[\Sigma]$ the associated 4d $\CN=2$ theory of class $\mathcal S$ \cite{Gaiotto-dualities, GMN}. An ideal triangulation $\mb t$ of $M^2$ leads to in IR description $T[\Sigma,\mb t]$ of $T[\Sigma]$ in some region of its Coulomb branch, determined by the choice of $\mb t$ \cite{GMN-Hitchin}. 
$T[\Sigma,\mb t]$ is described as an abelian gauge theory with a particular collection of BPS states \cite{SW}.

\item[$\bullet$\;d=3] A 3-manifold $M^3$, with $T[M^3]$ an associated 3d $\CN=2$ theory. If $M^3$ is a knot or link complement with an ideal triangulation $\mb t$, one can assemble a 3d $\CN=2$ theory $T[M^3,\mb t]$ of class $\mathcal R$ from the data of the triangulation \cite{DGG}. Different triangulations lead to IR-dual theories. It is expected that the $T[M^3,\mb t]$'s capture a subsector of $T[M^3]$, far out on its Coulomb branch.
\end{itemize}

\subsection{Warmup: $T[\Delta^3]$ as an interface}
\label{sec:TD3}

As a warmup, we briefly review how the abstract reasoning above can be used to recover the class-$\mathcal R$ theory $T[\Delta^3]$ of a single tetrahedron, just by knowing the 4d Seiberg-Witten theories associated to punctured surfaces with ideal triangulations. This particular interpretation of $T[\Delta^3]$ was developed in detail by \cite{CCV} and \cite{DGV} (and is also related to ideas in \cite{CNV, Yamazaki}).

Suppose that $T[\Sigma,\mb t]$ and $T[\Sigma,\mb t']$ are two Seiberg-Witten theories associated to a pair of triangulations $\mb t,\mb t'$ of the same surface $\Sigma$ that differ by a single flip --- \emph{i.e.} a single (2,2) Pachner move. The two theories $T[\Sigma,\mb t]$ and $T[\Sigma,\mb t']$ look almost identical. Indeed, they are merely related by a permutation of the fields in a 4d BPS hypermultiplet, which effectively swaps a particle and an antiparticle \cite{GMN-Hitchin}. There is a 3d $\CN=2$ duality interface between $T[\Sigma,\mb t]$ and $T[\Sigma,\mb t']$ that implements this permutation. The interface carries a single 3d $\CN=2$ chiral multiplet, coupled to hypers of both $T[\Sigma,\mb t]$ and $T[\Sigma,\mb t']$. The bulk-interface couplings effectively serve to equate 4d hypers across the 3d interface, modulo the expected permutation. 

The theory $T[\Delta^3]$ then becomes associated with this duality interface. Note that $T[\Delta^3]$ is independent of the choice of surface $\Sigma$ or particular triangulations $\mb t,\mb t'$. Just as the flip of an edge is a local move relating $\mb t$ and $\mb t'$, the interface $T[\Delta^3]$ couples ``locally'' to an isolated sector of $T[\Sigma,\mb t]$ and $T[\Sigma,\mb t']$ containing a single BPS hypermultiplet (corresponding to the flipped edge). The remaining gauge fields and BPS hypers of $T[\Sigma,\mb t]$ and $T[\Sigma,\mb t']$ remain unchanged when passing through the interface; \emph{i.e} the interface is transparent with respect to the remainder of these theories.

This description of $T[\Delta^3]$ as an interface serves to highlight another important feature. Despite the fact that one often describes $T[\Delta^3]$ as ``a 3d chiral multiplet,'' $T[\Delta^3]$ is \emph{not} truly an isolated 3d $\CN=2$ theory.
This is a direct consequence of the fact that the tetrahedron $\Delta^3$ itself has a boundary and is not a closed 3-manifold; thus one should not expect the compactification of the 6d (2,0) theory on $\Delta^3$ to make sense on its own. It is possible to extract a purely 3d theory from the interface $T[\Delta^3]$, at the cost of making additional, non-canonical choices. For example, one may choose boundary conditions for the 4d bulk theories $T[\Sigma,\mb t]$  and $T[\Sigma,\mb t']$ on either side of the interface. A particularly simple choice of boundary conditions kills all remaining bulk 4d degrees of freedom, leaving behind
\be T[\Delta^3] \quad\leadsto\quad \text{single 3d $\CN=2$ chiral multiplet} \label{TD-can} \ee
Other choices of boundary conditions lead to other 3d theories, such as chiral multiplets coupled to dynamical 3d $U(1)$ gauge fields, with various Chern-Simons levels.

\subsection{$T[\Delta^4]$ as an interface}
\label{sec:TD4}

We now wish to increase the dimension in the above analysis, to find a theory $T[\Delta^4]$ associated to the pentachoron $\Delta^4$. As in the case of a tetrahedron, $T[\Delta^4]$ will not be a stand-alone 2d theory; rather, it will appear as a duality interface.

Let us consider a pair of class-$\mathcal R$ theories associated to some 3-manifold $M^3$ (a knot or link complement), with ideal triangulations $\mb t$, $\mb t'$ that differ by a single (3,2) Pachner move. The Pachner move acts locally, replacing a part of the triangulation $\mb t$ that is isomorphic to a bipyramid composed of three tetrahedra (as on the right of \eqref{Pach23}), with a bipyramid in $\mb t'$ composed of two tetrahedra (as on the left of \eqref{Pach23}).

The corresponding transformation of class-$\mathcal R$ theories, described in \cite{DGG}, is also ``local,'' with respect to field content.
The Pachner move replaces a subsector of $T[M^3,\mb t]$ containing three chiral multiplets coupled by a cubic superpotential, with a new subsector in $T[M^3,\mb t']$ that contains two chiral multiplets charged under a $U(1)$ gauge field. In other words, the Pachner move replaces a 3d $\CN=2$ ``XYZ model'' with 3d $\CN=2$ SQED\,:
\be \overset{\displaystyle\text{XYZ model}}{\text{3 chirals, $W=X_{3d}Y_{3d}Z_{3d}$}}  \qquad\leftrightarrow\qquad   \overset{\displaystyle\text{SQED}}
{\text{$U(1)$ w/ 2 chirals $\Phi,\tilde\Phi$ of charge $\pm1$}} \,. \ee
These two simple 3d $\CN=2$ theories are dual to each other in the infrared \cite{AHISS}\,.

More concretely, we can isolate the sector of theories $T[M^3,\mb t]$ and $T[M^3,\mb t']$ that participate in the Pachner move by simply considering the bipyramid $\mb{B}$ itself. 
Since $\mb B$ is a 3-manifold with boundary, defining a stand-alone 3d theory associated to it requires (as always) an additional choice. In the formalism of \cite{DGG}, this choice was encoded in a ``polarization''  $\Pi$ of the boundary of~$\mb B$. 
Geometrically, the most important part of the data of a polarization $\Pi$ is a maximal subset of the edges of the triangulated boundary $\pd \mb B$, with the property that no two edges in the subset bound a common triangle on $\pd \mb B$. (Physically, the edges in this subset correspond to electric BPS states of the Seiberg-Witten theory $T[\pd \mb B]$.)

\begin{figure}[htb]
\centering
\includegraphics[width=5in]{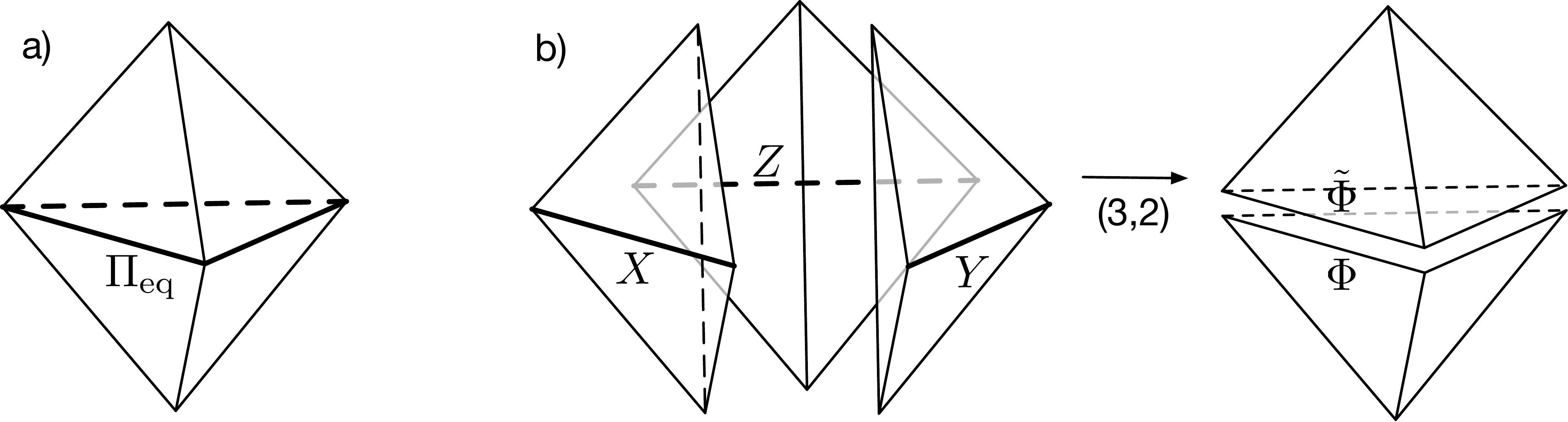}
\caption{a) The three ``electric'' edges in the polarization $\Pi_{\rm eq}$, shown in bold. b) Chiral multiplets associated to tetrahedra in the two triangulations of the bipyramid $\mb B$. }
\label{fig:BipEq}
\end{figure}

Here we take an equatorial polarization $\Pi_{\rm eq}$ of the bipyramid, whose subset of edges contains the three horizontal edges on the equator of the bipyramid, as in Figure \ref{fig:BipEq}a. Given this choice, the class-$\mathcal R$ theory associated to a bipyramid cut into three tetrahedra is precisely $T[\mb B,\mb t_3;\Pi_{\rm eq}]=\text{XYZ}$, and the theory associated to a bipyramid cut into two tetrahedra is $T[\mb B,\mb t_2;\Pi_{\rm eq}]=\text{SQED}$.

The superpotentials, flavor symmetries, gauge symmetries, and Chern-Simons levels (for both global and gauge symmetries) of these two bipyramid theories are all uniquely determined. We list them here for reference:
\be \label{XYZ-geom}
 \begin{array}{c|ccc|cc}
  & X & Y & Z & \Phi & \tilde\Phi  \\\hline
 U(1)_{\rm gauge} & 0 & 0 & 0 & 1 & -1 \\
 U(1)_y & -1 & 1 & 0 & -1 & 0 \\
 U(1)_z & -1 & 0 & 1 & 0 & -1 \\
 U(1)_R & 2 & 0 & 0 & 1 & 1
\end{array}
\qquad
\begin{array}{c}
W_{\rm XYZ} = XYZ \\[.3cm]
 \begin{array}{lcl}
\CA_{\rm bulk}[\text{XYZ}] &=& \mb y \mb z \\[.2cm]
\CA_{\rm bulk}[\text{SQED}] &=& \mb f(\mb z-\mb y)+\frac12 \mb y^2+\frac12 \mb z^2
\end{array}
\end{array}
\ee
We have denoted the chirals of the XYZ model (resp. SQED) as $X,Y,Z$ (resp. $\Phi,\,\tilde \Phi$). There are two flavor symmetries $U(1)_{y,z}$ and a $U(1)_R$ R-symmetry, in addition to the $U(1)$ gauge symmetry of SQED.
 We encode 3d Chern-Simons couplings in anomaly polynomials\footnote{Given a theory in $d= 2n+1$ dimensions with some Chern-Simons term that we may denote by $\CL_{CS}$, the anomaly polynomial $\CA_{CS}$ is given by ${\rm d} \CL_{CS}$. For example, given a 3d Chern-Simons term for a $U(1)$ gauge theory of the form $\int\CL_{CS} = k \int A \wedge F$ (appropriately quantized), we have an anomaly polynomial ${\rm d} \CL_{CS} = k F \wedge F$, which we denote by  $k \mb f^2$. Given a bulk $2n+1$-dimensional theory with a $2n$-dimensional boundary, the bulk current from such Chern-Simons term can cancel one-loop boundary anomalies arising from the presence of chiral fermions by the usual anomaly inflow mechanism. The anomaly polynomials are naturally interpreted as terms in $2n+2$-dimensional theories that extend to the $2n+1$-dimensional boundary and for compact groups $G$ are classified by elements of $H^{2n+2}(BG, \mathbb{Z})$ \cite{DW}.} $\CA_{\rm bulk}$, following our conventions in \cite{DGP-bdy}, where $\mb y,\mb z,\mb r,\mb f$ are field strengths for the $U(1)_{y,z}, U(1)_R, U(1)_{\rm gauge}$ symmetries, respectively. Note that both $U(1)_y$ and $U(1)_z$ function as axial symmetries in SQED. They are distinguished in SQED because their difference is the topological flavor symmetry that rotates the dual photon; this is reflected by the $\mb f(\mb y-\mb z)$ coupling in the anomaly polynomial $\CA_{\rm bulk}[\text{SQED}]$.

A duality interface between the XYZ model and SQED that preserves 2d $\CN=(0,2)$ supersymmetry was found in  \cite{DGP-bdy}. 
It resulted from analyzing dual boundary conditions, somewhat similar in spirit to analyses performed in  \cite{GGP-param} and generalized in \cite{GGP} (see \cite{BrunnerSchulzTabler, Rocek, Jockers} for other recent studies of 1/2-BPS codimension-1 defects in 3d $\CN=2$ theories). The duality interface is nontrivial in the UV, but was precisely engineered so as to flow to a trivial interface in the IR.

In the conventions of \cite{DGP-bdy}, the interface may be constructed as follows.
We envision the XYZ theory supported on the ``left'' half-space $\mathbb{R}^{1, 1} \times \mathbb{R}_{x^{\perp} \leq 0}$ and SQED supported on the ``right'' half-space $\mathbb{R}^{1, 1} \times \mathbb{R}_{x^{\perp} \geq 0}$, with the 2d interface at $x^{\perp} = 0$. The gauge multiplet of SQED is given a Neumann ($\mathcal N$) boundary condition, which preserves gauge symmetry at the interface, as well as 2d $\CN=(0,2)$ SUSY. The bulk chiral multiplets on the left and right may each be decomposed under the $\CN=(0,2)$ subalgebra of 3d $\CN=2$ SUSY into pairs of $\CN=(0,2)$ chiral and fermi multiplets, which we denote $(X,\Psi_X)$, $(Y,\Psi_Y)$, $(Z,\Psi_Z)$ and 
$(\Phi,\Psi)$, $(\tilde \Phi,\tilde \Psi)$. Initially, $(Y,\Psi_Y)$, $(Z,\Psi_Z)$, $(\Phi,\Psi)$, and $(\tilde \Phi,\tilde \Psi)$ are given Neumann (N) boundary conditions at the interface, which means all the fermis $\Psi_Y, \Psi_Z, $ etc. are set to zero while $Y,Z,$ etc. are left unconstrained. The pair $(X,\Psi_X)$ is given a Dirichlet (D) boundary condition, which means $X$ is set to zero at the interface, with $\Psi_X$ unconstrained.

So far, we have just described two independent boundary conditions at $x^\perp=0$. These boundary conditions are tied together at the interface by introducing
\begin{itemize}
\item an additional 2d $(0,2)$ fermi multiplet $\Gamma$, supported at $x^\perp=0$, with charges
\be  \begin{array}{c|cccc}
& U(1)_{\rm gauge} & U(1)_y & U(1)_z & U(1)_R \\\hline
\Gamma & 1 & -1 & 1 & 0 \end{array} \ee
\item $\CN=(0,2)$ superpotential couplings at $x^\perp=0$, of the form
\be \label{W-XYZ} \int d\theta^+\big[ Y \Gamma \tilde\Phi + \Psi_X \Phi\tilde \Phi  \big]\,,\qquad E_\Gamma = Z \Phi\,,\ee
where $\Gamma$ is given both J and E terms $J_\Gamma=Y\tilde\Phi$, $E_\Gamma=Z \Phi$. Given the charge assignments, these couplings are unique up to fermionic T-duality \cite{DGP-bdy}.
\end{itemize}
The interface superpotential deforms the initial boundary condition $X\big|_{x^\perp=0}=0$ to 
\be X\big|_{x^\perp=0} = J_{\Psi_X}=  \Phi \tilde\Phi \label{Xphi}\big|_{x^\perp=0}\,, \ee
whereupon the product of $J$ and $E$ terms for $\Gamma$ becomes
\be J_\Gamma E_\Gamma = (Y\Phi)(Z\tilde\Phi) \overset{\eqref{Xphi}}= XYZ = W_{XYZ}\,. \ee
Thus, the bulk superpotential is factorized at the interface, as required to solve an analogue of the ``Warner problem'' \cite{Warner, GGP} (further explored in the recent \cite{BrunnerSchulzTabler}).

This interface also satisfies a highly nontrivial constraint of anomaly cancellation. All UV gauge and 't Hooft anomalies at $x^\perp=0$ must vanish, in order for the interface to flow to a trivial interface in the IR. There are three sources of such anomalies: 1) the difference of bulk Chern-Simons terms $I_{\rm bulk}[\text{XYZ}]-I_{\rm bulk}[\text{SQED}]$; 2) anomalies $\CA_\pd$ from bulk fermions that survive at $x^\perp=0$ given the initial boundary conditions in the construction of the interface (noting that bulk fermions contribute exactly half the usual anomaly of a purely 2d fermion \cite{DGP-bdy}); and 3) the chiral anomaly $\CA_\Gamma$ of $\Gamma$. A careful calculation yields
\be \CA_\pd = \overbrace{\tfrac12(-\mb y-\mb z+\mb r)^2-\tfrac12(\mb y-\mb r)^2-\tfrac12(\mb z-\mb r)^2}^{\rm XYZ} + \overbrace{\tfrac12\mb r^2-\tfrac12(\mb f-\mb y)^2-\tfrac12(-\mb f-\mb z)^2}^{\rm SQED}\,, \ee
\be \CA_\Gamma = (\mb f-\mb y+\mb z)^2\,, \ee
with $\CA_{\rm bulk}[\text{XYZ}]-\CA_{\rm bulk}[\text{SQED}] +\CA_\pd +\CA_\Gamma = 0$ as required.

Altogether, we will represent the duality interface schematically as
\be T[\Delta^4] \;\simeq\; \overset{\text{XYZ}}{\text{(D,N,N)}} \big| \Gamma \big| \overset{\text{SQED}}{(\CN,\text{N,N})}\,. \label{XYZ-int1} \ee
Since the interface implements the (3,2) Pachner move, we identify it with the pentachoron theory.

\subsubsection{Reversed orientation}

There are several other versions of the duality interface (and hence the pentachoron theory), related to \eqref{XYZ-int1} by making some slightly different choices.

Above, we decided to look at a (3,2) Pachner move, rather than a (2,3) Pachner move. Geometrically, this meant splitting the boundary of the pentachoron $\Delta^4$ in a particular way. Had we considered the $(2,3)$ Pachner move instead, we would have found the inverse interface
\be T[\Delta^4]' \;\simeq\; \overset{\text{SQED}}{(\CN,\text{N,N})} \big| \Gamma' \big| \overset{\text{XYZ}}{\text{(D,N,N)}} \,. \label{XYZ-int2} \ee
The construction of this interface is virtually identical to that above. The only difference is that the fermi multiplet $\Gamma'$ now has charges
\be  \begin{array}{c|cccc}
& U(1)_{\rm gauge} & U(1)_y & U(1)_z & U(1)_R \\\hline
\Gamma' & 1 & 0 & 0 & 0 \end{array} \ee
and superpotential couplings $\int d\theta^+\big[ \tilde \Phi\Gamma'Z - \Psi_X \Phi \tilde\Phi\big]$\,, $E_{\Gamma'} = -\Phi Y$,  consistent with the charge assignments.%
\footnote{Some minus signs are introduced here to ensure that $J\cdot E=W_{\rm left}-W_{\rm right}$ factorizes the difference of bulk superpotentials. In this case, the coupling $-\Psi_X\Phi\tilde\Phi$ deforms the D b.c. on $X$ to $X = -(J_{\Psi_X}) = + \Phi\tilde\Phi$ (since $X$ is sitting on the right of the interface); and we correctly get $J\cdot E = (\tilde\Phi Z)(-\Phi Y) = - XYZ = -W_{\rm right}$.}
Anomaly cancellation now takes the form $\CA_{\rm bulk}[\text{SQED}]-\CA_{\rm bulk}[\text{XYZ}] +\CA_\pd +\CA_{\Gamma'} = 0$, with $\CA_{\Gamma'} = \mb f^2$\,.

\subsubsection{Alternate polarizations}

In order to isolate 3d bipyramid theories above, we also had to make a choice of polarization, namely $\Pi_{\rm eq}$. Other choices are possible. Changing the choice of polarization modifies the interface $T[\Delta^4]$ in \eqref{XYZ-int1} by gauging or ungauging various $U(1)$ symmetries. This is a generalization of Witten's $SL(2,\Z)$ action on 3d theories \cite{Witten-bdy}, now applied to 3d theories containing 2d interfaces. (Similar setups were considered in \cite{GGP-param}.)

We emphasize that the necessity of choosing a polarization stems from the fact that there is  no completely canonical way to isolate a sector of the class-$\mathcal R$ theory $T[M^3,\mb t]$ associated to a single bipyramid $\mb B\subset M^3$. Rather, there is a family of ways to do this, labelled by different polarizations.

One \emph{could} obtain a canonical, polarization-independent description of the pentachoron theory by coupling the entire bulk-interface system of \eqref{XYZ-int1} to a 4d $\CN=2$ $U(1)^2$ gauge theory (by gauging the $U(1)_y\times U(1)_z$ flavor symmetry). This leads to a characterization of $T[\Delta^4]$ as a 2d $\CN=(0,2)$ defect, on the 3d $\CN=2$ boundary of a 4d $\CN=2$ gauge theory. We will not pursue such 2d-3d-4d systems further in the present paper.

\begin{figure}[htb]
\centering
\includegraphics[width=5in]{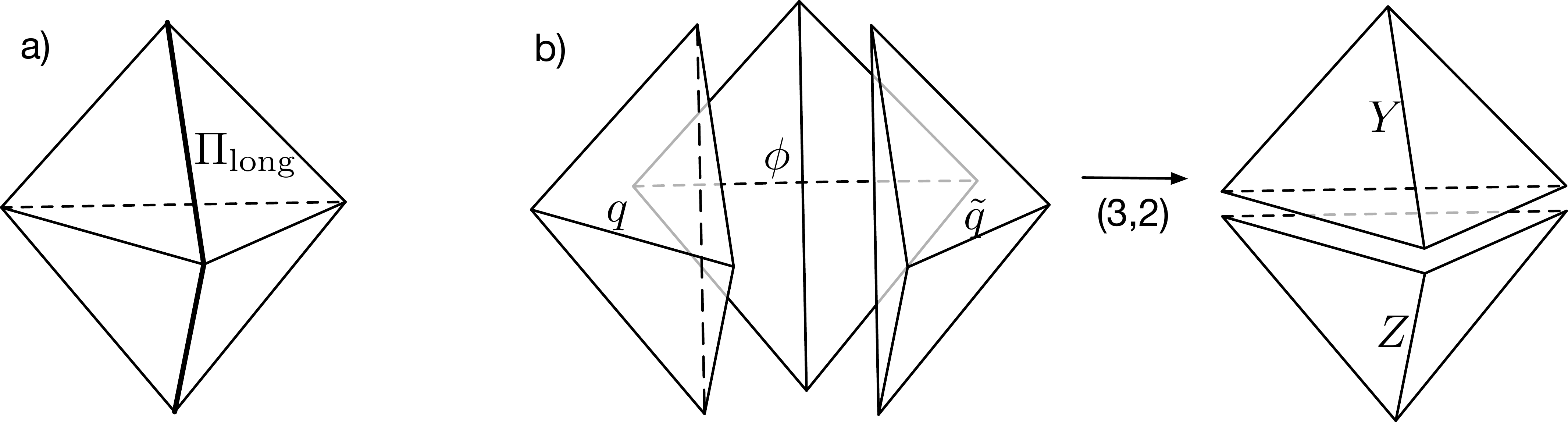}
\caption{a) The two ``electric'' edges in the polarization $\Pi_{\rm long}$, shown in bold. b) Chiral multiplets associated to tetrahedra in the two triangulations of the bipyramid $\mb B$, adapted for this polarization.}
\label{fig:BipLong}
\end{figure}

For future reference, we do mention one convenient, alternative choice of polarization for the bipyramid: a ``longitudinal polarization'' $\Pi_{\rm long}$, whose preferred subset of edges contains two vertical edges as in Figure \ref{fig:BipLong}. With this polarization (also considered in \cite{DGG}), the 3d theory $T[\mb B,\mb t_3,\Pi_{\rm long}]$ roughly becomes 3d $\CN=4$ SQED, and $T[\mb B,\mb t_2,\Pi_{\rm long}]$ roughly becomes a 3d $\CN=4$ hypermultiplet. We say ``roughly'' because these are still viewed as 3d $\CN=2$ theories, and their flavor symmetries (and background Chern-Simons couplings) are shifted slightly from standard 3d $\CN=4$ conventions.

The charges and Chern-Simons levels of the two new bipyramid theories are: 
\be \begin{array}{c|ccc|cc}
 & \phi & q & \tilde q  & Y &  Z \\\hline
U(1)_{\rm gauge} & 0 & -1 & 1 & 0 & 0 \\
U(1)_y & 1 & -1 & 0 & 1 & 0 \\
U(1)_z & 1 & 0 & -1 & 0 & 1 \\
U(1)_R & 0 & 1 & 1 & 0 & 0
\end{array}
\qquad
\begin{array}{c} W_{\text{$\CN=4$ SQED}} = \phi q \tilde q \\[.2cm]
\begin{array}{lcl}
 \CA_{\rm bulk}[\text{$\CN=4$ SQED}] &=& \mb f(\mb y-\mb z)-\mb y \mb z+\mb r(\mb y+\mb z)-\mb r^2  \\
 \CA_{\rm bulk}[\text{hyper}] &=&  -\frac12(\mb y-\mb r)^2-\frac12(\mb z-\mb r)^2
\end{array} \end{array}
\ee
In $\CN=2$ language, $T[\mb B,\mb t_3;\Pi_{\rm long}]$ contains the three chirals $\phi,q,\tilde q$, with a $U(1)$ gauge field and a superpotential $\phi q\tilde q$; while $T[\mb B,\mb t_2;\Pi_{\rm long}]$ just contains the two chirals $Y,Z$. The flavor symmetry is $U(1)_x\times U(1)_y$, and there is a $U(1)_R$ R-symmetry.

The duality interface corresponding to the (3,2) Pachner move now takes the form
\be \label{TD4-long} T[\Delta^4;\Pi_{\rm long}] \;\simeq\;  \overset{\text{$\CN=4$ SQED}}{(\CN, \rm D, N, N)}| \Gamma| \overset{\text{hyper}}{(\rm N,N)}\,, \ee
with an extra 2d $\CN=(0,2)$ fermi multiplet $\Gamma$ of charges 
\be  \begin{array}{c|cccc}
& U(1)_{\rm gauge} & U(1)_y & U(1)_z & U(1)_R \\\hline
\Gamma & 1 & 0 & 0 & 0 \end{array} \ee
and interface superpotential couplings $\int d\theta^+\big[ \Psi_\phi YZ+ q \Gamma Y\big]$, $E_\Gamma = \tilde q Z$\,. We encourage the reader to check that UV anomalies at the interface are perfectly cancelled, and that the bulk superpotential is factorized as $E_\Gamma J_\Gamma$.

In the opposite orientation, the (2,3) Pachner move corresponds to the interface
\be \label{TD4-long'} T[\Delta^4;\Pi_{\rm long}]' \;\simeq\;  \overset{\text{hyper}}{(\rm N,N)}| \Gamma'| \overset{\text{$\CN=4$ SQED}}{(\CN, \rm D, N, N)} \,, \qquad \begin{array}{c|cccc}
& U(1)_{\rm gauge} & U(1)_y & U(1)_z & U(1)_R \\\hline
\Gamma' & 1 & 1 & -1 & 0 \end{array} \ee
with superpotential couplings $\int d\theta^+\big[q \Gamma' Z-\Psi_\phi YZ\big]$, $E_{\Gamma'} = -\tilde q Y$\,.

\subsection{A purely 2d pentachoron theory}
\label{sec:pure2d}

One may wish two isolate a purely two-dimensional version of the pentachoron theory. Purely 2d versions of the pentachoron theory come from ``sandwiching'' the interfaces \eqref{XYZ-int1} or \eqref{TD4-long} between a pair of boundary conditions. As we have emphasized many times, there is no canonical way to choose the boundary conditions. There are, however, some especially simple and convenient choices.

Let us focus on the version \eqref{XYZ-int1} of the interface. One minimal choice of boundary conditions is
\begin{itemize}
\item (N,D,D) as a left b.c. for the XYZ model (meaning $\Psi_X$, $Y$, $Z$ are set to zero)
\item ($\CD$,D,D) as a right b.c. for SQED (meaning Dirichlet for the gauge multiplet, and $\Phi,\tilde \Phi$ set to zero)
\end{itemize}
We depict this schematically as
\be \big|{\rm N}_X{\rm D}_Y \overset{\rm XYZ}{{\rm D}_Z - {\rm D}_X} \overset{\rm interface}{{\rm N}_Y{\rm N}_Z\big|\Gamma\big|\CN_f {\rm N}_{\Phi}}\overset{\rm SQED}{{\rm N}_{\tilde \Phi} - \CD_f} {\rm D}_{\Phi}{\rm D}_{\tilde \Phi}\big| \ee
This choice kills all the bulk degrees of freedom, in both XYZ and SQED. Flowing to the infrared, we find a purely 2d $\CN=(0,2)$ theory consisting of the fermi multiplet $\Gamma$ alone.

Unfortunately, this choice of boundary conditions is a little \emph{too} minimal.
A useful 2d version of the pentachoron theory would have the property that it can be re-coupled to XYZ on one side, and SQED on the other, in order to recover the full duality interface \eqref{XYZ-int1}. The single fermi multiplet $\Gamma$ does not have this property. In order to recover the duality interface \eqref{XYZ-int1} one would somehow need to know to add a 2d superpotential $\Psi_X \Phi\tilde \Phi$ that couples together fields of XYZ and SQED; and this is not information carried by $\Gamma$.

A better choice of 2d pentachoron theory is the following. Consider
\be \big|{\rm D}_X{\rm D}_Y \overset{\rm XYZ}{{\rm D}_Z - {\rm D}_X} \overset{\rm interface}{{\rm N}_Y{\rm N}_Z\big|\Gamma\big|\CN_f {\rm N}_{\Phi}}\overset{\rm SQED}{{\rm N}_{\tilde \Phi} - \CD_f} {\rm D}_{\Phi}{\rm D}_{\tilde \Phi}\big| \ee
\begin{itemize}
\item (D,D,D) as a left b.c. for the XYZ model (setting $X,Y,Z$ to zero while keeping $\Psi_X,\Psi_Y,\Psi_Z$ unconstrained)
\item ($\CD$,D,D) as a right b.c. for SQED
\end{itemize}
In the infrared, the interface sandwiched between these boundary conditions flows to a 2d $\CN=(0,2)$ theory containing \emph{two} fermi multiplets: $\Gamma$ and $\Psi_X$. Let us call the second fermi multiplet $\eta$ instead of $\Psi_X$, to avoid confusion when re-coupling to the bulk. We have found
\be \boxed{T[\Delta^4] \;\leadsto\; \text{two $2d$ $\CN=(0,2)$ fermi multiplets $\Gamma,\eta$}} \ee

These two fermi multiplets carry the full $U(1)_y\times U(1)_z$ flavor symmetry of the bulk XYZ and SQED theories. In the current conventions, the flavor and R charges are
\be \begin{array}{c|cc} & \Gamma & \eta \\ \hline
U(1)_x & -1 & 1 \\ 
U(1)_y & 1 & 1 \\
U(1)_R & 0& -1 \end{array} \ee
Moreover, this simple 2d theory has the property that it can be re-coupled to XYZ on one side, and SQED on the other, to recover the full duality interface \eqref{XYZ-int1}. To implement this recoupling, we consider right b.c. (N,N,N) for XYZ and left b.c. ($\CN$,N,N) for SQED, with an interface superpotential
\be W = \int d\theta^+\big[ Y \Gamma \tilde\Phi + \eta (\Phi\tilde \Phi - X)\big]\,,\qquad E_\Gamma=Z\Phi\,,\quad E_\eta= -Y Z\,. \ee
Note that
\be J\cdot E = J_\Gamma E_\Gamma + J_\eta E_\eta = (Y\tilde\Phi)(Z\Phi)+(\Phi\tilde\Phi-X)(-YZ) = XYZ\,, \ee
factoring the bulk superpotential of the XYZ model, as required. Moreover, integrating out the fermi multiplet $\eta$ precisely reproduces our original duality interface \eqref{XYZ-int1}, with the couplings in \eqref{W-XYZ}.

Other versions of the duality interface for the (3,2) move, \emph{e.g.} \eqref{TD4-long}, can similarly be sandwiched between boundary conditions to produce isolated 2d theories. In all cases, the simplest choice of boundary conditions that still allows one to recover the interface (by re-coupling to the 3d bulk) leads to a 2d theory containing a pair of fermi multiplets.


\section{The (3,3) move and 2d $\CN=(0,2)$ dualities}
\label{sec:33}

Having identifed the interface theory corresponding to a pentachoron, we can begin to construct sequences of interfaces corresponding to clusters of glued pentachora. We expect to find equivalences among gluings corresponding to the four-dimensional Pachner moves.

In this section we will focus on the (3,3) Pachner move, deferring the (2,4) and (4,2) moves to Section \ref{sec:24}.
Let us recall from Section \ref{sec:vis} that (3,3) move can be interpreted as a relation between two \emph{sequences} of (2,3) and (3,2) moves. In the model of Figure \ref{fig:33sec2}, each sequence acts locally by changing the triangulation of a 3d octahedron.%
\footnote{As mentioned at the end of Section \ref{sec:vis}, there are other ways to model the (3,3) move. In  \cite{CarterKauffmanSaito, kashaev2015, kashaev2018} there appeared sequences with only (2,3) moves, and no (3,2) moves. Unfortunately, translating the model of \cite{CarterKauffmanSaito, kashaev2015, kashaev2018} to field theory necessarily involves 3d $\CN=2$ theories with monopole operators in their superpotentials. The octahedron model of Figure \ref{fig:33sec2} neatly avoids this complication.} %
The first sequence (containing two (2,3) moves and one (3,2) move) is depicted along the top of the figure, and the second sequence (again containing two (2,3) moves and one (3,2) move) is depicted along the bottom. 

Using the dictionary of Section \ref{sec:TD4}, we can translate the two sequences of 3d Pachner moves on the octahedron to two sequences of duality interfaces. If the anticipated correspondence between field theory and geometry is robust, the two sequences of interfaces must themselves be IR dual to each other.

Let us expand briefly on this statement. After choosing a polarization for the boundary of the octahedron (discussed further below), the six triangulations of the octahedron in Figure~\ref{fig:33sec2} produce six different 3d $\CN=2$ theories of class $\mathcal R$. We label the octahedra (and associated theories) I--VI, as in Figure \ref{fig:octtheories} on page \pageref{fig:octtheories}. These six 3d bulk theories are all IR dual.

The sequence of 3d Pachner moves along the top of Figure \ref{fig:octtheories} corresponds to a sequence of three duality interfaces connecting the 3d theories
$\mathcal T_{\rm I}\to \mathcal T_{\rm II} \to \mathcal T_{\rm III}\to \mathcal T_{\rm IV}$. Colliding these interfaces together produces a single duality interface $\mathcal I_{\rm top}$ between theories I and IV. The collision involves a partial flow to the IR, enough to make most modes of the 3d fields inbetween interfaces massive. We will obtain a description of $\CI_{\rm top}$ as a Lagrangian 2d $\CN=(0,2)$ theory with finitely many fields, coupled to the bulk 3d Lagrangians of theories I and IV.
Similarly, the Pachner moves along the bottom connect the 3d theories
$\mathcal T_{\rm I}\to \mathcal T_{\rm VI} \to \mathcal T_{\rm V}\to \mathcal T_{\rm IV}$ with a different set of interfaces, whose collision (with partial flow to the IR) produces a second duality interface $\mathcal I_{\rm bot}$ between theories I and IV.

Now, in the deep IR, both of the bulk theories $\CT_{\rm I}$ and $\CT_{\rm IV}$ should flow to the same 3d SCFT, say $\CT_{\rm IR}$. Similarly, we expect the interfaces $\CI_{\rm top}$ and $\CI_{\rm bot}$ to both flow to the identity (transparent) interface in $\CT_{\rm IR}$. This expectation ultimately relies on being able to commute two RG flows: 1) the flow implicit in the collision of duality interfaces between adjacent octahedron theories; and 2) the flows to the IR in the 3d bulks of the various octahedron theories. Assuming the flows do commute, the IR equivalence of $\CI_{\rm top}$ and $\CI_{\rm bot}$ becomes the most basic physical manifestation of the (3,3) Pachner move.
We will verify the equivalence of $\CI_{\rm top}$ and $\CI_{\rm bot}$ explicitly in Section \ref{sec:index}, by computing interface indices.

More interesting consequences of the IR duality between $\CI_{\rm top}$ and $\CI_{\rm bot}$ can be obtained by ``probing'' these interfaces with additional boundary conditions.
For example, given any left boundary condition%
\footnote{As usual, we restrict ourselves to boundary conditions that flow to superconformal b.c. in the infrared; in particular, we require boundary conditions to preserve 2d $\CN=(0,2)$ SUSY and a $U(1)_R$ symmetry} %
$\CB_{\rm I}$ for theory $\CT_{\rm I}$, we can collide it with either $\CI_{\rm top}$ or $\CI_{\rm bot}$ (using a partial IR flow) to produce a new boundary condition for theory $\CT_{\rm IV}$. We denote these collisions as $\CB_{\rm I}\circ \CI_{\rm top}$ and $\CB_{\rm I}\circ \CI_{\rm bot}$, respectively. Assuming again that collision commutes with bulk RG flow, the (3,3) Pachner move should manifest as a duality between pairs of boundary conditions for theory $\CT_{\rm IV}$:
\be \text{(3,3) move on b.c.:} \qquad   \CB_{\rm I}\circ \CI_{\rm top} \; \overset{\text{IR dual}}{\simeq} \; \CB_{\rm I}\circ \CI_{\rm bot}  \qquad \forall\, \CB_{\rm I}\,. \label{33-boundaries} \ee

Going one step further, we may also choose a right boundary condition $\CB_{\rm IV}$ for theory $\CT_{\rm IV}$. Then ``sandwiching'' interfaces between $\CB_{\rm I}$ on the left and $\CB_{\rm IV}$ on the right (doing a partial flow to collide) produces a pair of purely 2d $\CN=(0,2)$ theories, $\mathcal B_{\rm I}\circ \CI_{\rm top} \circ \mathcal B_{\rm IV}$ and $\mathcal B_{\rm I}\circ \CI_{\rm bot} \circ \mathcal B_{\rm IV}$. The Pachner move now manifests as an IR duality of 2d theories
\be \text{(3,3) move on 2d theories:} \qquad  \mathcal B_{\rm I}\circ \CI_{\rm top} \circ \mathcal B_{\rm IV} \; \overset{\text{IR dual}}{\simeq} \;  \mathcal B_{\rm I}\circ \CI_{\rm bot} \circ \mathcal B_{\rm IV} \qquad \forall\,\CB_{\rm I},\CB_{\rm IV} \,.  \label{33-sandwiches} \ee

We will consider scenarios of both types \eqref{33-boundaries} and \eqref{33-sandwiches} below, and will use supersymmetric indices to check the expected dualities.

\subsection{3d theories and interfaces}
\label{sec:oct-theories}

To proceed, we carefully identify the 3d class-$\mathcal R$ theories associated to the six octahedra, and the duality interfaces that link them. We choose a polarization for the boundary of the octahedron (independent of its internal triangulation) whose subset of edges form a great circle, shown in bold here:
\be \notag \raisebox{-.5in}{\includegraphics[width=1.3in]{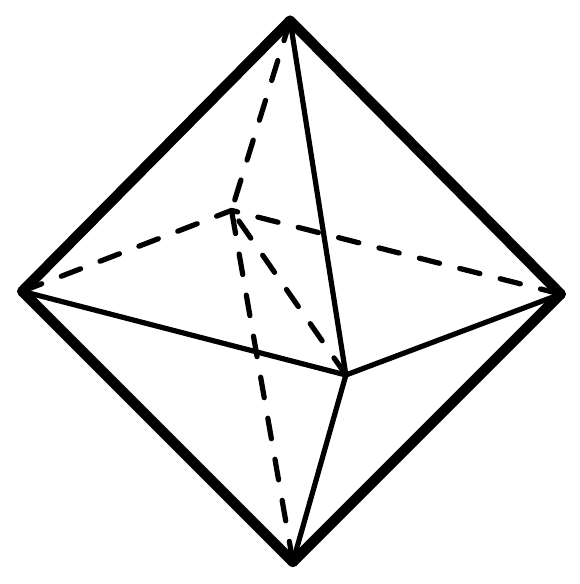}} \ee

\begin{figure}[htb]
\centering
\includegraphics[width=6in]{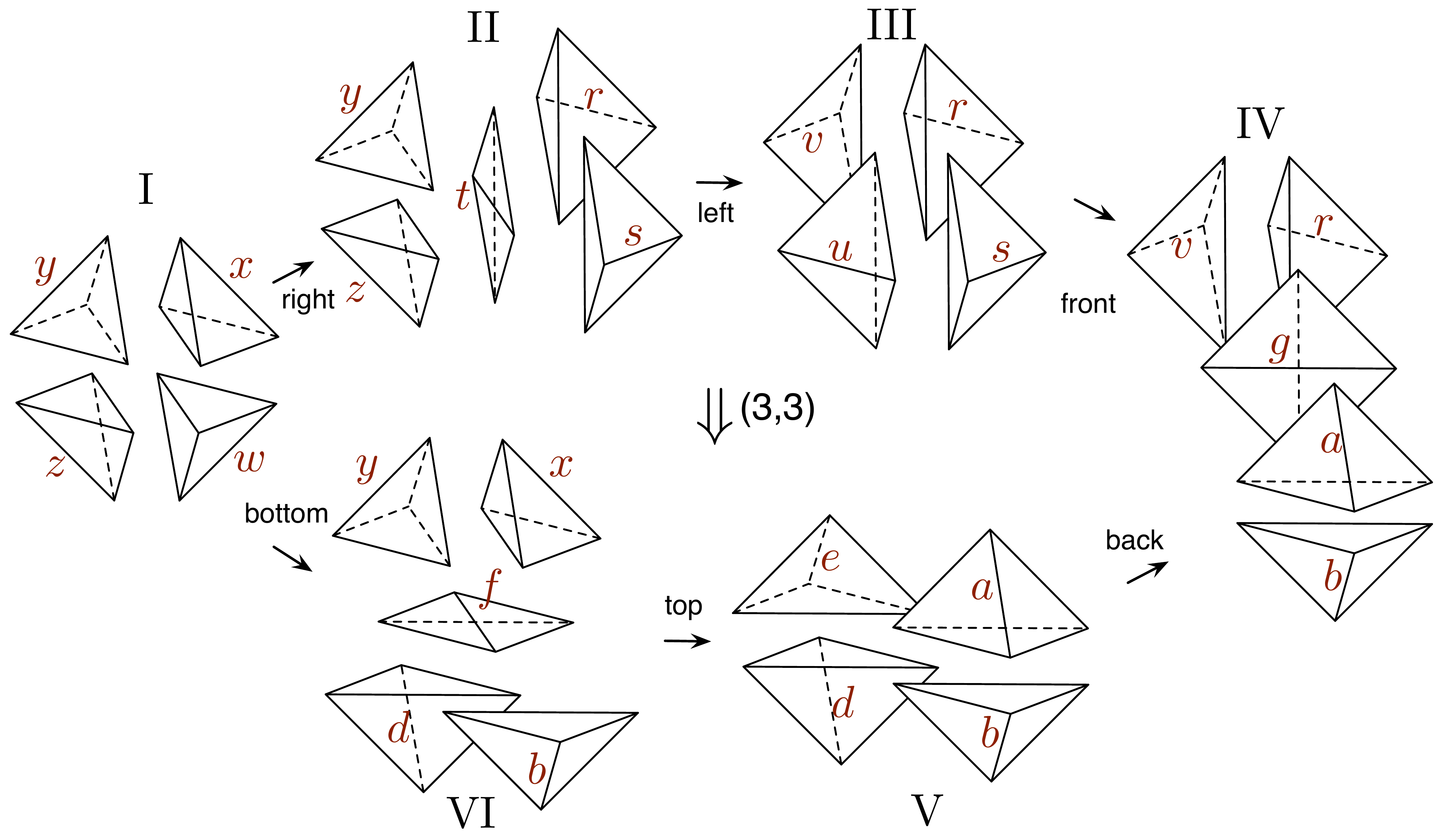}
\caption{Six theories of class $\mathcal R$, labelled I - VI, are associated to the triangulated octahedra in the (3,3) move. Each tetrahedron (labelled $w,x,y,z,...$) gives rise to a 3d chiral multiplet (labelled $\phi_w,\phi_x,\phi_y,\phi_z,...$) in one of these theories.}
\label{fig:octtheories}
\end{figure}

This choice of polarization makes the initial theory $\CT_{\rm I}$ as simple as possible: it contains four chiral multiplets 
coupled by a quartic superpotential. 
The other theories can be described schematically as follows. In Figure \ref{fig:octtheories} we label the tetrahedra corresponding to the various chiral multiplets that appear here.
\begin{itemize}
\item[$\CT_{\rm I}$:] Four chirals, $W_{\rm I}=\phi_w\phi_x\phi_y\phi_z$.
\item[$\CT_{\rm II}$:] Five chirals $\phi_r,\phi_s,\phi_t,\phi_y,\phi_z$ with charges $(1,-1,0,0,0)$ under a $U(1)_{g_1}$ gauge symmetry, with $W_{\rm II} = \phi_r\phi_s\phi_t + \phi_t\phi_y\phi_z$.
\item[$\CT_{\rm III}$:] Four chirals $\phi_r,\phi_s,\phi_v,\phi_u$ with charges $(1,-1,0,0)$, $(0,0,1,-1)$ under $U(1)_{g_1}\times U(1)_{g_2}$ gauge symmetry, coupled by $W_{\rm III}=\phi_r\phi_s\phi_u\phi_v$, \emph{i.e.} two coupled copies of $N_f=1$ SQED.
\item[$\CT_{\rm IV}$:] Five chirals $\phi_b,\phi_a,\phi_g,\phi_r,\phi_v$, with charges $(0,1,-1,1,0)$, $(1,0,-1,0,1)$, $(1,-1,0,0,0)$ under $U(1)_{g_1}\times U(1)_{g_2}\times U(1)_{g_3}$ gauge symmetry, with $W_{\rm IV}=\phi_a\phi_b\phi_g+\phi_g\phi_r\phi_v$, \emph{i.e.} two copies of $N_f=1$ SQED with a common axial symmetry gauged.
\item[$\CT_{\rm V}$:] Similar to $\CT_{\rm III}$, with chirals relabeled as $(\phi_r,\phi_s,\phi_v,\phi_u)\to(\phi_b,\phi_d,\phi_a,\phi_e)$.
\item[$\CT_{\rm VI}$:] Similar to $\CT_{\rm II}$, with chirals relabeled as $(\phi_r,\phi_s,\phi_t,\phi_y,\phi_z)\to(\phi_b,\phi_d,\phi_f,\phi_y,\phi_x)$.
\end{itemize}

\begin{figure}[htb]
\centering
\includegraphics[width=4in]{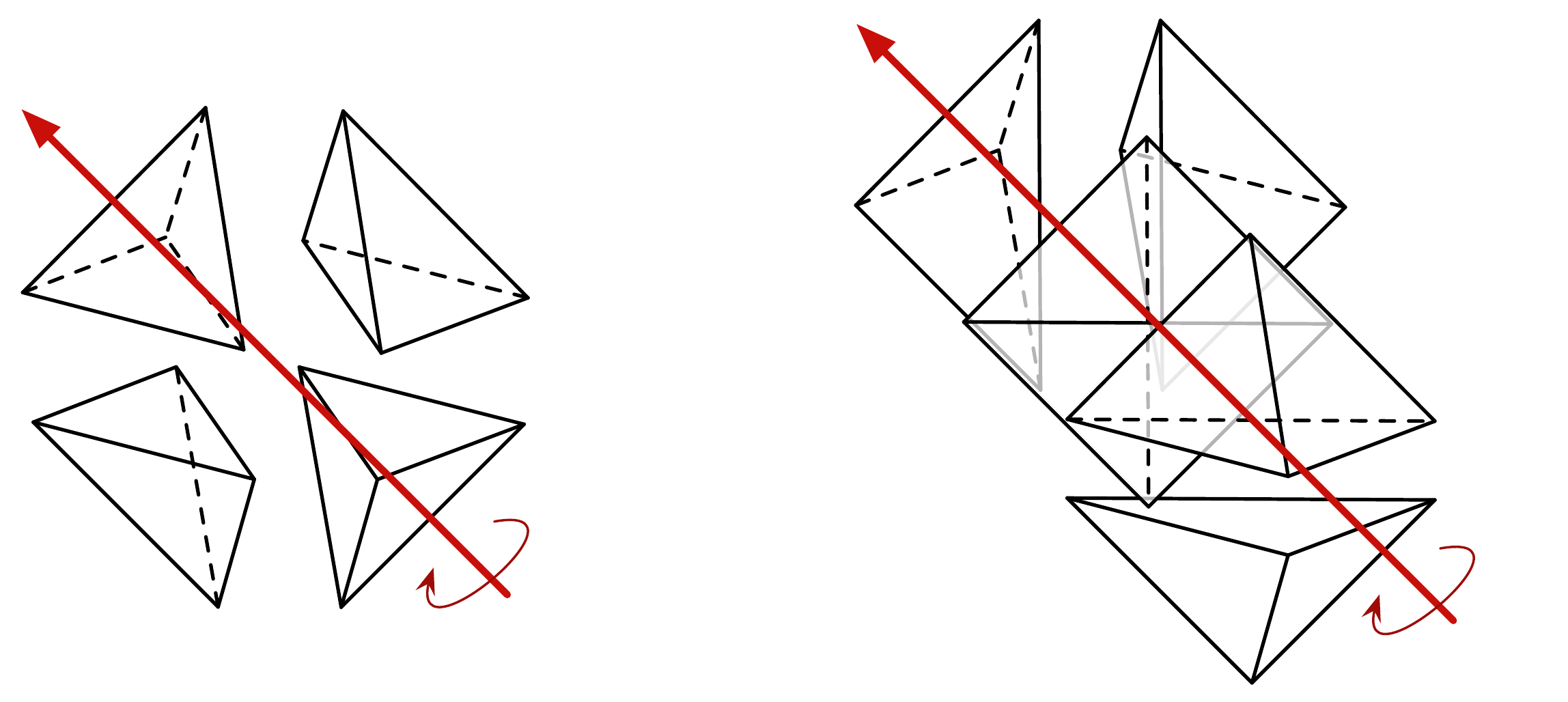}
\caption{A 180$^\circ$ rotation relates the top sequence of triangulated octahedra to the bottom sequence. Here the triangulated octahedra I and IV are depicted, for which the rotation is a symmetry. (In contrast, octahedra II$\leftrightarrow$VI and III$\leftrightarrow$V are interchanged.)}
\label{fig:rotation}
\end{figure}

Notice that the theories along the bottom and top halves of the loop look identical, up to a relabeling of fields. This simply reflects the geometric fact that the sequence of 3d Pachner moves along the top of the loop is related to the sequence of Pachner moves along the bottom of the loop by a rotation of 180$^\circ$ about a diagonal axis, as shown in Figure \ref{fig:rotation}. This rotation permutes various chirals associated to tetrahedra. Crucially, it also permutes the action of flavor symmetries, which we have yet to spell out. We shall see that matching flavor symmetries between the top and bottom sequences is where much of the nontriviality of the (3,3) move lies.

Each octahedron theory has a $U(1)^3$ flavor symmetry, which we denote as $U(1)_{v_1}\times U(1)_{v_2}\times U(1)_a$ below. (The notation is suggestive of ``vector'' and ``axial'' symmetries in the SQED-like theories along the top.) In theory $\CT_{\rm I}$, the flavor symmetry acts in a standard way by rotating the chiral multiplets; while in theory $\CT_{\rm IV}$ the flavor symmetry is entirely topological, and rotates the three dual photons. 
The flavor charges and Chern-Simons couplings for each octahedron theory are summarized in Appendix \ref{app:tables}. (The precise way in which the flavor theories become topological can be read off from the Chern-Simons couplings.)

\subsubsection{Interfaces along the top}
\label{sec:33int-top}

Let us now go through the top half of the loop in steps ($\CT_{\rm I} \rightarrow \CT_{\rm II} \rightarrow \CT_{\rm III} \rightarrow \CT_{\rm IV}$). We wish to describe the basic duality interfaces between each pair of theories, and to collide them together to obtain the composite interface $\CI_{\rm top}$ between $\CT_{\rm I}$ and $\CT_{\rm IV}$.

\begin{itemize}
\item[$\CT_{\rm I} \rightarrow \CT_{\rm II}$:] This interface is a copy of \eqref{TD4-long'}, implementing a (2,3) Pachner move on two of the chiral multiplets of $\CT_{\rm I}$.
It is defined by using N b.c. for $\phi_x,\phi_w$ (on the left) and N,N,D b.c. for $\phi_r,\phi_s,\phi_t$ (on the right); then introducing a 2d fermi multiplet $\Gamma$ with interface superpotential
\begin{subequations} \label{int-top1}
\be \mathcal{W}_{\rm I-II}= \int d\theta^+ \left(\phi_x \Gamma \phi_s - \phi_x \phi_w \Psi_{t}\right)\,,\qquad  E_{\Gamma} = -\phi_r \phi_w\,. \ee
The $-\phi_x\phi_w\Psi_t$ coupling sets $\phi_x\phi_w=\phi_t$ as desired, and the boundary E and J terms factor the difference of bulk superpotentials: $J_\Gamma E_\Gamma = -\phi_x\phi_s\phi_r\phi_w = -\phi_r\phi_s\phi_t = W_{\rm I}- W_{\rm II}$. We summarize the interface as
\be {\rm N}_x {\rm N}_w \big|\Gamma\big| \CN_{g_1}  {\rm D}_t {\rm N}_r {\rm N}_s\,. \ee
\end{subequations}
Note that the interface is completely invisible (transparent) to $\phi_y$ and $\phi_z$.

\item[$\CT_{\rm II} \rightarrow \CT_{\rm III}$:] The next interface is a copy of \eqref{XYZ-int1} involving $\phi_t,\phi_y,\phi_z$ on the left and $\phi_u,\phi_v$ on the right, with a 2d fermi multiplet $\Gamma'$ in the middle. It can be summarized as
\be \label{int-top2} \hspace{-.5in} {\rm D}_t{\rm N}_y {\rm N}_z \big|\Gamma'\big| \CN_{g_2} {\rm N}_u {\rm N}_v\,,\qquad
\mathcal{W}_{\rm II-III} = \int d\theta^+ \left(\phi_y \Gamma' \phi_v +  \phi_u \phi_v \Psi_t \right)\,,\qquad E_{\Gamma'} = \phi_z \phi_u\,. \ee
This interface is transparent to $\phi_r,\phi_s$ and to the gauge multiplet for $U(1)_{g_1}$.

\item[$\CT_{III} \rightarrow \CT_{IV}$:]  The final interface is again a copy of \eqref{TD4-long'}, involving a 2d fermi multiplet $\Gamma''$:
\be \label{int-top3} \hspace{-.5in} {\rm N}_u {\rm N}_s \big|\Gamma''\big| \CN_{g_3} {\rm D}_g {\rm N}_a {\rm N}_b\,,\qquad \mathcal{W}_{\rm III-IV} =   \int d\theta^+ \left(\phi_{u} \Gamma'' \phi_{b} -  \phi_{u} \phi_{s} \Psi_{g} \right)\,,\qquad E_{\Gamma''} =  -\phi_{s}\phi_{a}\,.\ee
\end{itemize}

When we collide these interfaces together to form $\CI_{\rm top}$, the bulk chiral multiplets $\phi_t,\phi_s,\phi_u$ get trapped, and potentially contribute additional 2d degrees of freedom. Note that these bulk chirals correspond to the three ``internal'' tetrahedra of \eqref{bdy3-}. To determine precisely what 2d degrees of freedom remain, we must examine the boundary conditions for $\phi_t,\phi_s,\phi_u$ on the interfaces that trap them:
\begin{itemize}
\item $\phi_t$ has Dirichlet b.c. on the left from \eqref{int-top1} and Dirichlet b.c. on the right form \eqref{int-top2}. Decomposing $\phi_t$ under the $\CN=(0,2)$ SUSY subalgebra into a chiral/fermi pair $(\Phi_t,\Psi_t)$, we find that $\Psi_t$ survives and contributes a purely 2d fermi multiplet to $\CI_{\rm top}$.
\item $\phi_s$ has Neumann b.c. on the left from \eqref{int-top1} and Neumann b.c. on the right from \eqref{int-top3}. It contributes a 2d chiral multiplet $\Phi_s$ to $\CI_{\rm top}$.
\item $\phi_u$ has N b.c. on the left from \eqref{int-top2} and N b.c. on the right from \eqref{int-top3}. It contributes a 3d chiral multiplet $\Phi_u$ to $\CI_{\rm top}$.
\end{itemize}

The composite interface $\CI_{\rm top}$ thus contains four 2d fermi multiplets ($\Gamma,\Gamma',\Gamma'',\Psi_t$) and two 2d chiral multiplets $(\Phi_s,\Phi_u)$.
Their couplings to the bulk matter of theories $\CT_{\rm I}$ and $\CT_{\rm IV}$ (and to each other) come from the sum of the three interface superpotentials above, as well as from $E$-terms. Notably, the fermi multiplet $\Psi_t$ also has an $E$-term $E_t = -\pd W_{\rm II}/\pd \phi_t  = -(\phi_y\phi_z+ \phi_r\Phi_s)$, induced by the bulk superpotential of theory $\CT_{\rm II}$. (As explained \emph{e.g.} in \cite{DGP-bdy}, the decomposition of 3d $\CN=2$ chirals into 2d $\CN=(0,2)$ chiral/fermi pairs turns 3d $F$-terms into 2d $E$-terms.) Altogether, the composite interface $\CI_{\rm top}$ may be summarized as:
\begin{subequations} \label{Itop}
\be \CI_{\rm top}\,:\qquad {\rm N}_w{\rm N}_x{\rm N}_y{\rm N}_z \big| \Gamma,\Gamma',\Gamma'',\Psi_t;\Phi_s,\Phi_u\big| \CN_{g_1}\CN_{g_2}\CN_{g_3} {\rm N}_a {\rm N}_b {\rm D}_g {\rm N}_v {\rm N}_r\,. \ee
The $J$ and $E$ terms of the 2d fermi multiplets are
\be \begin{array}{c|cccc|c} 
 & \Gamma & \Gamma' & \Gamma'' & \Psi_t & (\Psi_g) \\\hline
J & \phi_x\Phi_s & \phi_y\phi_v & \Phi_u\phi_b & \Phi_u\phi_v-\phi_x\phi_w  & (-\Phi_u\Phi_s) \\
E & -\phi_r\phi_w & \phi_z\Phi_u & -\Phi_s\phi_a & -\phi_y\phi_z- \phi_r\Phi_s & (-\phi_a\phi_b-\phi_r\phi_v) 
\end{array}
\ee
(Here we have also indicated the couplings involving the fermi half of the bulk $\phi_g$ from $\CT_{\rm IV}$. At the interface, these couplings set $\phi_g = \Phi_u\Phi_s$.) Rather beautifully, the interface $J$ and $E$ terms factorize the difference of bulks superpotentials,
\be J\cdot E = J_{\Gamma}E_{\Gamma} + J_{\Gamma'}E_{\Gamma'}+ J_{\Gamma''}E_{\Gamma''} + J_tE_t = \phi_w\phi_x\phi_y\phi_z - \phi_a\phi_b\phi_g - \phi_g\phi_r\phi_v = W_{\rm I}-W_{\rm IV} \ee

The charges of various fields under the $U(1)_{g_1}\times U(1)_{g_2}\times U(1)_{g_3}$ gauge group (on the right), the $U(1)_{v_1}\times U(1)_{v_2}\times U(1)_a$ flavor symmetry, and $U(1)_R$ R-symmetry are completely determined by the $J$ and $E$ couplings. For reference, these charges are:
\be
\begin{array}{c|cccc|cccccc|ccccc}
 & \phi_w & \phi_x & \phi_y & \phi_z & \Gamma & \Gamma' & \Gamma'' & \Psi_t  & \Phi_s & \Phi_u  & \phi_a & \phi_b &  \phi _g & \phi_v & \phi_r \\ \hline
 U(1)_{g_1} &0&0&0&0& 1&0&0& 0&-1&0& 1&0&-1&0&1 \\
 U(1)_{g_2} &0&0&0&0& 0&-1&0& 0&0&-1& 0&1&-1&1&0 \\
 U(1)_{g_3} &0&0&0&0& 0&0&-1& 0&0&0& -1&1&0&0&0 \\ \hline
 U(1)_{v_1} & 1 & -1 & 0 & 0 &1&0&0& 0&0&0& 0&0&0&0&0 \\
 U(1)_{v_2} & 0 & 1 & 0 & -1 &0&0&-1& -1&-1&1& 0&0&0&0&0 \\
 U(1)_{a} & 0 & 0 & 1 & -1 &0&-1&0& 0&0&0&  0&0&0&0&0 \\ \hline
 U(1)_R   &\frac12&\frac12&\frac12&\frac12  &0&0&0& 0&\frac12&\frac12 &\frac12&\frac12&1&\frac12&\frac12
\end{array}
\ee

\end{subequations}

\subsubsection{Interfaces along the bottom}

We can follow the same procedure to determine the interfaces along the bottom half of the loop. Alternatively, we can simply apply the symmetry of Figure \ref{fig:rotation} (discussed in greater detail in Appendix \ref{app:sym}) to the description of interfaces along the top, to obtain the interfaces along the bottom.

Applying the symmetry is more direct and revealing.
It amounts to 1) modifying charges of $\Gamma,\Gamma',\Gamma''$ and $\phi_t,\phi_s,\phi_u$ according to the field redefinition
\be {\bf v_2} \to {\bf v_1}-{\bf a}-{\bf v_2}\,,\qquad   {\bf g_1}\to {\bf g_2}+{\bf g_3}\,,\qquad {\bf g_2}\to {\bf g_1}-{\bf g_3}\,;  \label{topbot-sym}\ee
and 2) relabeling 
$\phi_t,\phi_s,\phi_u\to \phi_f,\phi_d,\phi_e$ and swapping the roles of $(\phi_x,\phi_a,\phi_b)\leftrightarrow (\phi_z,\phi_v,\phi_r)$ in all 2d or 3d superpotentials. (By a slight abuse of notation, we will denote the fermi multiplets on basic (2,3) and (3,2)  interfaces as $\Gamma$ once again.)

In particular, this leads to the following description of the composite interface $\CI_{\rm bot}$. As before, there are four 2d fermi multiplets and two 2d chiral multiplets, coupled to 3d bulk fields with boundary conditions
\begin{subequations} \label{Ibot}
\be \CI_{\rm bot}\,:\qquad {\rm N}_w{\rm N}_x{\rm N}_y{\rm N}_z \big| \Gamma,\Gamma',\Gamma'',\Psi_f;\Phi_d,\Phi_e\big| \CN_{g_1}\CN_{g_2}\CN_{g_3} {\rm N}_a {\rm N}_b {\rm D}_g {\rm N}_v {\rm N}_r\,. \ee
The $J$ and $E$ terms have become
\be \begin{array}{c|cccc|c} 
 & \Gamma & \Gamma' & \Gamma'' & \Psi_f & (\Psi_g) \\\hline
J & \phi_z\Phi_d & \phi_y\phi_a & \Phi_e\phi_v & \Phi_e\phi_a-\phi_z\phi_w  & (-\Phi_e\Phi_d) \\
E & -\phi_b\phi_w & \phi_x\Phi_e & -\Phi_d\phi_r & -\phi_y\phi_x- \phi_b\Phi_d & (-\phi_a\phi_b-\phi_r\phi_v) 
\end{array}
\ee
and the charges are now
\be
\begin{array}{c|cccc|cccccc|ccccc}
 & \phi_w & \phi_x & \phi_y & \phi_z & \Gamma & \Gamma' & \Gamma'' & \Psi_f  & \Phi_d & \Phi_e  & \phi_a & \phi_b &  \phi _g & \phi_v & \phi_r \\ \hline
 U(1)_{g_1} &0&0&0&0& 0&-1&0&0&0&-1& 1&0&-1&0&1 \\
 U(1)_{g_2} &0&0&0&0& 1&0&0&0&-1&0& 0&1&-1&1&0 \\
 U(1)_{g_3} &0&0&0&0& 1&1&-1&0&-1&1& -1&1&0&0&0 \\ \hline
 U(1)_{v_1} & 1 & -1 & 0 & 0 &1&0&-1&-1&-1&1& 0&0&0&0&0 \\
 U(1)_{v_2} & 0 & 1 & 0 & -1 &0&0&1&1&1&-1& 0&0&0&0&0 \\
 U(1)_{a} & 0 & 0 & 1 & -1 &0&-1&1&1&1&-1&  0&0&0&0&0 \\ \hline
 U(1)_R   &\frac12&\frac12&\frac12&\frac12  &0&0&0& 0&\frac12&\frac12 &\frac12&\frac12&1&\frac12&\frac12
\end{array}
\ee
\end{subequations}

\subsection{A functional identity for the (3,3) move}\label{sec:half33}

The interfaces $\CI_{\rm top}$ and $\CI_{\rm bot}$ are distinct in the UV. Though their 2d matter content is similar, both the charges under bulk flavor symmetries and the $J$ and $E$ couplings between 2d and 3d matter are manifestly different. As explained in the introduction to this section, the (3,3) Pachner move should correspond to an IR duality between $\CI_{\rm top}$ and $\CI_{\rm bot}$. We can check the expected duality by computing supersymmetric indices that count protected local operators bound to $\CI_{\rm top}$ and $\CI_{\rm bot}$.

The computation of ``interface indices'' in this case turns out to follow quickly from considering the collision of $\CI_{\rm top}$ and $\CI_{\rm bot}$ with a particular boundary condition for $\CT_{\rm I}$. Recall from \eqref{33-boundaries} that given a left boundary condition $\CB_{\rm I}$ for $\CT_{\rm I}$ (that preserves $\CN=(0,2)$ SUSY and $U(1)_R$), we also expect the collisions of $\CB_{\rm I}$ with $\CI_{\rm top}$ and $\CI_{\rm bot}$ to be IR-dual:
\be \CB_{\rm I} \circ \CI_{\rm top}\;\simeq\; \CB_{\rm I}\circ \CI_{\rm bot} \label{33-bdy2}\,. \ee
We can check an identity of the form \eqref{33-bdy2} by computing half-indices.

The half-index was introduced in \cite{GGP} and further explored in \cite{YoshidaSugiyama, DGP-bdy}. 
We will largely follow the conventions and definitions of \cite{DGP-bdy} for the half-index, except we use a $\mathbb{Z}/2 \mathbb{Z}$ homological grading given by fermion number $F$, rather than the grading given by $R$-symmetry. The half-index of a boundary condition $\CB$ for a 3d theory $\CT$ then becomes
\begin{equation}
\half_{\CT, \CB}(x;q):= {\rm Tr_{Ops}}_{\CB} (-1)^Fq^{J + R/2}x^{e}\,,
\end{equation} where the trace is taken over boundary local operators, $R$ is $U(1)_R$ charge, $J$ is the generator of $Spin(2)$ rotations in the plane of the boundary, and $x$ is a shorthand for any (and all) flavor fugacities with generators $e$.\footnote{For theories with integral R-charges, changing to this new grading amounts to replacing $q^{1/2} \mapsto -q^{1/2}$ in all formulas. We use $(-1)^F$ rather than $(-1)^R$ precisely because our theories are most naturally assigned non-integral R-charges.}
We record the basic functions used in the subsequent expressions in Appendix \ref{app:fns} and refer to \cite{DGP-bdy} for the systematics of how these expressions are obtained.

Now, consider the left boundary condition $\CB_{\rm I}=$\,(DDDD) for theory $\CT_{\rm I}$. In other words, we simply put Dirichlet b.c. on each of the four chirals $\phi_w,\phi_x,\phi_y,\phi_z$. The corresponding half-index may be immediately written down:
\begin{align}
\half_{\CT_{\rm I}, \CB_{\rm I}} &= \half_D(q^{\frac14} v_1)\half_D(q^{\frac14} v_2 v_1^{-1})\half_D(q^{\frac14} a)\half_D(q^{\frac14}/(a v_2)) \nonumber\\ 
&= (q^{\frac34}/v_1^{-1};q)_\infty (q^{\frac34}v_1v_2^{-1};q)_\infty (q^{\frac34}a^{-1};q)_\infty (q^{\frac34} av_2;q)_\infty \label{DDDD-index} \\
&= 1 + \left(-{1 \over a} - {1 \over v_1} - {v_1 \over v_2} - a v_2 \right)q^{3/4} + \left(a v_1 + {1 \over a v_1} + v_2 + {1 \over v_2} + {v_1 \over a v_2} + {a v_2 \over v_1}\right)q^{3/2} + O(q^{7/4}) \nonumber
\end{align} 
We use fugacities $v_1,v_2,a$ for flavor symmetries $U(1)_{v_1},U(1)_{v_2},U(1)_a$.
Note the manifest symmetry of this expression with respect to the exchange $v_2\to v_1/(av_2)$, corresponding to the 180$^\circ$ rotation of Figure \ref{fig:rotation}.

Next, consider the collisions $\CB_{\rm I}\circ \CI_{\rm top}$ and $\CB_{\rm I}\circ \CI_{\rm bot}$, which define boundary conditions for~$\CT_{\rm IV}$. Since $\CI_{\rm top}$ and $\CI_{\rm bot}$ should both be duality interfaces, both $\CB_{\rm I}\circ \CI_{\rm top}$ and $\CB_{\rm I}\circ \CI_{\rm bot}$ should be IR-dual to the simple boundary condition $\CB_{\rm I}$ for $\CT_{\rm I}$. In particular, $\CB_{\rm I}\circ \CI_{\rm top}$ and $\CB_{\rm I}\circ \CI_{\rm bot}$ should be IR-dual to each other.

Let us determine the matter content of $\CB_{\rm I}\circ \CI_{\rm top}$ and $\CB_{\rm I}\circ \CI_{\rm bot}$. Since $\CB_{\rm I} = {\rm D}_w {\rm D}_x {\rm D}_y {\rm D}_z$, colliding with $\CI_{\rm top} =  {\rm N}_w {\rm N}_x {\rm N}_y {\rm N}_z\big|...$ as in (\ref{Itop}a) completely kills all four 3d chirals $\phi_w,\phi_x,\phi_y,\phi_z$ from $\CT_{\rm I}$. (The D b.c. kills their 2d chiral parts and the N b.c. kills their 2d fermi parts.) We are simply left with
\be \label{BItop} \CB_{\rm I}\circ \CI_{\rm top} \,= \,\Gamma,\Gamma',\Gamma'',\Psi_t;\Phi_s,\Phi_u\big| \CN_{g_1}\CN_{g_2}\CN_{g_3} {\rm N}_a {\rm N}_b {\rm D}_g {\rm N}_v {\rm N}_r\,,\ee
\emph{i.e.} with 2d multiplets $\Gamma,\Gamma',\Gamma'',\Psi_t,\Phi_s,\Phi_u$, coupled to the bulk fields of $\CT_{\rm IV}$ via the $J$ and $E$ terms of (\ref{Itop}b), restricted to $\phi_w=\phi_x=\phi_y=\phi_z=0$. Similarly,
\be  \label{BIbot} \CB_{\rm I}\circ \CI_{\rm bot} \,=\, \Gamma,\Gamma',\Gamma'',\Psi_f;\Phi_d,\Phi_e\big| \CN_{g_1}\CN_{g_2}\CN_{g_3} {\rm N}_a {\rm N}_b {\rm D}_g {\rm N}_v {\rm N}_r\,, \ee
where now $\Gamma,\Gamma',\Gamma''$ refer to the 2d fermis of \eqref{Ibot}.

By using the charge assignments from \eqref{Itop} and \eqref{Ibot}, the two new half-indices are readily written down:
\begin{align} \half_{\CT_{\rm IV}, \CB_{\rm I}\circ \CI_{\rm top}} &= (q)_{\infty}^3 \oint {dg_1 \over (2 \pi i g_1)} \oint {dg_2 \over (2 \pi i g_2)} \oint {dg_3 \over (2 \pi i g_3)} \notag \\
 & \times \text{F}(g_1 v_1)  \text{F}(1/(a g_2)) \text{F}(1/(g_3 v_2)) \text{F}(1/v_2) \text{C}(q^{\frac14}/(g_1 v_2))\text{C}(q^{\frac14} v_2/g_2) \notag \\
 & \qquad \times \half_N(q^{\frac14} g_1) \half_N(q^{\frac14} g_2)\half_D(q^{\frac14}/(g_1 g_2))\half_N(q^{\frac14} g_2 g_3)\half_N(q^{\frac14} g_1/g_3) \,,
 \end{align}
\begin{align}\nonumber \half_{\CT_{\rm IV}, \CB_{\rm I}\circ \CI_{\rm bot}} &= (q)_{\infty}^3 \oint {dg_1 \over (2 \pi i g_1)} \oint {dg_2 \over (2 \pi i g_2)} \oint {dg_3 \over (2 \pi i g_3)} \notag \\
&\hspace{-.2in} \times \text{F}(v_1 g_2 g_3) \text{F}(g_3/(ag_1))\text{F}(av_2/(g_3 v_1)) \text{F}(a v_2/v_1) \text{C}(q^{\frac14} a v_2/(g_2 g_3 v_1)) \text{C}(q^{\frac14} v_1g_3/(a g_1 v_2)) \notag \\
 & \qquad  \half_N(q^{\frac14} g_1) \half_N(q^{\frac14} g_2)\half_D(q^{\frac14}/(g_1 g_2))\half_N(q^{\frac14} g_2 g_3)\half_N(q^{\frac14} g_1/g_3) \,.
\end{align}
Here F and C denote half-indices of 2d fermi and chiral multiplets --- see Appendix \ref{app:fns}.\footnote{Also note that the contour integrals, corresponding to 3d gauge fields in $\CT_{\rm IV}$, are \emph{not} to be evaluated via a Jeffrey-Kirwan prescription. They simply denote projections to invariants. The integrand should be expanded as a series in $q$, whose coefficients are polynomials in $g_1,g_2,g_3$; and the integrals pick out the terms independent of $g_1,g_2,g_3$.} %

Satisfyingly, one can expand both of the functions $\half_{\CT_{\rm IV}, \CB_{\rm I}\circ \CI_{\rm top}} $ and $\half_{\CT_{\rm IV}, \CB_{\rm I}\circ \CI_{\rm bot}}$, and check order-by-order in $q$ that they are both equal to $\half_{\CT_{\rm I}, \CB_{\rm I}}$. In particular, they are equal to each other! We have checked to $O(q^8)$. We suspect that equivalence
\be \half_{\CT_{\rm IV}, \CB_{\rm I}\circ \CI_{\rm top}} = \half_{\CT_{\rm I}, \CB_{\rm I}} = \half_{\CT_{\rm IV}, \CB_{\rm I}\circ \CI_{\rm bot}} \label{33halfequiv} \ee
can be proved using difference equations, similar to some of the examples in \cite{DGP-bdy}.

We also observe that $\half_{\CT_{\rm IV}, \CB_{\rm I}\circ \CI_{\rm top}}$ and $\half_{\CT_{\rm IV}, \CB_{\rm I}\circ \CI_{\rm bot}}$ are actually involved in a larger duality web. The simple boundary condition $\CB_{\rm I} = \text{DDDD}$ for $\CT_{\rm I}$ has a manifest $S_4$ symmetry, coming from permuting the four bulk multiplets $\phi_w,\phi_x,\phi_y,\phi_z$. As generators of $S_4$, we may take the three $\mathbb Z/2\mathbb Z$ symmetries that act on flavor fugacities as
\be \label{S4-TI} 1)\quad v_2\,\leftrightarrow\, \frac{v_1}{a v_2}\,,\qquad 2) \quad  v_1\leftrightarrow a\,,\; v_2\,\leftrightarrow \frac{1}{v_2}\,,\qquad 3) \quad v_1\,\leftrightarrow\, \frac{v_2}{v_1}\,.  \ee
The transformation (1) relates $\half_{\CT_{\rm IV},\CB_{\rm I}\circ \CI_{\rm top}}$ and $\half_{\CT_{\rm IV}, \CB_{\rm I}\circ \CI_{\rm bot}}$. (The integrands of $\half_{\CT_{\rm IV},\CB_{\rm I}\circ \CI_{\rm top}}$ and $\half_{\CT_{\rm IV}, \CB_{\rm I}\circ \CI_{\rm bot}}$ are related by (1) together with $(g_1,g_2)\leftrightarrow (g_2 g_3,g_1/g_3)$, which becomes trivial once the integration is performed.)
From $\half_{\CT_{\rm IV},\CB_{\rm I}\circ \CI_{\rm top}}(v_1,v_2,a) = \half_{\CT_{\rm I}, \CB_{\rm I}}(v_1,v_2,a)$ combined with \eqref{S4-TI} we find more generally that
\be \half_{\CT_{\rm IV},\CB_{\rm I}\circ \CI_{\rm top}}(v_1,v_2,a) =  \underset{=\half_{\CT_{\rm IV},\CB_{\rm I}\circ \CI_{\rm bot}}(v_1,v_2,a)}{\half_{\CT_{\rm IV},\CB_{\rm I}\circ \CI_{\rm top}}(v_1,v_1/(a v_2),a)}=\half_{\CT_{\rm IV},\CB_{\rm I}\circ \CI_{\rm top}}(a,1/v_2,v_1) =  \half_{\CT_{\rm IV},\CB_{\rm I}\circ \CI_{\rm top}}(v_1/v_1,v_2,a) \notag \ee
\be = \text{any other $S_4$ permutation of flavor symmetries}\,.\ee

\subsubsection{Upgrade to interface indices}\label{sec:index}

We come back to the question of whether the interfaces $\CI_{\rm top}$ and $\CI_{\rm bot}$ themselves are IR dual. We would like to test this by computing an index that counts local operators bound to them. We expect the result to be equal to the ``interface index'' of the identity interface in either theory $\CT_{\rm I}$ or theory $\CT_{\rm IV}$ --- \emph{a.k.a.} the full bulk 3d index \cite{IY, KapustinWillett} of theories $\CT_{\rm I}$ or $\CT_{\rm IV}$. 

The bulk 3d index of $\CT_{\rm I}$ is almost trivial to calculate, since there are no gauge fields. Comparing the formulas of \cite{KapustinWillett, DGG-index} to Appendix \ref{app:fns}, we find that the bulk index factorizes:
\begin{align} & I_{\CT_{\rm I}} = \half_N(q^{\frac14} v_1)\half_N(q^{\frac14} v_2 v_1^{-1})\half_N(q^{\frac14} a)\half_N(q^{\frac14}/(a v_2)) \\
 & \hspace{.25in} \times \half_D(q^{\frac14} v_1)\half_D(q^{\frac14} v_2 v_1^{-1})\half_D(q^{\frac14} a)\half_D(q^{\frac14}/(a v_2))\,. \notag\end{align}
This reflects the fact that one can ``factorize'' the identity interface in $\CT_{\rm I}$ by coupling a right (NNNN) boundary condition to a left (DDDD) boundary condition, using a canonical quadratic 2d superpotential. More succinctly:
\be I_{\CT_{\rm I}} = \half_{{\CT_{\rm I},{\rm NNNN}}} \times  \half_{{\CT_{\rm I},{\rm DDDD}}}\,. \ee

The interface indices for $\CI_{\rm top},\CI_{\rm bot}$ behave in a similar way. Using a standard  \textit{folding trick}, we can think of an interface between the two theories $\CT_{\rm I},\CT_{\rm IV}$ as a boundary condition for the tensor product $\CT_{\rm I}\otimes\CT_{\rm IV}$. This allows us to compute interface indices using standard formulas for half-indices.%
\footnote{One should in general take care to distinguish contributions from bulk fields coming from the right and from the left; however, in practice this only becomes important in the presence of Dirichlet b.c. for gauge fields. In the current setup, all boundary conditions on gauge fields are Neumann.}
Moreover, the left sides of $\CI_{\rm top}$ and $\CI_{\rm bot}$ are especially simple, in that there are no nontrivial gauge degrees of freedom involved in coupling to $\CT_{\rm I}$. Correspondingly, the interface indices factorize
\be \begin{array}{c} \half_{\CI_{\rm top}} =  \half_{{\CT_{\rm I},{\rm NNNN}}} \times  \half_{\CT_{\rm IV}, \CB_{\rm I}\circ \CI_{\rm top}}  \\[.1cm]
 \half_{\CI_{\rm bot}} =  \half_{{\CT_{\rm I},{\rm NNNN}}} \times  \half_{\CT_{\rm IV}, \CB_{\rm I}\circ \CI_{\rm bot}}\,. 
 \end{array}
\ee 
Therefore, the equivalence of interface indices
\be  \half_{\CI_{\rm top}}  =  I_{\CT_{\rm I}} =  \half_{\CI_{\rm top}}  \ee
becomes a direct consequence of \eqref{33halfequiv}.

\subsection{2d dualities from sandwiches}\label{sec:full33}

It is satisfying to find dualities of interfaces, and half-index identities, that encode the geometry of the (3,3) move. 
However, one might also hope for dualities of purely 2d theories that are in some way intrinsic to the 4d geometry. As explained in \eqref{33-sandwiches}, this can is achieved by choosing a \emph{pair} of boundary conditions $\CB_{\rm I}$, $\CB_{\rm IV}$ for $\CT_{\rm I}$, $\CT_{\rm IV}$, and sandwiching the duality interfaces between them.%
\footnote{Some choices of sandwiching boundary conditions may have interesting geometric origins.
In particular, $\CB_{\rm I}$ and $\CB_{\rm IV}$ may themselves correspond to 4-manifolds with a single 3d boundary component (homeomorphic to an octahedron), which are glued onto the complex of pentachora in the (3,3) move, closing it off. 
Analogous proposals in the smooth context were discussed in \cite{GGP}. We hope to flesh out the geometric counterpart of sandwiching in future work. Here we content ourselves by presenting a two-dimensional duality arising from boundary conditions with simple field-theoretic definitions.} %
We consider one particularly simple sandwich here.

We choose $\CB_{\rm I} = \text{(DDDD)}$ as in Section \ref{sec:half33}, and choose the right boundary condition
\be  \CB_{\rm IV} = \CN_{g_1}\CN_{g_2} \CD_{g_3} {\rm N}_a{\rm D}_b {\rm N}_g {\rm N}_v {\rm D}_r\big|\eta,\eta' \ee
Here $U(1)_{g_1}\times U(1)_{g_2}$ remain gauge symmetries on the boundary (due to Neumann b.c. on the corresponding gauge multiplets) while $U(1)_{g_3}$ is broken to a boundary flavor symmetry (by Dirichlet b.c. $\CD_{g_3}$).
This turns out to be one of the simplest boundary conditions for $\CT_{\rm IV}$ that is free from gauge (or mixed gauge-flavor) anomalies.
There are two extra 2d fermi multiplets $\eta,\eta'$ on the boundary, whose role is to cancel boundary anomalies.
Explicitly, the charges of $\eta,\eta'$ and the total boundary anomaly are
\be \begin{array}{c|ccc|ccc|c}
& U(1)_{g_1}& U(1)_{g_2} & U(1)_{g_3} & U(1)_{v_1} & U(1)_{v_2} & U(1)_a & U(1)_R \\\hline
\eta & 1&0&-1&-1&1&0&0 \\
\eta' & 0&1&0&0&-1&-1&0 
 \end{array}
\ee
\be 
{\CA[\CT_{\rm IV}]} +\frac12\mb r^2 -\tfrac12(\mb g_1-\mb g_3-\tfrac12\mb r)^2+\tfrac12(\mb g_2+\tfrac12 \mb g_3- \tfrac12 \mb r)^2-\tfrac12(-\mb g_1-\mb g_2)^2-\tfrac12(\mb g_2-\tfrac12\mb r)^2 + \tfrac12(\mb g_1-\tfrac12 \mb r)^2  \notag \ee
\be + (\mb g_1-\mb g_3-\mb v_1+\mb v_2)^2 + (\mb g_2-\mb a-\mb v_2)^2 = \tfrac12 \mb r^2 +\mb g_3^2 +\mb g_3(2\mb v_1-\mb r)+\mb v_1^2+\mb v_2^2+\mb a^2 -\mb v_1\mb v_2+\mb a\mb v_2\,. \notag  \ee
This anomaly polynomial is independent of $\mb g_1$ and $\mb g_2$, indicating the absence of gauge and mixed gauge-flavor anomalies (there remain flavor 't Hooft anomalies).

\subsubsection{Top sandwich}

To identify the result of sandwiching the top interface $\CI_{\rm top}$ between $\CB_{\rm I}$ and $\CB_{\rm IV}$, we can proceed in two steps. We first collide with $\CB_{\rm I}$ on the left, obtaining a left boundary condition \eqref{BItop} for $\CT_{\rm IV}$, and then collide with $\CB_{\rm IV}$ on the right to get a 2d theory
\begin{align} 
\CT_{\rm top} &= \CB_{\rm I}\circ \CI_{\rm top} \circ \CB_{\rm IV} \notag \\
 &= \Gamma,\Gamma',\Gamma'',\Psi_t;\Phi_s,\Phi_u\big| \CN_{g_1}\CN_{g_2}\CN_{g_3} {\rm N}_a {\rm N}_b {\rm D}_g {\rm N}_v {\rm N}_r-\CN_{g_1}\CN_{g_2} \CD_{g_3} {\rm N}_a{\rm D}_b {\rm N}_g {\rm N}_v {\rm D}_r\big|\eta,\eta' \label{top-sandwich} \\
& \simeq U(1)_{g_1}\times U(1)_{g_2} + \Gamma,\Gamma',\Gamma'',\Psi_t,\eta,\eta';\, \Phi_s,\Phi_u,\Phi_a,\Phi_v\,.  \notag
\end{align}
The bulk multiplets $\phi_a$ and $\phi_v$ of $\CT_{\rm IV}$ survive, as they are sandwiched between N b.c. on either side, and give rise to 2d chiral multiplets $\Phi_a$ and $\Phi_v$. In addition, 2d vector multiplets associated to the $U(1)_{g_1}\times U(1)_{g_2}$ gauge symmetry survive, while $U(1)_{g_3}$ has become a 2d flavor symmetry. 
Altogether, we find a 2d theory with $U(1)^2$ gauge symmetry, six fermi multiplets, and four chiral multiplets. Its charges and remaining couplings are summarized in Appendix \ref{app:sandwich}.

Here it is more instructive to observe that the theory $\CT_{\rm top}$ \emph{factorizes} into two completely decoupled subsectors, $\CT_{\rm top} = \CT_{\rm top}^1\otimes \CT_{\rm top}^2$\,.
The factorization is most manifest if we redefine (gauge and) flavor charges according to the field redefinitions $\mb g_1\to \mb g_1+\mb a-\mb v_2$, $\mb g_2\to -\mb g_2$, $\mb g_3\to \mb g_3+\mb a-\mb v_1-\mb v_2$, $\mb v_1\to \mb v_1+\mb v_2-\mb a$, $\mb v_2\to\mb v_2-\mb a$. We also swap the fermi multiplets $\Gamma''$ and $\eta'$ with their fermionic T-duals $\tilde\Gamma''$, $\tilde\eta'$ (essentially their conjugates, \emph{cf.} \cite{DGP-bdy}). Then we find that both $\CT_{\rm top}^1$ and $\CT_{\rm top}^2$ are $U(1)$ gauge theories with three fermi multiplets and two chiral multiplets, namely
\be \CT_{\rm top}^1\,:\;\;
\begin{array}{c}
\begin{array}{c|ccc|cc}
& \Gamma & \tilde\Gamma'' & \eta & \Phi_a & \Phi_s \\\hline
U(1)_{g_1} & 1&0&1&1&-1 \\ \hline
U(1)_{g_3} & 0&1&-1&-1&0 \\
U(1)_{v_1} &1&-1&0&1&0 \\
U(1)_R &0&0&0&\frac12&\frac12
\end{array} \\[.5cm]
J_{\tilde \Gamma''} = \Phi_a\Phi_s \end{array}
\qquad\qquad
\CT_{\rm top}^2\,:\;\;
\begin{array}{c}
\begin{array}{c|ccc|cc}
& \tilde\eta' &\Psi_t & \Gamma'  & \Phi_v & \Phi_u \\\hline
U(1)_{g_2} & 1&0&1&1&-1 \\ \hline
U(1)_{a} & 0&1&-1&-1&0 \\
U(1)_{v_2} &1&-1&0&1&0 \\
U(1)_R &0&0&0&\frac12&\frac12
\end{array} \\[.5cm]
J_{\Psi_t} = \Phi_v\Phi_u \end{array} \label{T12top}
\ee
We see that $\CT_{\rm top}^1$ and $\CT_{\rm top}^2$ are essentially identical. One carries the flavor symmetry $U(1)_{g_3}\times U(1)_{v_1}$ and the other carries $U(1)_{a}\times U(1)_{v_2}$.

The only $J$ and $E$ couplings obviously induced from the sandwich \eqref{top-sandwich} are the two $J$-terms above. Curiously, the flavor charges of the various fields allow for additional 2d superpotential terms that are cubic in fermi multiplets (and their T-duals),%
\footnote{To see that these couplings preserve $\CN=(0,2)$ SUSY, we must check that the total $\ol\CD$ variation of the superpotential vanishes. (This is a generalization of the usual $E\cdot J=0$ constraint.) The couplings themselves correct the fermion variations to $\ol \CD \tilde \Gamma''=0$, $\ol \CD \Gamma = -\tilde\Gamma''\tilde\eta(\Phi_a)^2$, $\ol \CD\tilde\Gamma=-\tilde\Gamma''\eta(\Phi_s)^2$, $\ol\CD \eta = \tilde\Gamma''\tilde\Gamma(\Phi_a)^2$, $\ol\CD\tilde\eta = \tilde \Gamma''\Gamma(\Phi_s)^2$. Notably, these are all proportional to $\tilde\Gamma''$. Then, simply due to $(\tilde\Gamma'')^2=0$, we will have $\ol\CD\big[ \tilde \Gamma''\Phi_a\Phi_s + \tilde \Gamma'' \Gamma\eta (\Phi_s)^2 + \tilde \Gamma'' \tilde \Gamma \tilde \eta (\Phi_a)^2\big]=0$ as required.
}
\be \mathcal W^1 = \int d\theta^+\big[ \tilde \Gamma''\Phi_a\Phi_s + \tilde \Gamma'' \Gamma\eta (\Phi_s)^2 + \tilde \Gamma'' \tilde \Gamma \tilde \eta (\Phi_a)^2 \big] \label{W-3F} \ee
in $\CT_{\rm top}^1$, and similarly in $\CT_{\rm top}^2$. We suspect that these additional terms will be generated during the collisions of boundaries and interfaces.

The elliptic genus of $\CT_{\rm top}^1$ is very simple. Using the Jeffrey-Kirwan contour prescription of  \cite{benini2014elliptic, benini2015elliptic} the evaluate the contour integral for the gauge multiplet (\emph{not} a direct projection to invariants, as would be appropriate for 3d gauge fields) we find
\begin{align} \label{Z2F}
 \CZ[\CT_{\rm top}^1] &= (q)_\infty^2 {\rm F}(g_3/v_1) \int_{JK} \frac{dg_1}{2\pi i g_1}  {\rm F}(g_1 v_1) {\rm F}(g_1/g_3) {\rm C}(q^{\frac14}g_1v_1/g_3) {\rm C}(q^{\frac14}/g_1) \\
 &= {\rm F}(q^{\frac14}v_1){\rm F}(q^{\frac14}g_3^{-1})\,,  \notag
\end{align}
suggesting that $\CT_{\rm top}^1$ flows to two free fermi multiplets in the IR. Similarly, $\CZ[\CT_{\rm top}^2]$ evaluates to ${\rm F}(q^{\frac14}v_2){\rm F}(q^{\frac14}a^{-1})$. Thus, the elliptic genus of the full sandwiched theory is
\be \CZ[\CT_{\rm top}] =  {\rm F}(q^{\frac14}v_1){\rm F}(q^{\frac14}g_3^{-1}){\rm F}(q^{\frac14}v_2){\rm F}(q^{\frac14}a^{-1})\,, \label{Ztop} \ee
and we suspect that the full sandwiched theory flows to four free fermi multiplets in the IR.

We observe that the dualities between $\CT^{1, 2}_{\rm top}$ to two free fermi multiplets are precisely an instance of the abelian duality from \cite{GGP-trialities} with $N_f=2$. As in \cite{GGP-trialities}, we have the $\CT^1_{\text{top}}$ duality map $\Psi'_1\equiv \Phi_s\Gamma, \Psi'_2 \equiv \Phi_s\eta$, where $\Psi'_1, \Psi'_2$ are the two dual fermi multiplets (and similarly for $\CT^2_{\text{top}}$). Notice that $(\Gamma, \eta)$ and $(\Psi'_1, \Psi'_2)$ both transform in the fundamental of an $SU(2)$ flavor symmetry whose Cartan is generated by ${\bf v_1} + {\bf g_3}$, under which the other fields are singlets. There is also an additional $U(1)$ flavor symmetry generated by ${1\over 2}({\bf v_1} - {\bf g_3})$ under which $\tilde{\Gamma}''$ transforms in the antifundamental and $\Phi_a$ transforms in the fundamental. These fields are the counterparts of $(P_a, \Gamma_a), a = 1,\ldots,N_f-1$ in \cite{GGP-trialities}.

\subsubsection{Bottom sandwich}

A nice feature of the boundary condition $\CB_{\rm IV}$ is that it is invariant under the ``180$^\circ$ rotation'' of Figure \ref{fig:rotation}. In particular, it treats the pairs of chirals $(\phi_a,\phi_v)$ and $(\phi_b,\phi_r)$ corresponding to tetrahedra exchanged by the rotation in a symmetric way, giving them both N b.c. or both D b.c. Thus we may expect that sandwiching the bottom interface $\CI_{\rm bot}$ between $\CB_{\rm I}$ and $\CB_{\rm IV}$ will produce a 2d $\CN=(0,2)$ theory $\CT_{\rm bot}$ that looks the same as $\CT_{\rm top}$ above --- up to an important permutation of fields and flavor symmetries. This is indeed what we find.

The sandwich with $\CI_{\rm bot}$ defines a 2d $\CN=(0,2)$ theory $\CT_{\rm bot}$ with $U(1)_{g_1}\times U(1)_{g_2}$ gauge symmetry, six fermi multiplets, and four chirals 
\begin{align}
 \CT_{\rm bot} &=  \CB_{\rm I}\circ \CI_{\rm bot}\circ \CB_{\rm IV}  \notag \\
  &= \Gamma,\Gamma',\Gamma'',\Psi_f;\Phi_d,\Phi_e\big| \CN_{g_1}\CN_{g_2}\CN_{g_3} {\rm N}_a {\rm N}_b {\rm D}_g {\rm N}_v {\rm N}_r - \CN_{g_1}\CN_{g_2} \CD_{g_3} {\rm N}_a{\rm D}_b {\rm N}_g {\rm N}_v {\rm D}_r\big|\eta,\eta'  \label{bot-sandwich} \\
  &\simeq \; U(1)_{g_1}\times U(1)_{g_2} + \Gamma,\Gamma',\Gamma'',\Psi_f,\eta,\eta';\, \Phi_d,\Phi_e,\Phi_a,\Phi_v \notag
\end{align}
This theory is distinguished from $\CT_{\rm top}$ by the flavor charges of $ \Gamma,\Gamma',\Gamma'',\Psi_f;\Phi_d,\Phi_e$, summarized in Appendix \ref{app:sandwich}.

As before, we find a factorization $\CT_{\rm bot} = \CT_{\rm bot}^1\otimes \CT_{\rm bot}^2$, with two almost identical decoupled subsectors. After redefining charges $\mb g_1\to \mb g_1+\mb g_3-\mb v_2, \mb g_2\to -\mb g_2, \mb g_3\to \mb g_3+\mb a-\mb v_1-\mb v_2$, $\mb v_1\to \mb v_1+\mb v_2-\mb a$, $\mb v_2\to\mb v_2-\mb a$ \emph{exactly} the same way as above \eqref{T12top},%
\footnote{There is one exception: $\mb g_1\to \mb g_1+\mb g_3-\mb v_2$ here, while $\mb g_1\to \mb g_1+\mb a-\mb v_2$ in \eqref{T12top}. The difference is a shift of flavor charges by multiples of gauge charges, which is not physically meaningful.}
we obtain the simplified description
\be \label{T12bot}
 \CT_{\rm bot}^1\,:\;\;
\begin{array}{c}
\begin{array}{c|ccc|cc}
& \tilde \Gamma' & \Psi_f & \eta & \Phi_a & \Phi_e \\\hline
U(1)_{g_1} & 1&0&1&1&-1 \\ \hline
U(1)_{a} & 0&1&-1&-1&0 \\
U(1)_{v_1} &1&-1&0&1&0 \\
U(1)_R &0&0&0&\frac12&\frac12
\end{array} \\[.5cm]
J_{\Psi_f} = \Phi_a\Phi_e \end{array}
\qquad\qquad
\CT_{\rm bot}^2\,:\;\;
\begin{array}{c}
\begin{array}{c|ccc|cc}
& \tilde\eta' &\tilde \Gamma'' & \tilde\Gamma  & \Phi_d & \Phi_v \\\hline
U(1)_{g_2} & 1&0&1&1&-1 \\ \hline
U(1)_{g_3} & 0&1&-1&-1&0 \\
U(1)_{v_2} &1&-1&0&1&0 \\
U(1)_R &0&0&0&\frac12&\frac12
\end{array} \\[.5cm]
J_{\tilde \Gamma''} = \Phi_d\Phi_v \end{array}
\ee
The flavor charges allow a more general 2d superpotential $\mathcal W^1 = \int d\theta^+ \big[\Psi_f\Phi_a\Phi_e+ \Psi_f\tilde \Gamma'\eta(\Phi_e)^2 + \Psi_f\Gamma'\tilde\eta (\Phi_a)^2\big]$ in $\CT_{\rm bot}^1$, and a similar superpotential for $\CT_{\rm bot}^2$.

The elliptic genera are computed as in \eqref{Z2F} to be
\be \begin{array}{c} \CZ[\CT_{\rm bot}^1] = {\rm F}(q^{\frac14}v_1){\rm F}(q^{\frac14}a^{-1})\,,\qquad \CZ[\CT_{\rm bot}^2] = {\rm F}(q^{\frac14}v_2){\rm F}(q^{\frac14}g_3^{-1})\,, \\[.2cm]
 \CZ[\CT_{\rm bot}] =  {\rm F}(q^{\frac14}v_1){\rm F}(q^{\frac14}a^{-1}){\rm F}(q^{\frac14}v_2){\rm F}(q^{\frac14}g_3^{-1})\,. \end{array}
\ee
Again we find four free fermions. This matches \eqref{Ztop}, but in a nontrivial way: the pairs of fermions to which $\CT_{\rm top}^1,\CT_{\rm top}^2$ individually flow are distinct from the pairs of fermions to which $\CT_{\rm bot}^1,\CT_{\rm bot}^2$ flow. The difference is detected by the flavor symmetry. (The two setups are related by swapping $a\leftrightarrow g_3$.) Only after taking a product of sectors do we match elliptic genera and recover a plausible IR duality
\be \CT_{\rm top}^1\otimes \CT_{\rm top}^2 \;\simeq\; \CT_{\rm bot}^1\otimes \CT_{\rm bot}^2\,.  \ee

\subsection{Toward more systematic gluing rules}

In Section \ref{sec:pure2d}, we posited a minimal model for a purely two-dimensional pentachoron theory $T[\Delta^4]$, involving two 2d fermi multiplets $\Gamma,\eta$. Moreover, we showed how this theory could be re-coupled to bulk XYZ and SQED theories (say), to recover the duality interface governing (2,3) and (3,2) Pachner moves. A nice feature of constructing the basic interface this way was that \emph{every} bulk multiplet started off with Neumann boundary conditions, which then got deformed by coupling to the pair of 2d fermi multiplets.

We can use the 2d description of $T[\Delta^4]$ from Section \ref{sec:pure2d} to make the sequence of interfaces in the (3,3) Pachner look a bit more canonical. Every pentachoron mediating a (2,3) or (3,2) move will contribute a pair of fermi multiplets; and every internal tetrahedron (trapped between pentachora) will contribute a 2d chiral multiplet.

For example, from this perspective, the composite interface $\CI_{\rm top}$ takes the form
\begin{subequations}  \label{Itop-can}
\be \CI_{\rm top}\,:\qquad {\rm N}_w{\rm N}_x{\rm N}_y{\rm N}_z \big| \Gamma,\eta,\Gamma',\eta',\Gamma'',\eta'';\Phi_t,\Phi_s,\Phi_u\big| \CN_{g_1}\CN_{g_2}\CN_{g_3} {\rm N}_a {\rm N}_b {\rm N}_g {\rm N}_v {\rm N}_r\,. \ee
All bulk chirals and gauge multiplets have Neumann b.c. All interface couplings are encoded in the (somewhat intricate) $E$ and $J$ terms for the six 2d fermi multiplets. A bit of careful work reveals them to be
\be \begin{array}{c|ccc|ccc} 
 & \Gamma & \Gamma' & \Gamma'' & \eta & \eta' & \eta ''  \\\hline
J & \phi_x\Phi_s & \phi_y\phi_v & \Phi_u\phi_b & \Phi_t-\phi_x\phi_w  & \Phi_u \phi_v-\Phi_t&\phi_g-\Phi_u\Phi_s \\
E & -\phi_r\phi_w & \phi_z\Phi_u & -\Phi_s\phi_a & -\phi_y\phi_z- \phi_r\Phi_s &-\phi_y\phi_z- \phi_r\Phi_s& -\phi_a\phi_b-\phi_r\phi_v 
\end{array}
\ee
\end{subequations}
These satisfy $J\cdot E = W_{\rm I}-W_{\rm IV}$ on the nose, with no additional relations required.

We emphasize that the description \eqref{Itop-can} is completely equivalent to the slightly simpler \eqref{Itop}. To get from \eqref{Itop-can} to \eqref{Itop}, we can integrate out $\eta'$ (and thus solve for $\Phi_t$), identify $\eta$ with $\Psi_t$, and integrate out $\eta''$ to effectively ``flip'' the Neumann b.c. on $\phi_g$ to Dirichlet. (Such ``flips'' were discussed at length in \cite{DGP-bdy}.)

We suspect that this more canonical characterization of interfaces will be useful when investigating 2d $\CN=(0,2)$ theories associated to more general triangulated four-manifolds.

\section{The (2,4) and (4,2) moves}
\label{sec:24}
\begin{figure}[htb]
\centering
\includegraphics[width=5.8in]{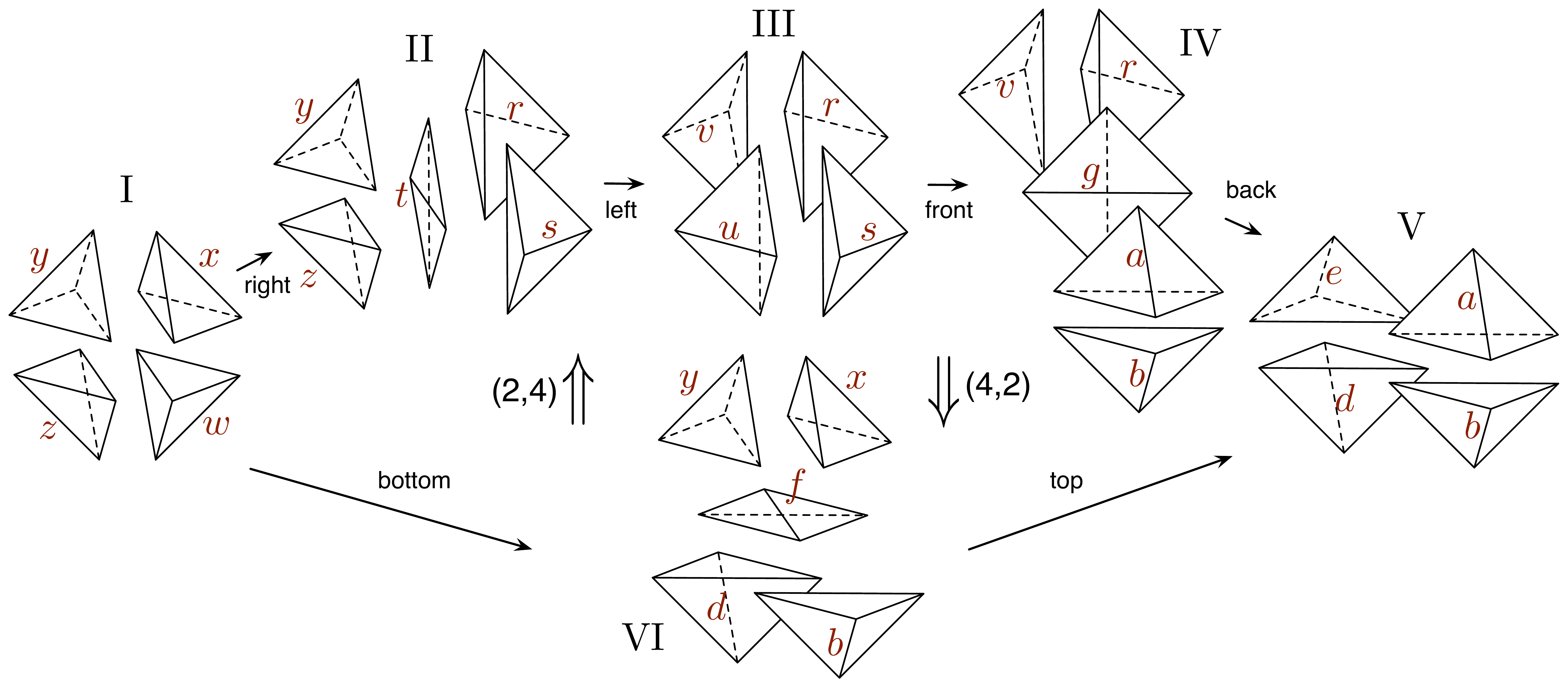}
\caption{Sequences of (2,3) and (3,2) moves that encode the (2,4) and (4,2) Pachner moves.}
\label{fig:oct24}
\end{figure}

It is simple to modify the above discussion slightly to obtain the interface for the (4,2) and (2,4) moves. The geometric picture is nearly identical to the one for the (3,3) move, except we need to reverse the arrow between $\CT_{\rm IV}$ and $\CT_{\rm V}$ so that the top sequence gains one step and the bottom sequence loses one step, as in Figure \ref{fig:oct24}.
We will then obtain two different composite interfaces between $\CT_{\rm I}$ and $\CT_{\rm V}$: an interface $\CI_2$ from collisions along the bottom, and an interface $\CI_4$ from collisions along the top.

We start with the two-step sequence for simplicity. The composite bottom interface is simply the interface obtained by truncating the bottom sequence in the (3,3) move after the second step:

\begin{subequations} \label{Ibot42}
\be \CI_{\rm 2}\,:\qquad {\rm N}_w{\rm N}_x{\rm N}_y{\rm N}_z \big| \Gamma,\Gamma', \Psi_f \big| \CN_{g_1'}\CN_{g_2'} {\rm N}_a {\rm N}_b {\rm N}_d {\rm N}_e\,. \ee
The $J$ and $E$ terms of the 2d fermi multiplets are
\be \begin{array}{c|cccc|c} 
 & \Gamma & \Gamma' & \Psi_f \\\hline
J & \phi_z\phi_d & \phi_y\phi_a & \phi_e\phi_a - \phi_z\phi_w \\
E & -\phi_b\phi_w & \phi_x\phi_e  & -\phi_y\phi_x - \phi_b\phi_d 
\end{array}
\ee
which immediately factorize the superpotential:
\be J\cdot E = J_{\Gamma}E_{\Gamma} + J_{\Gamma'}E_{\Gamma'}+ J_{\Psi_f}E_{\Psi_f} = \phi_x\phi_y\phi_w\phi_z - \phi_a\phi_b\phi_c\phi_d = W_{\rm I}-W_{\rm V}. \ee

For simplicity in what follows, we will use conventions for gauge charges that are better adapted to the bottom sequence --- corresponding to fugacities and field strengths denoted  $g_1',g_2',g_3'$ and $\mb g_1',\mb g_2',\mb g_3'$ in Appendix~\ref{app:tables}. The charges of fields involved in $\CI_{\rm 2}$ are

\be
\begin{array}{c|cccc|ccc|cccc}
 & \phi_w & \phi_x & \phi_y & \phi_z & \Gamma & \Gamma' & \Psi_f & \phi_a & \phi_b & \phi_d & \phi_e \\ \hline
 U(1)_{g_1'} &0&0&0&0 & 1&0&0  & 0& 1& -1 & 0 \\
 U(1)_{g_2'} &0&0&0&0 & 0&-1&0   & 1&0& 0& -1\\ \hline
 U(1)_{v_1} &1&-1&0&0& 1& 0& -1& 0& 0& -1& 1 \\
 U(1)_{v_2} & 0 & 1 & 0 & -1 &0&0&1& 0&0&1&-1 \\
 U(1)_{a} & 0 & 0 & 1 & -1 &0&-1&1& 0&0&1&-1 \\ \hline
 U(1)_R   &\frac12&\frac12&\frac12&\frac12  &0&0& 0 &\frac12&\frac12&\frac12&\frac12
\end{array}
\ee
\end{subequations}

Alternatively, we may reach $\CT_{\rm V}$ by passing through interfaces $\CT_{\rm I}\to \CT_{\rm II}\to \CT_{\rm III}\to \CT_{\rm IV}$ just as in Section \ref{sec:33int-top}, and then applying one final (2,3) move $\CT_{\rm IV}\to \CT_{\rm V}$ using a basic interface \eqref{TD4-long}, in the form
\be \nonumber \label{intYZrev} \hspace{-.5in} \CN_{g_3'}{\rm D}_g {\rm N}_r {\rm N}_v \big|\Gamma'''\big|{\rm N}_e {\rm N}_d\,,\qquad
\mathcal{W}_{\rm IV-V} = \int d\theta^+ \left(\phi_v \Gamma''' \phi_e  + \phi_e \Psi_g \phi_d \right)\,,\qquad E_{\Gamma'''} = \phi_r \phi_d\,. \ee
The interface is transparent to the fields $\phi_a, \phi_b$.

The composite interface for this four-step sequence can be obtained by colliding our previous $\CI_{\rm top}$ from \eqref{Itop} with \eqref{intYZrev}. By the same reasoning outlined in constructions of Section \ref{sec:33}, the final collision traps two 2d chirals $\Phi_r, \Phi_v$, and a 2d fermi $\Psi_g$. In addition, the 3d vector multiplet for the $U(1)_{g_3'}$ symmetry of $\CT_{\rm IV}$ has Neumann b.c. on either side, and reduces to a 2d vector multiplet $\CV_{g_3'}$ for a ``trapped'' purely 2d gauge symmetry.
Summarizing, the composite four-step interface is:
\begin{subequations} \label{Itop42}
\be \CI_{\rm 4}\,:\qquad {\rm N}_w{\rm N}_x{\rm N}_y{\rm N}_z \big| \Gamma,\Gamma',\Gamma'', \Gamma''', \Psi_t, \Psi_g;\Phi_u, \Phi_s,\Phi_r, \Phi_v; \mathcal{V}_{g_3'}\big| \CN_{g_1'}\CN_{g_2'} {\rm N}_a {\rm N}_b {\rm N}_d {\rm N}_e\,. \ee
The $J$ and $E$ terms of the 2d fermi multiplets are
\be \begin{array}{c|cccccc} 
 & \Gamma & \Gamma' & \Gamma'' & \Gamma''' & \Psi_t & \Psi_g \\\hline
J & \phi_x\Phi_s & \phi_y\phi_v & \Phi_u\phi_a & \phi_e\Phi_v & \Phi_u\Phi_v - \phi_x\phi_w & -\Phi_u\Phi_s +\phi_e\phi_d\\
E & -\Phi_r\phi_w & \phi_z\Phi_u & -\Phi_s\phi_b & \phi_d\Phi_r & -\phi_y\phi_z -\Phi_r\Phi_s & -\phi_a\phi_b - \Phi_r\Phi_v
\end{array}
\ee
Again, the interface $J$ and $E$ terms factorize the difference of bulk superpotentials,
\be \notag J\cdot E = J_{\Gamma}E_{\Gamma} + J_{\Gamma'}E_{\Gamma'}+ J_{\Gamma''}E_{\Gamma''} + J_{\Gamma'''}E_{\Gamma'''}+ J_tE_t + J_gE_g = \phi_w\phi_x\phi_y\phi_z -\phi_a\phi_b\phi_d\phi_e = W_{\rm I}-W_{\rm V} \ee
The charges of the various fields are\footnote{For neatness of presentation, we perform an additional redefinition: $\mb g_3'\to \mb g_3'+\mb v_1-2\mb v_2-\mb a$.}:
\be
\begin{array}{c|cccc|cccccccccc|cccc}
 & \phi_w & \phi_x & \phi_y & \phi_z & \Gamma & \Gamma' & \Gamma'' & \Gamma''' & \Psi_t & \Psi_g & \Phi_s & \Phi_u & \Phi_r & \Phi_v  & \phi_a & \phi_b &  \phi _d & \phi_e \\ \hline
 U(1)_{g_1'} & 0&0&0&0& 0&-1&0&-1&0&1&0&-1&0&1& 0&1&-1&0\\
 U(1)_{g_2'} &0&0&0&0&1&0&0&1&0&1&-1&0&1&0&1&0&0&-1\\
 U(1)_{g_3'} &0&0&0&0&1&1&-1&1&0&0&-1&1&1&-1&0&0&0&0\\ \hline
 U(1)_{v_1} & 1 & -1 & 0 & 0 & 1&0&0&-1&0&0&0&0&0&0&0&0&-1&1 \\
 U(1)_{v_2} & 0 & 1 & 0 & -1 & 0&0&-1&1&-1&0&-1&1&0&0&0&0&1&-1 \\
 U(1)_{a} & 0 & 0 & 1 & -1 & 0&-1&0&1&0&0&0&0&0&0&0&0&1&-1 \\ \hline
 U(1)_R   &\frac12&\frac12&\frac12&\frac12 &0&0&0&0&0&0&\frac12&\frac12&\frac12&\frac12 &\frac12&\frac12&\frac12&\frac12
\end{array}
\ee
\end{subequations}
Note again that $U(1)_{g_3'}$ is a purely 2d gauge symmetry.

\subsection{Half-indices}

Just as we did for the (3,3) move, we can collide the $\CT_{\rm I} \mapsto \CT_{\rm V}$ interfaces with an all-Dirichlet boundary condition $\CB_I = (\text{DDDD})$ in $\CT_I$, to get an expected dual pair of left boundary conditions for $\CT_{\rm V}$:
\be \CB_{\rm I} \circ \CI_2 \simeq \CB_{\rm I}\circ \CI_{4}\,. \ee
%
Of course, since both $\CI_2$ and $\CI_4$ are duality interfaces, we expect both of these to simply be dual to $\CB_{\rm I}$ itself, in the deep IR where the bulk theories $\CT_{\rm I}$ and $\CT_{\rm IV}$ flow to the same 3d SCFT.

The expected dualities are borne out by half-index calculations. The following nontrivial-looking expressions are both equal to the simple half-index $\half_{\CT_{\rm I},\CB_{\rm I}}$ from \eqref{DDDD-index}, and in particular are equal to each other:
\begin{align} \half_{\CT_{\rm V}, \CB_{\rm I}\circ \CI_2} = & (q)_{\infty}^2 \oint {dg_1' \over (2 \pi i g_1')} \oint {dg_2' \over (2 \pi i g_2')}\half_N(g_2' q^{\frac14})\half_N(g_1'q^{\frac14})\half_N(q^{\frac14}v_1/(a g_2' v_2))\half_N(a q^{\frac14}v_2/(g_1' v_1)) \notag \\ 
 &\times \text{F}(g_1' v_1) \text{F}(a v_2/v_1) \text{F}(a^{-1}g_2'^{-1}) 
\end{align}
and
\begin{align} \half_{\CT_{\rm V}, \CB_{\rm I}\circ \CI_4} =& (q)_{\infty}^4 \oint {dg_1' \over (2 \pi i g_1')} \oint {dg_2' \over (2 \pi i g_2')}\int_{JK}{dg_3' \over 2\pi i g_3'} \notag \\
 & \half_N(q^{\frac14}g_1')\half_N(q^{\frac14}g_2')\half_N(q^{\frac14}v_1/(a g_2' v_2))\half_N(a q^{\frac14}v_2/(g_1' v_1)) \\ 
\nonumber &\times \text{C}(q^{\frac14}/(g_2' g_3' v_2))\text{C}(q^{\frac14}g_3' v_2/g_1')\text{C}(q^{\frac14} g_2' g_3')\text{C}(q^{\frac14}g_1'/g_3') \\ \nonumber
&\times \text{F}(g_2' g_3' v_1)\text{F}(g_3'/(a g_1'))\text{F}(1/(g_3' v_2))\text{F}(1/v_2)\text{F}(g_2' g_3' a v_2/(g_1' v_1))\text{F}(1/(g_1' g_2'))\,.
\end{align}
The contour integral of the (now 2d) gauge field $g_3'$ is performed first with the Jeffrey-Kirwan prescription (say, by evaluating and summing the residues at $g_3' = g_1'/(v_2 q^{1/4})$ and $g_3'= 1/(g_2' q^{1/4})$), followed by the contour integrals for $g_1', g_2'$ over the unit circle, which project to $g_1'$- and $g_2'$-invariants \cite{DGP-bdy}.

Since these half-indices are both equal to $\half_{\CT_{\rm I},\CB_{\rm I}}$, they possess they same non-manifest $v_2 \mapsto v_1/(v_2 a)$ symmetry discussed in Section \ref{sec:half33}.

\subsection{Sandwiches for the (4,2) move}

As before, we can make sandwiches for various choices of boundary conditions. We will describe two interesting possibilities. Many others are also possible.

\subsubsection{Mostly Neumann for gauge fields}

To start, we again choose left boundary condition $\CB_{\rm I} = \text{(DDDD)}$ as above, and the right boundary condition
\be  \CB_{\rm V} = \CN_{g_1'}\CD_{g_2'} {\rm N}_a{\rm N}_b {\rm N}_d {\rm D}_e \text{ + 1 Fermi multiplet $\rho$} \ee

We summarize the resulting 2d theory below. The anomaly polynomial of $\CB_{V}$ is given by
\begin{align*}
&\CA[\CT_{\rm V}] + ({\bf g_1'} - {\bf a} - {\bf v_2})^2 \\ & - \tfrac12\left( {\bf g_2'} - \tfrac12{\bf r}^2\right)^2 - \tfrac12\left( {\bf g_1'} - \tfrac12{\bf r}^2\right)^2  + \tfrac12\left( -{\bf g_2'} - {\bf a} + {\bf v_1} - {\bf v_2} - \tfrac12{\bf r}^2\right)^2 - \tfrac12\left( -{\bf g_1'} + {\bf a} - {\bf v_1} + {\bf v_2} - \tfrac12{\bf r}^2\right)^2\\
&= {\bf a}^2 + {\bf g_2'}({\bf r} + 2 {\bf a}) - \tfrac14{\bf r}^2 + {\bf r}({\bf v_2} - {\bf v_1}) + {\bf v_1 v_2} + {\bf a}({\bf r} + {\bf v_2}).
\end{align*}
Notice that the Fermi multiplet (which contributes the second term) has charges chosen to cancel the pure and mixed ${\bf g_1'}$ anomalies, as required.

The collision with $\CB_{\rm I}$ and $\CB_{\rm V}$ will kill all the degrees of freedom coming from the bulk fields in $\CT_{\rm I}$, while trapping 2d chirals associated to the three 3d chirals $\phi_a,\phi_b,\phi_d$ in $\CT_{\rm V}$, in addition to trapping a 2d gauge field $\CV_{g_1'}$ and contributing a new Fermi multiplet $\rho$ from the right boundary. The bulk gauge symmetry $U(1)_{g_2'}$ becomes a 2d flavor symmetry.

Let us denote the 2d $\CN=(0,2)$ theories resulting from these sandwiches as
\be \CT_2 = \CB_{\rm I} \circ \CI_2 \circ \CB_{\rm V}\,,\qquad \CT_4 = \CB_{\rm I} \circ \CI_4 \circ \CB_{\rm V}\,. \ee
We propose that these theories are IR-dual to one another. Summarizing their field content, we have
\begin{subequations} \label{42sand1}
\be
\begin{array}{c} \CT_{2}\,:\qquad \Gamma,\Gamma', \Psi_f, \rho; \Phi_a, \Phi_b, \Phi_d; \CV_{g_1'}\,,\\[.2cm]
 \CT_{4}\,:\qquad  \Gamma,\Gamma',\Gamma'', \Gamma''', \Psi_t, \Psi_g, \rho;\Phi_u, \Phi_s,\Phi_r, \Phi_v,\Phi_a, \Phi_b, \Phi_d; \CV_{g_1'} \mathcal{V}_{g_3'}\,. \end{array} \ee
The basic $J$ and $E$ terms we can infer from the collision are
\be \CT_2\,:\quad  \begin{array}{c|ccccc} 
 & \Gamma & \Gamma' & \Psi_f & \rho \\\hline
J & 0 & 0 & 0 & 0  \\
E & 0 & 0  & - \Phi_b\Phi_d & 0 
\end{array}
\qquad \CT_4\,:\quad \begin{array}{c|cccccccc} 
 & \Gamma & \Gamma' & \Gamma'' & \Gamma''' & \Psi_t & \Psi_g& \rho \\\hline
J & 0 & 0 & \Phi_u\Phi_a & 0 & \Phi_u\Phi_v & -\Phi_u\Phi_s & 0\\
E & 0 & 0 & -\Phi_s\Phi_b & \Phi_d\Phi_r &  -\Phi_r\Phi_s & -\Phi_a\Phi_b - \Phi_r\Phi_v &0.
\end{array}\ee In both cases, $J\cdot E = 0$, as required for supersymmetry-preservation.
Finally, the charges for the two theories are
\be
\begin{array}{c|cccc|ccc}
 & \Gamma & \Gamma' & \Psi_f &\rho & \Phi_a & \Phi_b & \Phi_d  \\ \hline
 U(1)_{g_1'}& 1&0&0& 1 &  0 & 1 & -1  \\\hline 
 U(1)_{g_2'} & 0&-1&0 & 0 & 1& 0& 0 \\
 U(1)_{v_1} & 1& 0& -1& 0 & 0 & 0& -1\\
 U(1)_{v_2} &0&0&1& -1& 0 & 0& 1 \\
 U(1)_{a} &0&-1&1 & -1 & 0& 0& 1\\ \hline
 U(1)_R  &0&0& 0 & 0& {1 \over 2}&{1\over 2}&{1 \over 2}
\end{array}
\ee
and 
\be
\begin{array}{c|ccccccccccc|ccc}
 &  \Gamma & \Gamma' & \Gamma'' & \Gamma''' & \Psi_t & \Psi_g & \rho & \Phi_s & \Phi_u & \Phi_r & \Phi_v  & \Phi_a & \Phi_b &  \Phi _d \\ \hline
 U(1)_{g_1'} & 0&-1&0&-1&0&1& 1&0&-1&0&1& 0&1&-1\\
 U(1)_{g_3'} &1&1&-1&1&0&0& 0&-1&1&1&-1&0&0&0\\ \hline
 U(1)_{g_2'} &1&0&0&1&0&1& 0&-1&0&1&0&1&0&0\\
 U(1)_{v_1} & 1&0&0&-1&0&0&0&0&0&0&0&0&0&-1 \\
 U(1)_{v_2}  & 0&0&-1&1&-1&0& -1&-1&1&0&0&0&0&1 \\
 U(1)_{a}  & 0&-1&0&1&0&0&-1&0&0&0&0&0&0&1 \\ \hline
 U(1)_R    &0&0&0&0&0&0&0&\frac12&\frac12&\frac12&\frac12 &\frac12&\frac12&\frac12
\end{array}
\ee

\end{subequations}

As before, we can check the proposed duality of these 2d theories by computing their elliptic genera. 
The elliptic genera can be written as 
\begin{eqnarray}\nonumber \CZ[\CT_2] &= (q)_{\infty}^2 \int_{JK} {dg_1' \over (2 \pi i g_1')} \text{C}(g_2' q^{1/4})\text{C}(g_1'q^{1/4})\text{C}(a q^{1/4} v_2/(g_1' v_1))\text{F}(g_1' v_1) \text{F}(a v_2/v_1) \text{F}(a^{-1}g_2'^{-1})\text{F}(g_1'/(a v_2)) \\ \nonumber
&= \text{C}(g_2' q^{1/4})\text{C}(q^{1/2} a v_2/v_1)\text{F}(v_1/q^{1/4})\text{F}(a v_2/v_1)\text{F}(1/(a g_2'))\text{F}(1/a v_2 q^{1/4})\\ \nonumber
&= \text{C}(g_2' q^{1/4})\text{F}(v_1/q^{1/4})\text{F}(1/(a g_2'))\text{F}(1/(a v_2 q^{1/4}))\\ \nonumber
&= 1 + q^{1/4}\left( g_2' - v_1 - {1 \over a v_2} \right) + q^{1/2}\left(-{1 \over a g_2'} - a g_2' + g_2'^2 - g_2' v_1 - {g_2' \over a v_2} + {v_1 \over a v_2}\right) + \text{O}(q^{3/4})\end{eqnarray}
and (evaluating the residues at, for example, $g_1' = q^{-1/4}, g_3' = g_2'^{-1} q^{-1/4}$)
\begin{eqnarray}\nonumber \CZ[\CT_4] &= (q)_{\infty}^4 \int_{JK} {dg_1' \over (2 \pi i g_1')} \int_{JK}{dg_3' \over 2\pi i g_3'}\text{C}(q^{1/4}g_2')\text{C}(q^{1/4}g_1')\text{C}(q^{1/4}v_2 a/(g_1' v_1))\text{C}(q^{\frac14}/(g_2' g_3' v_2))\text{C}(q^{\frac14}g_3' v_2/g_1')\text{C}(q^{\frac14} g_2' g_3') \\ \nonumber &\times\text{C}(q^{\frac14}g_1'/g_3')\text{F}(g_2' g_3' v_1)\text{F}(g_3'/(a g_1'))\text{F}(1/(g_3' v_2))\text{F}(1/v_2)\text{F}(g_2' g_3' a v_2/(g_1' v_1))\text{F}(1/(g_1' g_2'))\text{F}(g_1'/(v_2 a))  \\ \nonumber
&= \text{C}(g_2' q^{1/4})\text{F}(v_1/q^{1/4})\text{F}(1/(a g_2'))\text{F}(1/(a v_2 q^{1/4}))\\ \nonumber
&= 1 + q^{1/4}\left( g_2' - v_1 - {1 \over a v_2} \right) + q^{1/2}\left(-{1 \over a g_2'} - a g_2' + g_2'^2 - g_2' v_1 - {g_2' \over a v_2} + {v_1 \over a v_2}\right) + \text{O}(q^{3/4}),
 \end{eqnarray} where the last equality follows from a large number of pairwise cancellations of chiral and fermi multiplet contributions.

Notice that the elliptic genera reduce to that of a single chiral multiplet and three fermi multiplets. This is apparent from reexamining the two theories. In the case of $\CT_2$, $\Gamma'$ and $\Phi_a$ are decoupled from the rest of the theory%
\footnote{This is clear if we redefine ${\bf v_2} \rightarrow {\bf v_2} - {\bf a}$.} 
and simply carry the $U(1)_{g_2'}, U(1)_a$ flavor symmetries. The rest of $\CT_2$ furnishes another example of the basic abelian duality of \cite{GGP-trialities}, exactly as in Section \ref{sec:full33}. Repeating the argument therein yields the contribution of the two fermi multiplets $\Gamma'_1, \Gamma'_2$ dual to the rest of the fields, with the identification $\Psi'_1 \equiv \Phi_d \rho, \tilde{\Psi'_2} \equiv \Phi_d \Gamma$, where we write the T-dual of $\Psi'_2$ in the last equality to match the sign of flavor charges of $\Psi'_2$ in the index identity. Similar manipulations apply to $\CT_4$ which contains, in addition to the 5 fields dual to those in $\CT_2$, 4 chiral/fermi pairs that acquire masses along the RG flow and can be integrated out.

\subsubsection{All Dirichlet for gauge fields}

We consider one more sandwich to obtain a particularly nice identity of elliptic genera. Consider the sandwich by the boundary conditions $\CB_I= \text{(DDDD)}$ on the left and
\be 
\CB'_{\rm V} = \CD_{g_1'}\CD_{g_2'} {\rm D}_a{\rm D}_b {\rm D}_d {\rm D}_e 
\ee on the right.
This sandwich completely removes all bulk modes in $\CT_{\rm I}$ and $\CT_{\rm V}$ alike. The data for the 2d theories is now simply 

\begin{subequations} \label{42sand1}
\be \CT_{2}'\,:\qquad \Gamma,\Gamma', \Psi_f\, \ee
\be \CT_{4}'\,:\qquad  \Gamma,\Gamma',\Gamma'', \Gamma''', \Psi_t, \Psi_g;\Phi_u, \Phi_s,\Phi_r, \Phi_v; \mathcal{V}_{g_3'}\,. \ee
The $J$ and $E$ terms are
\be \begin{array}{c|ccc} 
 & \Gamma & \Gamma' & \Psi_f \\\hline
J & 0 & 0 & 0   \\
E & 0 & 0  &0
\end{array}
\ee
and 
\be \begin{array}{c|cccccc} 
 & \Gamma & \Gamma' & \Gamma'' & \Gamma''' & \Psi_t & \Psi_g \\\hline
J & 0 & 0 & 0 & 0 & \Phi_u\Phi_v & -\Phi_u\Phi_s \\
E & 0 & 0 & 0 & 0 &  -\Phi_r\Phi_s & - \Phi_r\Phi_v.
\end{array}\ee
In both cases, the condition $J\cdot E = 0$ is immediately satisfied.

Finally, the charges in the two theories (where $g_1', g_2'$ are now flavor symmetries) are
\be
\begin{array}{c|ccc}
 & \Gamma & \Gamma' & \Psi_f  \\ \hline
U(1)_{g_1'} & 1 & 0   & 0\\
U(1)_{g_2'}& 0 & -1  & 0  \\
U(1)_{v_1} & 1& 0& -1\\
 U(1)_{v_2} &0&0&1 \\
 U(1)_{a} &0&-1&1 \\ \hline
 U(1)_R  &0&0& 0 
\end{array}
\ee
and  
\be
\begin{array}{c|ccccccccccc}
 &  \Gamma & \Gamma' & \Gamma'' & \Gamma''' & \Psi_t & \Psi_g & \Phi_s & \Phi_u & \Phi_r & \Phi_v \\ \hline
 U(1)_{g_3'} &1&1&-1&1&0&0&-1&1&1&-1\\ \hline
 U(1)_{g_1'} & 0&-1&0&-1&0&1&0&-1&0&1\\
 U(1)_{g_2'} &1&0&0&1&0&1&-1&0&1&0\\
 U(1)_{v_1} & 1&0&0&-1&0&0&0&0&0&0 \\
 U(1)_{v_2}  & 0&0&-1&1&-1&0&-1&1&0&0 \\
 U(1)_{a}  & 0&-1&0&1&0&0&0&0&0&0 \\ \hline
 U(1)_R    &0&0&0&0&0&0&\frac12&\frac12&\frac12&\frac12
\end{array}
\ee
\end{subequations}

The resulting elliptic genera are

\begin{align}\nonumber
\CZ[\CT_2']&= \text{F}(g_1' v_1)\text{F}(av_2/v_1)\text{F}(1/(a g_2')) \\\nonumber
&= 1 + \left(-{1 \over a g_2'} - a g_2' - {1 \over g_1' v_1} - g_1' v_1 - {v_1 \over a v_2} - {a v_2 \over v_1} \right)q^{1/2} + \ldots\\\nonumber
\CZ[\CT_4']&= (q)^2_{\infty}\int_{JK}{dg_3' \over 2\pi i g_3'}\text{C}(q^{\frac14}/(g_2' g_3' v_2))\text{C}(q^{\frac14}g_3' v_2/g_1')\text{C}(q^{\frac14} g_2' g_3')\text{C}(q^{\frac14}g_1'/g_3') \\ \nonumber &\ \ \ \ \ \ \ \times \text{F}(g_2' g_3' v_1)\text{F}(g_3'/(a g_1'))\text{F}(1/(g_3' v_2))\text{F}(1/v_2)\text{F}(g_2' g_3' a v_2/(g_1' v_1))\text{F}(1/(g_1' g_2')) \\ \nonumber
 &=\text{C}(v_2/(g_1' g_2'))\text{F}(1/(a g_1' g_2' q^{1/4}))\text{F}(v_1 q^{-1/4})\text{F}(g_2' q^{1/4}/v_2)\text{F}(a v_2/(g_1' q^{1/4} v_1)) \\ \nonumber
& \ \ \ \ \ \ \ \ + \text{C}(g_1' g_2'/v_2)\text{F}(q^{1/4}/g_1')\text{F}(a g_2'/(q^{1/4} v_1))\text{F}(1/(a q^{1/4}v_2))\text{F}(g_1' g_2' v_1/(q^{1/4}v_2)) \\ \nonumber
&=1 + \left(-{1 \over a g_2'} - a g_2' - {1 \over g_1' v_1} - g_1' v_1 - {v_1 \over a v_2} - {a v_2 \over v_1} \right)q^{1/2} + \ldots 
\end{align}where the latter integral can be evaluated by summing over the contribution from the poles at, say, $g_3'= g_1'/(v_2 q^{1/4})$ and $g_3' =1/(q^{1/4} g_2')$. Notice also that although these functions apparently have five flavor fugacities, we do not actually have that number of flavor symmetries, since they only appear in the combinations $g_1' v_2, a v_2/v_1, 1/(a g_2')$.

\section{Discussion}
We conclude this note by describing some new opportunities for future progress in view of our results, as well as a discussion of some of the expected challenges and subtleties.

In this note we have proposed that a certain 2d $\CN=(0,2)$ duality interface can be associated with a 4-simplex, viewed with a certain 3+2 splitting of its boundary components. As mentioned earlier, this somewhat restricted way of viewing the 4-simplex is already sufficient, in principle, to describe a certain class of 4-manifolds that admit layered ideal triangulations (see \cite{JacoRubinstein, issa} for discussions of layered triangulations).
Topologically, these 4-manifolds are bundles over $S^1$ (or over an interval), whose fibers are the complement of a knot or link in a closed 3-manifold $\ol M_3$. 
In particular, sweeping over $S^1$ (or the interval), one views the 4-manifolds of interest as a sequence of triangulated 3-manifolds, related by 3d Pachner moves. Physically, this translates to a sequence of class-$\mathcal R$ theories connected by duality interfaces, much as in our study of 4d Pachner moves. 

Though we have the machinery in hand to discuss such triangulated 4-manifolds, we are currently able to do so only on a case-by-case basis, keeping careful track of boundary polarizations, etc. It would be useful to find a more algorithmic way of producing $\CN=(0, 2)$ theories and interfaces associated to this class of manifolds. 

One may also hope to find 4-manifold invariants (which, from this viewpoint, will entail chiral algebras, elliptic genera and half-indices, or entire $\CN=(0, 2)$ theories and interfaces) associated to more general ideal triangulations of 4-manifolds. This requires making the symmetries of the pentachoron more manifest in field theory. It turns out that some symmetries, such as a $\Z/3\Z$ subgroup of the pentachoron's $A_5$ rotational symmetry admit simple interpretations in the duality interface. To realize other subgroups, by contrast, requires studying IR-dual 3d theories arising from a rather involved chain of manipulations involving integrating chiral multiplets in and out, and rotating polarizations.
We defer a full discussion of penatachoral symmetries to future work, and leave systematizing the requisite gluing rules for general ideal triangulations as a future objective. 

As first discussed in \cite{GGP}, one expects a 4d-2d dictionary that associates observables of the 2d $\CN=(0, 2)$ theory $T[M_4; \mathfrak{g}]$ to observables in the Vafa-Witten twisted theory on $M_4$ \cite{VW}. For example, \cite{GGP, GGP-param} proposed that the Vafa-Witten partition function should equal the 2d elliptic genus. The categorification of the latter is a vertex algebra, sometimes called $\text{VOA}[M_4]$, which has recently been explored in detail in the abelian case for smooth (toric) 4-manifolds in \cite{GGP, FeiginGukov}; general properties of $\text{VOA}[M_4; \mathfrak{g}]$ have also been discussed in these works. These vertex algebras are expected to act on the cohomology of the VW instanton moduli spaces via correspondences, generalizing the seminal works of Nakajima and Grojnowski \cite{Nakajima, Grojnowski}. The 4d-2d dictionary should also dovetail with 3d-3d correspondence and AGT correspondence when the 4-manifold has boundaries, corners, etc. \cite{GGP}. In other words, we anticipate that the duality interfaces have as their 4d counterparts Vafa-Witten theories with a boundary condition inherited from the 3d-3d correspondence \cite{Yamazaki, DGG}; the latter dictates that the 3d $\CN=2$ theory should capture $SL(2, \C)$ Chern-Simons theory on the boundary 3-manifold. In the presence of such boundary conditions, we expect that the Vafa-Witten theory should capture a moduli space of \textit{ramified} instantons on $M_4$. It would be of great interest to study the theory with these boundary conditions directly for (smooth or PL) $M_4$.

Returning to the framework of ideal triangulations, we also recall that the 3d-3d prescription of \cite{DGG, dimofte2016k} captures only a subsector of $T[M_3; \mathfrak{g}]$ \cite{CDGS} (though see \cite{GangYonekura} for recent developments). It is already a fascinating and challenging open problem to understand if one can augment the triangulated 3d-3d story to recover the full $T[M_3]$. With this in mind, we anticipate that our $\CN=(0,2)$ theories will also capture only a subsector of the full $T[M_4]$. The virtues, of course, in elucidating these subsectors as we have done include their explicit Lagrangian descriptions and the eminent computability of their indices, anomaly polynomials, etc---analogous to the virtues of Seiberg-Witten descriptions of 4d $\CN=2$ theories. Understanding this subsector directly on the Vafa-Witten side and relating our findings to $\text{VOA}[M_4]$ will be insightful, and help us precisely demarcate the scope of our prescription in 4-manifold geometry. We expect, however, that recovering the complete $T[M_4]$ by enhancing the existing triangulation prescription will be another long-term challenge (if it can be done at all). Nonetheless, on the geometry side, triangulations have been used to give an expedited, computational proof that classes of Cappell-Shaneson homotopy spheres are standard (i.e. do not possess exotic smooth structures, as initially hoped) \cite{issa}, which was previously demonstrated laboriously using Kirby moves; we believe that exploring triangulations further in both geometry and physics will complement our current tools for studying 4-manifold invariants.

It would be interesting to further explore the uplift of the field theory side to 4d $\CN=2$ theories as discussed in the main text: viewing duality interfaces in 3d as junctions of defects in 4d \cite{GGP}. Boundary conditions in abelian 4d $\CN=2$ (e.g. Seiberg-Witten) theories transform into one another under a symplectic group action \cite{Witten-bdy}. It would be insightful to understand what remnant of this duality action, if any, is induced on the junction and check this at the level of junction indices. Analogous configurations in 4d $\CN=4$ and the associated ``quarter-indices'' were recently considered in \cite{GaiottoOkazaki}.

Finally (and a priori independently of any geometry), one can consider putative dual 2d $\CN=(0, 2)$ theories arising from numerous sandwiches of IR-dual 3d $\CN=2$ theories. We are currently exploring the resulting dualities and their relation to the trialities of \cite{GGP-trialities}. 

\acknowledgments

We would like to thank F. Costantino, A. Issa, and  R. Kashaev for helpful correspondence regarding 4d triangulations; A. Gadde, S. Gukov, and P. Putrov for useful discussions regarding the 4d-2d correspondence; and K. Costello and D. Gaiotto for collaboration on related projects.
The work of T.D. was supported by NSF CAREER Grant No. 1753077 and in part by ERC Starting Grant No. 335739; part of this work was performed at the Aspen Center for Physics, which is supported by National Science Foundation grant PHY-1607611.
 N.P. acknowledges support from a Sherman Fairchild Postdoctoral Fellowship. N.P. also gratefully acknowledges the following institutions for hospitality and support while this work was being completed, and where this work was presented: Rutgers University, OIST, Kavli IPMU, Copenhagen University, Stanford University, Princeton University, Perimeter Institute, and UC Davis/QMAP. This material is based upon work supported by the U.S. Department of Energy, Office of Science, Office of High Energy Physics, under Award Number DE-SC0011632.

\appendix
\section{Charges \& Chern-Simons levels}\label{app:tables}

Below we summarize the charges and bulk Chern-Simons levels of the 3d $\CN=2$ class-$\mathcal R$ theories associated to the six triangulated octahedra.

To be very precise, these theories are derived from the geometry of triangulated octahedra, using the rules and orientation conventions of \cite{DGG-index} (there is a difference in orientations between \cite{DGG} and \cite{DGG-index}). The initial polarizations of all tetrahedra correspond to the edges labelled by $w,x,y,z,r,s,t,u,v,a,b,g$ in Figure \ref{fig:octtheories}. The final boundary polarization for all octahedra is chosen to contain the four edges $w,x,y,z$ in octahedron I. At the end, we have shifted R-charges slightly (mixing them with flavor symmetries) in order to obtain a more symmetric and non-negative R-charge assignment.

The charges of the theories along the top of the loop are:
\be \CT_{\rm I}:\quad  \begin{array}{c|cccc}
 & \phi_w & \phi_x & \phi_y & \phi_z \\\hline
 U(1)_{v_1} & 1 & -1 & 0 & 0 \\
 U(1)_{v_2} & 0 & 1 & 0 & -1 \\
 U(1)_{a} & 0 & 0 & 1 & -1 \\\hline
 U(1)_R & \frac12 & \frac12 & \frac12 & \frac12 \end{array} \hspace{.5in} \CT_{\rm II}: \quad \begin{array}{c|ccccc}
   & \phi_t & \phi_r & \phi_s & \phi_y &  \phi_z \\\hline
 U(1)_{g_1} & 0 & 1 & -1 & 0 & 0 \\\hline
 U(1)_{v_1} & 0 & 0 & 0 & 0 & 0 \\
 U(1)_{v_2} & 1 & 0 & -1 & 0 & -1 \\
 U(1)_{a}   & 0 & 0& 0 & 1 & -1 \\\hline
 U(1)_R     & 1 & \frac12 & \frac12 & \frac12 & \frac12 \end{array}
\ee

\be \CT_{\rm III}:\quad  \begin{array}{c|cccc}
 & \phi_r & \phi_s & \phi_v & \phi_u \\\hline
 U(1)_{g_1} & 1 & -1 & 0 & 0 \\
 U(1)_{g_2} & 0 & 0 & 1 & -1  \\\hline
 U(1)_{v_1} & 0 & 0 & 0 & 0  \\
 U(1)_{v_2} & 0 & -1 & 0 & 1 \\
 U(1)_{a} &  0 & 0 & 0 & 0 \\\hline
 U(1)_R &  \frac12 & \frac12 & \frac12 & \frac12 \end{array}
  \hspace{.5in} \CT_{\rm IV}: \quad  \begin{array}{c|ccccc}
 & \phi_b & \phi_a & \phi_g & \phi_r & \phi_v \\\hline
 U(1)_{g_1} & 0 & 1 & -1 & 1 & 0 \\
 U(1)_{g_2}& 1 & 0 & -1 & 0 & 1 \\
 U(1)_{g_3}& 1 & -1 & 0 & 0 & 0 \\\hline
 U(1)_{v_1}& 0 & 0 & 0 & 0 & 0 \\
 U(1)_{v_2}& 0 & 0 & 0 & 0 & 0 \\
 U(1)_{a}&0 & 0 & 0 & 0 & 0 \\\hline
 U(1)_R& \frac12 & \frac12 & 1 & \frac12 & \frac12 \end{array} 
\ee

Similarly, along the bottom of the loop:
\be \CT_{\rm I}:\quad  \text{as above}
  \hspace{1.4in} \CT_{\rm VI}: \quad  \begin{array}{c|ccccc}
 & \phi_f & \phi_b & \phi_d & \phi_y &  \phi_x \\\hline
 U(1)_{g_1'} & 0 & 1 & -1 & 0 & 0 \\\hline
 U(1)_{v_1} & 1 & 0 & -1 & 0 & -1 \\
 U(1)_{v_2} & -1 & 0 & 1 & 0 & 1 \\
 U(1)_{a} & -1 & 0 & 1 & 1 & 0 \\\hline
 U(1)_R & 1 & \frac12 & \frac12 & \frac12 & \frac12 \end{array} 
\ee
\be \CT_{\rm V}:\quad  \begin{array}{c|cccc}
 & \phi_b & \phi_d & \phi_a & \phi_e \\\hline
 U(1)_{g_1'} & 1 & -1 & 0 & 0 \\
 U(1)_{g_2'} & 0 & 0 & 1 & -1  \\\hline
 U(1)_{v_1} & 0 & -1 & 0 & 1  \\
 U(1)_{v_2} & 0 & 1 & 0 & -1 \\
 U(1)_{a} & 0 & 1 & 0 & -1  \\\hline
 U(1)_R &  \frac12 & \frac12 & \frac12 & \frac12 \end{array} \hspace{.5in} \CT_{\rm IV}': \quad  \begin{array}{c|ccccc}
 & \phi_b & \phi_a & \phi_g & \phi_r & \phi_v \\\hline
 U(1)_{g_1'} & 1 & 0 & -1 & 0 & 1 \\
 U(1)_{g_2'}& 0 & 1 & -1 & 1 & 0 \\
 U(1)_{g_3'}& 0 & 0 & 0 & 1 & -1 \\\hline
 U(1)_{v_1}& 0 & 0 & 0 & 0 & 0 \\
 U(1)_{v_2}& 0 & 0 & 0 & 0 & 0 \\
 U(1)_{a}& 0 & 0 & 0 & 0 & 0 \\\hline
 U(1)_R& \frac12 & \frac12 & 1 & \frac12 & \frac12 \end{array} 
\ee
The two instances of theory IV that appear here ($\CT_{\rm IV}$ and $\CT_{\rm IV}'$) are identical, as they must be. They are simply related by a redefinition of the gauge fields. Writing the field strengths as $\mb g_i$, $\mb g_i'$, etc, the necessary redefinition is:
\be \label{eq:isoA}
{\bf g_{1}'} = {\bf g_2} + {\bf g_3}\,,\qquad 
{\bf g_{2}'} = {\bf g_1} - {\bf g_3}\,,\qquad 
{\bf g_{3}'} = {\bf g_3}
\ee

We present the polynomials in the conventions of \cite{DGP-bdy}. For each octahedron theory $\CT$, we write $\CA_{\CT} := \CA - \CA_{\CT_{I}}$, where $\CA$ is the full anomaly polynomial of $\CT$ and we subtract the contribution from the initial theory $\CT_{I}$ for neatness of presentation. The shift by the contribution of the four ungauged chirals is easily written down, but does not affect any of our computations.
\begin{align*}\label{eq:anomalies}
\CA[\CT_{\rm I}] &= 0 \qquad \text{(convention)} \\
\CA[\CT_{\rm II}]&= 2 {\bf g_1 v_1} - {\bf g_1 v_2} + {\bf v_1 v_2} - \frac12{\bf v_2}^2\\
\CA[\CT_{\rm III}]&= 2 {\bf a g_2} 
+ 2 {\bf g_1 v_1} - {\bf a v_2} - {\bf g_1 v_2} + {\bf g_2 v_2} 
 + {\bf v_1 v_2} - {\bf v_2}^2 \\
\CA[\CT_{\rm IV}]&= -\frac12{\bf g_1}^2 + 2{\bf a} {\bf g_2} + {\bf g_1 g_2} - \frac12{\bf g_2}^2 + {\bf g_1 g_3} - {\bf g_2 g_3} 
+ 2{\bf g_1 v_1} - {\bf a v_2} \\ & \ \ \ \ \ - 2 {\bf g_1 v_2} + 2{\bf g_2 v_2} +  2{\bf g_3 v_2} 
+ {\bf v_1 v_2} - {\bf v_2}^2\\
\CA[\CT_{\rm VI}] &= -\frac12{\bf a}^2 + {\bf a g_{1}'} 
+ {\bf g_{1}' v_1} + \frac12{\bf v_1}^2 - {\bf a v_2} + {\bf g_{1}' v_2} 
 - \frac12{\bf v_2}^2\\
\CA[\CT_{\rm V}]&= {\bf a} {\bf g_{1}}' + {\bf a} {\bf g_{2}}' 
 + {\bf g_{1}' v_1} + {\bf g_{2}' v_1} 
 - {\bf a v_2} + {\bf g_{1}' v_2} - {\bf g_{2}' v_2} 
 + {\bf v_1 v_2} - {\bf v_2}^2\\
 \CA[\CT_{\rm IV}'] &= 2{\bf a}{\bf g_1'}-\frac12 {\bf g_1'}^2 + {\bf g_1'} {\bf g_2'} - \frac12 {\bf g_2'}^2 - 2 {\bf a} {\bf g_3'} +{\bf g_1'} {\bf g_3'} - {\bf g_2'} {\bf g_3'} +2 {\bf g_2'} {\bf v_1} + 2 {\bf g_3'} {\bf v_1} \\ &  - {\bf a} {\bf v_2} + 2 {\bf g_1'} {\bf v_2} -2 {\bf g_2'} {\bf v_2} - 2 {\bf g_3'} {\bf v_2} + {\bf v_1} {\bf v_2} - {\bf v_2}^2 
\end{align*}

\subsection{Symmetry}
\label{app:sym}

The sequence of (2,3) and (3,2) Pachner moves performed in the bottom part of the loop is related to the sequence in the top part of the loop by a simple symmetry. The symmetry is evident in the geometry of Figure \ref{fig:octtheories}: if we permute tetrahedra
\be \label{tet-permute} \begin{array}{ccc}
\text{top} && \text{bottom} \\
 w,x,y,z & & w,z,y,x \\
 t,s,u   &\;\leftrightarrow\;& f, d, e \\
 a,b,g,v,r && v,r,g,a,b \end{array}
 \ee
then the bottom looks just like the top. 

The symmetry becomes manifest at the level of associated class-$\mathcal R$ theories, \emph{provided that} we reparameterize gauge and flavor symmetries to match the permutations of chirals. The reparameterization of gauge symmetries is just the relation between ${\bf g_i}$ and ${\bf g_i}'$ in \eqref{eq:isoA} above. The reparameterization of flavor symmetries is nontrivial; it is determined by $w,x,y,z\leftrightarrow w,z,y,x$ to be
\be {\bf v_1'}={\bf v_1}\,,\qquad {\bf v_2'}={\bf v_1}-{\bf a}-{\bf v_2}\,,\qquad {\bf a'}={\bf a}\,. \ee
For example one can easily check that after rewriting the charges of the chirals in $\CT_{\rm VI},\CT_{\rm V},\CT_{\rm IV}'$ entirely in terms of ${\bf v_1'},{\bf v_2'} ,{\bf a'}, {\bf g_i'}$, and implementing the permutation \eqref{tet-permute}, one recovers the charges of $\CT_{\rm II},\CT_{\rm III},\CT_{\rm IV}$. The same is true of Chern-Simons levels.

\section{Useful functions for (half-)indices}
\label{app:fns}

We define the basic functions that appear in the 3d $\CN=2$ half-index and the 2d $\CN=(0, 2)$ elliptic genera, mostly following notation introduced in \cite{DGP-bdy}. We refer to the latter for the prescription to compute the half-index from these building blocks and, \emph{e.g.}, \cite{benini2014elliptic, benini2015elliptic, GaddeGukov} for the corresponding computation of elliptic genera.
\begin{align*}
(x;q)_{\infty} &:= \prod_{n = 0}^{\infty}(1 - q^n x) \\
(q)_{\infty} &:= (q; q)_{\infty} \\
\half_D(x; q) &:= (q/x;q)_{\infty} \\
\half_N(x; q) &:= (x;q)_{\infty}^{-1} \\
\text{F}(x; q) &:= (x q^{1/2};q)_{\infty}(q^{1/2}/x;q)_{\infty} \\
\text{C}(x; q) &:= {1 \over (x;q)_{\infty} (q x^{-1};q)_{\infty}}
\end{align*}
We will typically suppress the second, $q$ argument of $\Pi_{D, N}(x; q)$ and $\text{F}(x; q), \text{C}(x; q)$ in our index formulas for concision. 

\section{Sandwich theories for the (3,3) move}
\label{app:sandwich}

The sandwich $\CT_{\rm top} = \CB_{\rm I}\circ \CI_{\rm top}\circ \CB_{\rm IV}$ described in \eqref{top-sandwich}, which is a purely 2d $\CN=(0,2)$ theory encoding (half of) the (3,3) Pachner move, is described in detail as follows. The 2d gauge group is $U(1)_{g_1}\times U(1)_{g_2}$, and the flavor symmetry is $U(1)_{g_3}\times U(1)_{v_1}\times U(1)_{v_2}\times U(1)_a$. There are six fermi multiplets and four chiral multiplets with charges
\be
\begin{array}{c|cccccc|cccc}
 &  \Gamma & \Gamma' & \Gamma'' & \Psi_t & \eta & \eta'  & \Phi_s & \Phi_u  & \Phi_a & \Phi_v  \\ \hline
 U(1)_{g_1} & 1&0&0& 0 &1&0 &-1&0& 1&0 \\
 U(1)_{g_2} & 0&-1&0& 0 &0&1 &0&-1& 0&1 \\ \hline
 U(1)_{g_3} & 0&0&-1& 0 &-1&0 &0&0& -1&0 \\ 
 U(1)_{v_1} &1&0&0& 0 &-1&0 &0&0& 0&0 \\
 U(1)_{v_2} &0&0&-1& -1 &1&-1 &-1&1& 0&0 \\
 U(1)_{a}  &0&-1&0& 0 &0&-1 &0&0&  0&0 \\ \hline
 U(1)_R    &0&0&0& 0 &0&0 &\frac12&\frac12 &\frac12&\frac12
\end{array}
\ee
There are also $J$ and $E$ terms induced from the collision of boundaries, given by
\be \begin{array}{c|cccccc} 
 & \Gamma & \Gamma' & \Gamma'' & \Psi_t & \eta&\eta' \\\hline
J & 0&0 & 0 & \Phi_u\Phi_v  &0&0 \\
E & 0&0&-\Phi_s\Phi_a&0&0&0 
\end{array}
\ee

The sandwich $\CT_{\rm bot} = \CB_{\rm I}\circ \CI_{\rm bot}\circ \CB_{\rm IV}$ described in \eqref{bot-sandwich} has essentially identical field content, but different charges under flavor and gauge symmetries:
\be
\begin{array}{c|cccccc|cccc}
 &  \Gamma & \Gamma' & \Gamma'' & \Psi_f & \eta & \eta'  & \Phi_d & \Phi_e  & \Phi_a & \Phi_v  \\ \hline
 U(1)_{g_1} & 0&-1&0& 0 &1&0 &0&-1& 1&0 \\
 U(1)_{g_2} & 1&0&0& 0 &0&1 &-1&0& 0&1 \\ \hline
 U(1)_{g_3} & 1&1&-1& 0 &-1&0 &-1&1& -1&0 \\ 
 U(1)_{v_1} &1&0&-1& -1 &-1&0 &-1&1& 0&0 \\
 U(1)_{v_2} &0&0&1& 1 &1&-1 &1&-1& 0&0 \\
 U(1)_{a}  &0&-1&1& 1 &0&-1 &1&-1&  0&0 \\ \hline
 U(1)_R    &0&0&0& 0 &0&0 &\frac12&\frac12 &\frac12&\frac12
\end{array}
\ee
The $J$ and $E$ terms induced from the collision of boundaries are given by
\be \begin{array}{c|cccccc} 
 & \Gamma & \Gamma' & \Gamma'' & \Psi_f & \eta&\eta' \\\hline
J & 0&0 & 0 & \Phi_e\Phi_a  &0&0 \\
E & 0&0&-\Phi_d\Phi_v&0&0&0 
\end{array}
\ee

\bibliographystyle{JHEP}
\bibliography{33move}

\end{document}